\documentclass[12pt,epsf,amssymb]{article}

\usepackage{graphicx}
\usepackage{amsfonts}
\usepackage{amsmath}
\usepackage[usenames]{color}
\usepackage{amssymb}
\usepackage[mathscr]{eucal}
\usepackage{url}

\usepackage{hyperref}
\hypersetup{
 setpagesize=false,
 bookmarksnumbered=true,%
 colorlinks=false,%
}

\usepackage[all]{xy}

\def\N{\mathbb{N}}
\def\Z{\mathbb{Z}}
\def\Q{\mathbb{Q}}
\def\R{\mathbb{R}}
\def\C{\mathbb{C}}
\def\P{\mathbb{P}}

\def\Im{\mathrm{Im}}

\usepackage{here}

\newcommand{\UH}{\mathcal{H}}
\newcommand{\const}{\mathrm{const.}}

\newcommand{\Li}{\mathrm{Li}}

\newcommand{\parren}[1]{\left( #1 \right)}
\newcommand{\braces}[1]{\left\{ #1 \right\}}
\newcommand{\sqbra}[1]{\left[ #1 \right]}
\newcommand{\abs}[1]{\left| #1 \right|}

\newcommand{\pdiff}[3][]{\dfrac{\partial^{#1}{#2}}{\partial{#3}^{#1}}}

\newcommand{\rank}{\operatorname{rank}}
\newcommand{\tensor}{\otimes}
\newcommand{\isom}{\cong}

\newcommand{\NRS}{\widetilde{\partial}^S}

\newcommand{\sgn}{\operatorname{sgn}}
\newcommand{\Mod}{\operatorname{Mod}}
\newcommand{\Tr}{\operatorname{Tr}}

\newcommand{\mymid}{\mathrel{}\middle|\mathrel{}}

\renewcommand{\Im}{\operatorname{Im}}

\usepackage{amsthm}
\theoremstyle{definition}

\begin{document}

\baselineskip 0.6cm
\newcommand{\vev}[1]{ \left\langle {#1} \right\rangle }
\newcommand{\bra}[1]{ \langle {#1} | }
\newcommand{\ket}[1]{ | {#1} \rangle }
\newcommand{\Dsl}{\mbox{\ooalign{\hfil/\hfil\crcr$D$}}}
\newcommand{\nequiv}{\mbox{\ooalign{\hfil/\hfil\crcr$\equiv$}}}
\newcommand{\nsupset}{\mbox{\ooalign{\hfil/\hfil\crcr$\supset$}}}
\newcommand{\nni}{\mbox{\ooalign{\hfil/\hfil\crcr$\ni$}}}
\newcommand{\nin}{\mbox{\ooalign{\hfil/\hfil\crcr$\in$}}}
\newcommand{\Slash}[1]{{\ooalign{\hfil/\hfil\crcr$#1$}}}
\newcommand{\EV}{ {\rm eV} }
\newcommand{\KEV}{ {\rm keV} }
\newcommand{\MEV}{ {\rm MeV} }
\newcommand{\GEV}{ {\rm GeV} }
\newcommand{\TEV}{ {\rm TeV} }

\def\diag{\mathop{\rm diag}\nolimits}
\def\tr{\mathop{\rm tr}}

\def\Spin{\mathop{\rm Spin}}
\def\SO{\mathop{\rm SO}}
\def\SU{\mathop{\rm SU}}
\def\U{\mathop{\rm U}}
\def\Sp{\mathop{\rm Sp}}
\def\SL{\mathop{\rm SL}}

\def\change#1#2{{\color{blue}#1}{\color{red} [#2]}\color{black}\hbox{}}

\begin{titlepage}

\begin{flushright}
IPMU19-0168
\end{flushright}

\vskip 1cm
\begin{center}

  {\large \bf Modular Forms as Classification Invariants of 4D {\cal N}=2 Heterotic--IIA Dual Vacua}

 \vskip 1.2cm

Yuichi Enoki and Taizan Watari,

 \vskip 0.4cm

  {\it Kavli Institute for the Physics and Mathematics of the Universe (WPI),
     the University of Tokyo, Kashiwa-no-ha 5-1-5, 277-8583, Japan}

 \vskip 1.5cm

 \abstract{
 We focus on 4D $\mathcal{N}=2$ string vacua described both by
perturbative Heterotic theory and by Type IIA theory; a Calabi--Yau three-fold
$X_{\rm IIA}$ in the Type IIA language is further assumed to have a regular K3-fibration.
It is well-known that one can assign a modular form $\Phi$ to such a vacuum
by counting perturbative BPS states in Heterotic theory or collecting
Noether--Lefschetz numbers associated with the K3-fibration of $X_{\mathrm{IIA}}$.
In this article, we expand the observations and ideas (using gauge threshold
correction) in the literature and formulate a modular form $\Psi$
with full generality for the class of vacua above, which can be used
along with $\Phi$ for the purpose of classification of those vacua.
Topological invariants of $X_{\mathrm{IIA}}$ can be extracted from $\Phi$ and
$\Psi$, and even a pair of diffeomorphic Calabi--Yau's with different
K\"{a}hler cones may be distinguished by introducing the notion of
``the set of $\Psi$'s for Higgs cascades/for curve classes''.
We illustrated these ideas by simple examples.
 }

\end{center}
\end{titlepage}

\tableofcontents

\section{Introduction}

The duality between the Heterotic string theory and the Type IIA string theory
has been known for a long time. The duality with $\SO(5,1)$ symmetry and
16 supersymmetry charges---the duality at 6D---comes with just one piece
of moduli space
$[{\rm Isom}({\rm II}_{4,20}) \backslash \SO(4,20)/\SO(4) \times \SO(20)] \times \R_{>0}$,
and its various aspects are understood very well \cite{HetIIA-6D}.
The duality with $\SO(3,1)$ symmetry and 8 supersymmetry
charges \cite{KV}---the duality at 4D---is less understood.
The moduli space of Heterotic--IIA dual 4D vacua forms
a complicated network of branches. It is desirable that those individual
branches are characterized both in the languages of Heterotic string
and Type IIA string, and the dictionary between the
branch-characterizing data on both sides are understood. At the moment,
we do not have one for the 4D Heterotic--IIA duality\footnote{
An exception is for the case $\Lambda_S=U$ (see main text for what it is),
when the duality lifts to the Heterotic--F-theory duality at 6D.}
as clear as Batyrev's dual polyhedra for mirror symmetry.

For a systematic approach, we need to find invariants characterizing
the branches of the moduli space.
A lattice pair $\widetilde{\Lambda}_S \oplus \Lambda_T$ that fits into
${\rm II}_{4,20}$ is assigned for those
branches \cite{Ferrara:1995yx, KLM, VW-95, Aspinwall:1995vk},
and a modular form $\Phi$ of certain type that depends on $\Lambda_S$
is also assigned \cite{HM, MP}. It is known, however, that there are
physically distinct branches of vacua that cannot be distinguished by
the triple of invariants $(\Lambda_S, \Lambda_T, \Phi)$.
The primary purpose of this article is to introduce more invariants
by using modular forms to improve the state of affairs.

From the perspective of pure mathematics, this task is equivalent to
classification of Calabi--Yau three-folds with a K3-fibration.
The modular forms introduced in this article can be used therefore
for study of such a geometry classification. It should be mentioned,
however, that we consider only regular K3-fibrations in this article.

\bigskip

The organization of this article is as follows.
We begin in section \ref{ssec:Het-II-review} with a short review on
the Heterotic--Type IIA duality and a summary of technical limitations
on the class of vacua to be considered in this article.
%
%
The modular form $\Phi$ is parametrized by low-energy BPS indices, which are
bounded from below as we argue in
section \ref{ssec:gamma_0-level-quantization-condition};
in the Heterotic language, the bounds come from the quantization of the level
of current algebra and the spin under the $SU(2)$ action. In sections \ref{ssec:examples-of-Phi} and \ref{ssec:bd-EulerN}, those bounds and the modular property
are combined to constrain the possibilities of $\Phi$ and the Euler number
of a Calabi--Yau three-fold $X_{\mathrm{IIA}}$ that compactifies the Type IIA theory.

We formulate in section \ref{sec:finerC} a modular form $\Psi$ for $X_{\rm IIA}$
and a little more data, and use it to define new invariants for $X_{\rm IIA}$
in addition to $(\Lambda_S,\Lambda_T, \Phi)$.
%
Section \ref{sssec:Higgs-cascade} includes the basic definition of $\Psi$, which follows from the idea of \cite{HM} and subsequent works.
This modular form appears in the integrand of the 1-loop gauge threshold correction in the Heterotic string.
In section \ref{sssec:space-of-Psi} we comment on some restrictions on the degree of freedom of $\Psi$, that comes from
its modular property and some physical constraints.
We introduce a map $({\rm diff}_{\rm coarse},{\rm diff}_{\rm fine})$
in section \ref{sssec:matching-summary}, which extracts from $(\Phi, \Psi)$
the full information in specifying the diffeomorphism
class of a Calabi--Yau three-fold $X_{\rm IIA}$.
%
%
Combining this map with the degree-of-freedom study of $\Phi$ and $\Psi$
in sections \ref{sec:coarseC} and \ref{sssec:space-of-Psi}, we can use
modular property of $\Phi$ and $\Psi$ to obtain non-trivial results
in the diffeomorphism classes of real six-dimensional manifolds realized
by Calabi--Yau three-folds. We also propose to use the notion
of ``the set of $\Psi$'s for Higgs cascades'' or
of ``the set of $\Psi$'s for curve classes'' as an
invariant that may resolve a diffeomorphic pair of Calabi--Yau three-folds
with different cone of curves.
In section 3.2, all those ideas are illustrated by simple examples.

In section 4 we comment on some open questions.

The appendix A contains basics about (vector-valued) modular forms and explicit Fourier expansions of modular forms in the main text. In appendix B
we review the lattice unfolding method and the embedding trick of Borcherds's \cite{Bor-Grass}, presented in a form we need for
threshold calculations in the Heterotic theory.
The embedding trick is used in explicitly evaluating  integrals, for example in the case of $\Lambda_S = \vev{+2}$ in appendix \ref{sssec:unfold-deg2}.
%

\section{Coarse Classification}
\label{sec:coarseC}

\subsection{A Brief Review}
\label{ssec:Het-II-review}

Let us first review what is known in the literature about
the classification using the new supersymmetric index / the
generating function of the Noether--Lefschetz number.

\subsubsection{Heterotic Description: the New Supersymmetric Index}
\label{sssec:review-Het}

A Heterotic string compactification to 3+1-dimensions has an unbroken
${\cal N}=2$ supersymmetry (8 supersymmetry charges), if and only if
the right-mover of the internal worldsheet CFT
contains an $N=4$ superconformal algebra (SCA)
with central charge $\tilde{c}=6$ and an $N=2$ free SCA with $\tilde{c}=3$
corresponding to a flat space of one complex dimension \cite{BD,LLT}.
We restrict our attention in this article only to compactifications without
an NS5-brane or its generalizations discussed in \S5 of \cite{BW-16}.

Let $\rho$ be the number of free chiral bosons in the left-mover
in such a compactification. There are vertex operators of the form
$e^{i p_L \cdot X_L + i p_R \cdot X_R}$ in the $(c,\tilde{c}) = (22,9)$ CFT, where $X_L$ and $X_R$
are the $\rho+2$ chiral bosons in the left-mover and right-mover;
the set of U(1) charges $\{ (p_L,p_R) \} = \widetilde{\Lambda}_S$
forms a lattice with the quadratic form given\footnote{
We use the convention $\alpha' = 2$. } by $p_R^2/2-p_L^2/2$, so its
signature is $(+,-)=(2,\rho)$.
%
%
%
This lattice $\widetilde{\Lambda}_S$ should be even, $p_R^2/2-p_L^2/2 \in \Z$
for any element of $\widetilde{\Lambda}_S$, since the contribution of the state $e^{i p_L \cdot X_L + i p_R \cdot X_R}$ to the partition function should be invariant under $T : \tau \to \tau + 1$.
The $U(1)$ charge of any worldsheet operator should lie in\footnote{
\label{fn:all-charge-exist-assume}
We assume any charge $v \in \widetilde{\Lambda}_S^\vee$ is realized by
some state. The Type IIA counter part of this assumption is that the
pairing $(H^2(X;\Z)/\Z D_s) \times [H_2(X;\Z)]^{\rm vert} \rightarrow \Z$
is represented by the unit matrix; see page \pageref{pg:IIA-notation}
for notations.}
$\widetilde{\Lambda}_S^\vee$, the dual lattice of $\widetilde{\Lambda}_S$.
Note that we deal with cases where $\widetilde{\Lambda}_S$ is not necessarily unimodular, so that the discriminant group $G_S := \widetilde{\Lambda}_S^\vee/\widetilde{\Lambda}_S$ may be non-trivial.
We assume, however, that $\widetilde{\Lambda}_S$ is a primitive sublattice
of\footnote{The even unimodular lattice of signature (1,1) is denoted by $U$.
We use the same notation ${\cal R}$ in this article for one of ADE types,
its Lie algebra, and its root lattice with positive definite signature.
} ${\rm II}_{4,20}=U^{\oplus 4} \oplus E_8[-1]^{\oplus 2}$; the orthogonal
complement $[\widetilde{\Lambda}_S^\perp \subset {\rm II}_{4,20}]$ is denoted
by $\Lambda_T$.

For example, when we compactify the Heterotic theory on $K3 \times T^2$ with
instantons in $\mathfrak{g} \subset E_8 \times E_8$, the lattice
$\widetilde{\Lambda}_S$ is equal to $U^{\oplus 2} \oplus W$, where
$W = [\mathfrak{g}^\perp \subset E_8[-1]^{\oplus 2}]$.

The Hilbert space of the internal CFT can be decomposed using the action of the free boson algebra and the $N=4$ right-mover SCA \cite{AFGNT, HM}:
\begin{align}
    \mathcal{H}^{\mathrm{int}} =
    \bigoplus_{(\gamma,\tilde{h},\tilde{I})}
      \mathcal{H}^{(22-\rho,0)}_{\gamma,(\tilde{h},\tilde{I})} \tensor
      \mathcal{H}^{(\rho,3)}_{\gamma} \tensor
      \mathcal{H}^{(0,6)}_{\tilde{h},\tilde{I}},
\end{align}
where superscripts indicate the central charge $(c,\tilde{c})$.
The rank-$(\rho+2)$ $U(1)$ charge in $\widetilde{\Lambda}_S$ is denoted
by $v$, and $\gamma$ is its corresponding element
$v+ \widetilde{\Lambda}_S$
in the discriminant group $G_S$.
The vector space $\mathcal{H}^{(\rho,3)}_{\gamma}$ consists of states of charge $v \in \gamma$
with any free bosonic/fermionic oscillator excitations. A pair
$(\tilde{h},\tilde{I})$ of conformal weight and $SU(2)$ spin labels a
unitary irreducible representation $\mathcal{H}^{(0,6)}_{\tilde{h},\tilde{I}}$
of the $N=4$ SCA.
The spectrum of the free sector $\mathcal{H}^{(\rho,3)}_{w}$ is specified
by the central charge $p_R^\C: \widetilde{\Lambda}_S^\vee \rightarrow \C$,
which appears in the 4D $\mathcal{N}=2$ supersymmetry algebra.
At the Heterotic string perturbative level, $p_R^\C$ is governed by the
Coulomb branch moduli space
\begin{align}
  D(\widetilde{\Lambda}_S) := \P \left\{ \mho \in \widetilde{\Lambda}_S \otimes \C \; |
  \; (\mho,\mho)_{\widetilde{\Lambda}_S} = 0, \; ( \mho, \overline{\mho} )_{\widetilde{\Lambda}_S} > 0 \right\},
  \label{eq:def-Narain-period-dom}
\end{align}
which constitutes the special geometry along with the dilaton complex
scalar $s:= 4\pi i S$; the weak coupling limit\footnote{The real and
imaginary components of $s$ are denoted by $s_{1,2}$. Similar notations
($t_2$, $\tau_{1,2}$, $\rho_2$ etc.) are used throughout this article. } is
$s_2 \gg 1$.
The rest of spectrum information (i.e. the spectrum of
$\mathcal{H}^{(22-\rho,0)}_{w,(\tilde{h},\tilde{I})}$ and
$\mathcal{H}^{(0,6)}_{\tilde{h},\tilde{I}}$
for each possible
$(w;\tilde{h},\tilde{I})$) depends also on the hypermultiplet moduli space.

The new supersymmetric index \cite{newSUSY, HM} of the internal CFT is
defined by\footnote{To be more precise, only the trace part is called as
the new supersymmetric index of the internal CFT. The $1/\eta^2$ factor
is included within $Z_{\rm new}$ here because the trace part appears
in $\Delta_{\rm grav}$ in the combination (\ref{eq:def-Znew}); the $1/\eta^2$
factor is from the 4D Minkowski part in the light-cone gauge.
}
\begin{align}
  Z_{\mathrm{new}}(\tau,\bar\tau) :=
     \frac{-i}{\eta(\tau)^2} \Tr_{\text{R-sector}}^{(22,9)} \left[ e^{\pi i F_R}F_R\; q^{L_0-\frac{c}{24}}\bar{q}^{\tilde{L}_0-\frac{\tilde{c}}{24} }  \right],
  \label{eq:def-Znew}
\end{align}
where $F_R : = 2(\tilde{J}^{\tilde{c}=6}_3)_0 + (\tilde{J}^{\tilde{c}=3})_0$ is
the zero mode of the total $U(1)$ current in the right-mover; $\tilde{J}^{\tilde{c}=6}_3$ is
the $SU(2)$ Cartan current in the $N=4$ SCA and $\tilde{J}^{\tilde{c}=3}$
the $U(1)$ current in the $N=2$ SCA.
An important point is that this index does not depend on any continuous deformations of the hypermultiplet moduli, although it does on that of vector multiplet moduli. In addition, it is used in computing the 1-loop correction $\Delta_{\rm grav}$ to the gravitational
coupling $\sqrt{-g}R^2$ in the 4D effective theory \cite{ANT-letter, AGN-1, AGNT-threshold};
\begin{align}
\Delta_{\rm grav} = \int \frac{d\tau_1 d\tau_2}{\tau_2}
    \left( {\cal B}_{\rm grav} - b_{\rm grav} \right),
        \qquad   {\cal B}_{\rm grav} = Z_{\mathrm{new}} \hat{E}_2,
   \label{eq:def-Delta-grav}
\end{align}
where the integration is over the fundamental region of ${\rm SL}(2;\Z)$ in the upper complex
half plane (of the torus world sheet complex structure $\tau$), and
$
 \hat{E}_2 := E_2 - \frac{3}{\pi \tau_2}
$
is a non-holomorphic modular form of weight $(2,0)$.
The constant $b_{\rm grav}$ is set to the $q^0$ coefficient of ${\cal B}_{\rm grav}$,
to cut off the IR divergent 1-loop contributions and brings the massless degrees of
freedom back into the path integration in low energy effective theory.
The index $Z_{\mathrm{new}}$ must have weight $(-1,1)$ due to the modular invariance of the integrand.

The action of the free boson algebra on $\mathcal{H}^{\mathrm{int}}$ leads to the following decomposition
\begin{align}
 & Z_{\mathrm{new}}(\tau,\bar\tau) = \sum_{\gamma \in G_S} \theta_{\widetilde{\Lambda}_S[-1]+\gamma}(\tau, \bar{\tau})
   \frac{\Phi_{\gamma}(\tau)}{\eta(\tau)^{24}},
  \label{eq:Het-Isusy=Z-Phi-eta24}
 \\ &
  \theta_{\widetilde{\Lambda}_S[-1]+\gamma}(\tau,\bar{\tau}) :=
  \sum_{v \in \widetilde{\Lambda}_S+\gamma}
  q^{p_L^2(v)/2} \bar{q}^{p_R^2(v)/2},
\end{align}
where
$\theta_{\widetilde{\Lambda}_S[-1]+\gamma}$
is the Siegel theta function, which describes the whole continuous dependence of the index $Z_{\mathrm{new}}$ on the vector multiplet moduli, while $\Phi_\gamma$ allows only discrete choices as seen below.
This Siegel theta function is a vector-valued modular form of weight $(\rho/2,1)$ and type $\rho_{\widetilde{\Lambda}_S[-1]}$.
Similarly, $\Phi = \sum_{\gamma \in G_S} e_\gamma \Phi_\gamma$ lies in $\Mod(11-\rho/2, \rho_{\widetilde{\Lambda}_S})$.\footnote{
In fact, we impose a stronger restriction on $\Phi$. See \eqref{eq:restriction-3-Het}.
}
In particular, $\Phi$ is holomorphic at cusps, i.e. has no negative power of $q$ in its expansion. Transformation law under $T:\tau\to\tau+1$ fixes the fractional part of power of $q$, so $\Phi_\gamma/\eta^{24}$ can be expanded as\footnote{
Note that $c_\gamma(\nu) = c_{-\gamma}(\nu)$ follows from properties of the Weil representation. See appendix \ref{apdx:A1}.
}
\begin{align}
    \frac{\Phi_\gamma(\tau)}{\eta^{24}} = \sum_{\nu \in \gamma^2/2 + \Z} c_\gamma(\nu) q^\nu, \qquad c_\gamma(\nu) = 0 \quad\text{for}\quad \nu < -1.
    \label{eq:def_of_c}
\end{align}

Non-zero contributions to $Z_{\mathrm{new}}$ comes only from those states whose right-movers are given by
the (Ramond) ground states of the $\tilde{c}=3$ sector and the short representations of the $N=4$ SCA in the $\tilde{c}=6$ sector.
There are two types of short representations of the $N=4$ SCA: vector-type and hyper-type.
Their quantum numbers $(\tilde{h},\tilde{I})$ are given by the following table.
\begin{center}
    \begin{tabular}{ccc}
                     & R-sector    & NS-sector \\ \hline
        vector-type: & $(1/4,1/2)$ & $(0,0)$ \\
        hyper-type:  & $(1/4,0)$   & $(1/2,1/2)$
    \end{tabular}
\end{center}
Their contributions can be written as \cite{HM}
\begin{align}
    \frac{\Phi_\gamma}{\eta^{24}} =
    -2 \times \frac{1}{\eta^2} \Tr_{\mathcal{H}^{(22,0)}_{\gamma, V}}
    \sqbra{ q^{L_0-\frac{c}{24}} }
    + 1 \times \frac{1}{\eta^2} \Tr_{\mathcal{H}^{(22,0)}_{\gamma, H}}
    \sqbra{ q^{L_0-\frac{c}{24}} },
    \label{eq:ContributionsToPhi}
\end{align}
where
$\mathcal{H}_{\gamma,(\tilde{h},\tilde{I})}^{(22,0)}
:= \mathcal{H}_{\gamma,(\tilde{h},\tilde{I})}^{(22-\rho,0)} \tensor \mathcal{H}^{(\rho,0)}$, with
$\mathcal{H}^{(\rho,0)}$ being the space of the (neutral) left-moving oscillations in the $c=\rho$ sector, and
\begin{align}
    & \mathcal{H}_{\gamma,V}^{(22,0)} := \mathcal{H}_{\gamma,(1/4,1/2)}^{(22,0)} = \mathcal{H}_{\gamma,(0,0)}^{(22,0)},
    \\
    & \mathcal{H}_{\gamma,H}^{(22,0)} := \mathcal{H}_{\gamma,(1/4,0)}^{(22,0)} = \mathcal{H}_{\gamma,(1/2,1/2)}^{(22,0)}.
\end{align}
The second equality in each line comes from
the spectral flow of the $N=4$ SCA that brings the representations
in the R-sector to those in the NS-sector.
The coefficients $-2$ and $+1$ in \eqref{eq:ContributionsToPhi} correspond to the Witten index of the representations $(1/4,1/2)$ and $(1/4,0)$, respectively \cite{ET,EOTY}:
\begin{align}
    \Tr_{(1/4,1/2)} (-1)^{2(\tilde{J}^{\tilde{c}=6}_3)_0} = -2,
    \qquad
    \Tr_{(1/4,0)} (-1)^{2(\tilde{J}^{\tilde{c}=6}_3)_0} = +1.
    \label{eq:Witten_index_of_short_rep}
\end{align}

It follows from \eqref{eq:ContributionsToPhi}, in particular, that all the Fourier coefficients $c_\gamma(\nu)$ are integers, so we have only discrete choices of $\Phi$. In fact, since $\Phi/\eta^{24}$ has a negative weight as a modular form, it can be uniquely specified\footnote{
In fact, we assume some $n_\gamma$ to be zero. See the comments around \eqref{eq:restriction-3-Het}.
}
by the coefficients $c_\gamma(\nu)$ with $\nu < 0$, i.e.
\begin{align}
    n_\gamma := -2 n_\gamma^V + n_\gamma^H
    = c_\gamma([\gamma^2/2]_{\mathrm{frac}}-1),
    \quad
    \gamma \in G_S.
\end{align}
Here the fractional part $[x]_{\mathrm{frac}}$ of $x \in \R$ is defined by $[x]_{\mathrm{frac}} \equiv x \bmod \Z$ and $0 \leq [x]_{\mathrm{frac}} < 1$. $n_\gamma^{V/H}$ is the number of states with conformal weight $[\gamma^2/2]_{\mathrm{frac}}$ in $\mathcal{H}_{\gamma,V/H}^{(22,0)}$.

We can deduce $n_{\gamma = 0} = -2$ from the supersymmetry constraints.
First, the uniqueness of the ground state forces $n_0^V = 1$;
$n_0^V$ is the number of states in $\mathcal{H}_{\gamma = 0,V}^{(22,0)}$ with conformal weight $0$.
Such state, tensored with the right-moving highest weight state of $(\tilde{h},\tilde{I})=(0,0)$, gives the ground state in the Hilbert space of the internal CFT.
Second,\footnote{
$X^{\mu=2,3}$ are the scalar operators representing  Minkowski coordinates under the light cone gauge.
}
$\partial X^{\mu=2,3}$ tensored with the right-moving Ramond ground states (in the $\tilde{c}=3$ sector) and the highest weight states of $(\tilde{h},\tilde{I})=(1/4,1/2)$ (in the $\tilde{c}=6$ sector) gives the exactly required number of gravitino states for 4D $\mathcal{N} = 2$ supersymmetry.
If $n^H_0 > 0$, the 4D effective theory would have ${\cal N} = 2+n_0^H$ supersymmetry in a similar way.
Since we focus on the case with only 8 supercharges, $n^H_0$ should be zero, so $n_0 = -2$.
For the explanation of $n_0 = -2$ from the Type IIA perspective,
see section \ref{sssec:review-IIA}.

The BPS index $n_\gamma$ for $\gamma \neq 0$ also has a simple interpretation in terms of spacetime effective theory: for each $v \in \gamma \subset \widetilde{\Lambda}_S^\vee$ such that $-2 \leq v^2 < 0$,
\footnote{
The condition $-2 \leq v^2$ comes from the left-right matching of conformal weights.
}
there exist $n_\gamma^V$ BPS vector multiplets and $n_\gamma^H$ BPS half-hypermultiplets,\footnote{
If $2\gamma \neq 0 \in G_S$, these $n_\gamma^H + n_{-\gamma}^H = 2n_\gamma^H$ half-hypermultiplets are combined to become $n^H_\gamma$ full hypermultiplets.
} both of which have $U(1)$-charge $v$ and BPS mass-square $p_R^2(v)$.
$v^2 < 0$ implies that these states\footnote{
In section \ref{ssec:gamma_0-level-quantization-condition}, we show that there are non-trivial constraints about $\gamma$ for BPS massless vector multiplet states to exist, coming from charge and level quantization conditions.}
 will be massless at some points in the (weak coupling) Coulomb branch moduli space $D(\widetilde{\Lambda}_S)$.
For this reason, we call $n_\gamma$'s the low-energy BPS indices in this article.

In this article, we impose some technical constraints on $\widetilde{\Lambda}_S$ and $\Phi$ for simplicity.
First, we only consider the case
\begin{align}
    \widetilde{\Lambda}_S = U[-1] \oplus \Lambda_S
    \label{eq:restriction-2-Het}
\end{align}
for some even lattice $\Lambda_S$ of signature $(1,\rho-1)$.
Note that the direct summand $U[-1]$ does not contribute to the discriminant group: $G_S = \widetilde{\Lambda}_S^\vee/\widetilde{\Lambda}_S = \Lambda_S^\vee/\Lambda_S$. The factor $U[-1]$ corresponds to $H^0(\mathrm{K3};\Z) \oplus H^4(\mathrm{K3};\Z)$ of the generic fibre of a K3-fibred Calabi-Yau manifold that compactifies the Type IIA theory. See also section \ref{sssec:review-IIA}.\footnote{
A choice of $\widetilde{\Lambda}_S$ that does not have $U[-1]$ as a direct summand may correspond to some non-geometric background of in the Type IIA language, such as mirror-folds, etc.
}

Second, we assume that if $\gamma^2/2 \equiv 0 \bmod \Z$ and $\gamma \neq 0$ then $n_\gamma = c_\gamma(-1) = 0$.
In other words, $\Phi$ has an expansion of the form $-2 q^0e_0 + (\text{strictly higher power of $q$})$.
We denote this condition\footnote{
This condition corresponds to the claim \eqref{eq:Phi'-modform-IIA} in the IIA side. Even when there is a Heterotic construction that is not subject to \eqref{eq:restriction-3-Het}, there would probably be no Type IIA dual with a geometric phase. See also section \ref{sssec:review-IIA}.
}\raisebox{4pt}{,}\footnote{This is an extra non-trivial condition only
in a lattice $\Lambda_S$ with a non-zero isotropic element $\gamma$.
Examples of such lattices include $\Lambda_S = \vev{2n^2m}$ for any $n,m\in\N$. Indeed, let $e$ be
a generator of the rank-1 abelian group $\Z$ underlying $\Lambda_S$, so
$(e,e)_{\Lambda_S}=2n^2m$. Elements such as $\gamma = e/n +\Lambda_S \in G_S$
give $\gamma^2/2 = m \equiv 0 \bmod \Z$. }
as
\begin{align}
    \Phi\in \Mod_0(11-\rho/2,\rho_{\widetilde{\Lambda}_S}).
    \label{eq:restriction-3-Het}
\end{align}
We define $h_{\mathrm{min}}(\gamma)$ so that the Fourier expansion of $\Phi_\gamma$ begins with $\mathcal{O}(q^{h_{\mathrm{min}}(\gamma)})$ (or higher):
\begin{align}
    h_{\mathrm{min}}(\gamma) \in \gamma^2/2 + \Z,
    \quad h_{\mathrm{min}}(0) = 0,
    \quad 0 < h_{\mathrm{min}}(\gamma) \leq 1 \text{ for } \gamma \neq 0.
\end{align}
We denote
\begin{align}
  G_S^< = \{ \gamma \in G_S \mid h_{\mathrm{min}}(\gamma) < 1 \},
  \quad
  d^< = \abs{G_S^</\pm}.
\end{align}
After all, the modular form $\Phi$ is specified by $n_0 = -2$ and at most $(d^<-1)$ integers
\begin{align}
    n_{\abs{\gamma}} = n_{\pm\gamma} \geq -2,
    \quad \abs{\gamma} \in G_S^</\pm,
    \quad \gamma \neq 0.
\end{align}
Sometimes, the modular properties predict linear relations among the
low-energy BPS indices $\{n_{\abs{\gamma}}\}$; see section \ref{sssec:lin-reltn}.

\subsubsection{Type IIA Description: the Generating Function of the Noether--Lefschetz Numbers}
\label{sssec:review-IIA}

In this article, we consider the Type IIA string theory compactified on a non-singular
Calabi--Yau three-fold $X=X_{\rm IIA}$ that has a K3-fibration $\pi: X \rightarrow \P^1 = \P^1_{\mathrm{IIA}}$; complexified K\"{a}hler
parameters may be analytically continued out of a geometric phase, but
otherwise we remain in a geometric phase.\footnote{
Both $R^4\pi_*\Z$ and $R^0\pi_*\Z$ are trivial rank-1 local systems on $\P^1$.}
This restriction means, in particular, that we do not treat T-folds or
mirror folds \cite{Tfold}.
This restriction corresponds to \eqref{eq:restriction-2-Het} in the
Heterotic side. We also assume that
\begin{align}
  h^{1,0}(X)=h^{2,0}(X)=0.
    \label{eq:h10=0}
\end{align}
The effective theory on 3+1-dimensions has strictly ${\cal N}=2$ supersymmetry,
not more, not less.

K3-fibres in $\pi: X \rightarrow \P^1$ may degenerate at isolated points
on the base $\P^1$. Degenerations of a K3-fibration are classified (by allowing
a base change locally) into Type I, Type II, and Type III \cite{Kulikov}.
When a K3-fibration $\pi: X \rightarrow \P^1$ only has degenerations classified
as Type I,\footnote{We avoid saying ``only has Type I fibres'' here,
because such expressions as ``Type I fibre'' are reserved only for the central
fibres after the local geometry of $\pi: X \rightarrow \P^1$ around a degeneration point $p \in \P^1$ is brought into a Kulikov model by a base change around $p$.} such a K3-fibration is said to be {\it regular}. In this article, we
only consider regular K3-fibrations, because that is when one can find
Heterotic dual descriptions without (generalization of) NS5-branes \cite{BW-16}.

Let $\pi: X \rightarrow \P^1$ be a regular K3-fibration. Then the cohomology
groups of $X$ have the following filtration:
\begin{align}
&  H^2(X;\Z) \supset \Z D_s \supset \{0\}, \qquad H^2(X;\Z)/\Z D_s =: \Lambda_S,   \label{eq:filtr-coH2} \\
&  H_2(X;\Z) \supset [H_2(X;\Z)]^{\rm vert} \supset \{0\},
\end{align}
where $D_s$ is the total K3-fibre divisor class, and $[H_2(X;\Z)]^{\rm vert}$ the subgroup generated by curves that are projected to points on $\P^1$.
\label{pg:IIA-notation}
The free abelian group $\Lambda_S$ is a subgroup of the Neron--Severi lattice
$L_S$ of $X_p$, the fibre K3 surface over a generic point $p \in \P^1$.
So, an intersection form is introduced on $\Lambda_S$ by restricting
the intersection form of $L_S$; $\Lambda_S$ is now regarded as a lattice.
The natural pairing between $\Lambda_S$ and $[H_2(X;\Z)]^{\rm vert}$ is
non-degenerate, and we have an isomorphism
$[H_2(X;\Z)]^{\rm vert} \cong_{ab} \Lambda_S^\vee$ as abelian groups.
We reserve $\rho$ for the rank of $\Lambda_S$, not for $L_S$.
$\Lambda_S$ [resp. $L_S$] is a primitive sublattice of
${\rm II}_{3,19} \cong H^2(K3;\Z)$; the orthogonal complement lattice is
denoted by $\Lambda_T$ [resp. $L_T$].

For a regular K3-fibration $\pi:X \rightarrow \P^1$, one can think of
a generating function $\Phi$ of the number of Noether--Lefschetz
points on the base $\P^1_{\rm IIA}$. When it is defined appropriately
(see below until \eqref{eq:Phi'-modform-IIA}), it is known to be a modular form \cite{MP}. First,
there is a holomorphic map
\begin{align}
  \iota_\pi : \P^1_{\rm IIA} \longrightarrow  D(\Lambda_T)/\Gamma_T,
\end{align}
where
$\Gamma_T := {\rm Ker}\left( {\rm Isom}(\Lambda_T) \rightarrow {\rm Isom}(G_T) \right)$; this is because $D(\Lambda_T)/\Gamma_T$ is the coarse moduli space
of $\Lambda_S$-polarized K3 surfaces.
At points on the base $\P^1_{\rm IIA}$ where the $\iota_\pi$-image hits the
Noether--Lefschetz divisor $D_{NL(F_\perp)}$ of $D(\Lambda_T)$ for
$F_\perp \in \Lambda_T^\vee$,
the transcendental cycles $(F_\perp + \Lambda_S^\vee) \cap {\rm II}_{3,19}$
become algebraic.

Second, think of the Heegner divisor \cite{MP}
\begin{align}
 \Phi_{\rm pre} := \sum_{F_\perp \in \Lambda_T^\vee} D_{NL(F_\perp)} \, q^{-F_\perp^2/2} \,
  e_{[F_\perp]} \, / \, \Gamma_T
    \;\in\;  {\rm Pic}(D(\Lambda_T)/\Gamma_T)[\![q^{1/N}]\!] \otimes \C[G_T],
\end{align}
where $\{e_{\gamma}\}_{\gamma \in G_T}$ is the set of formal basis elements of
the vector space $\C[G_T]$, $q=e^{2\pi i \tau}$ a formal variable, and
$N$ the level of the quadratic discriminant form of the lattice $\Lambda_T$.
The sum over $F_\perp \in \Lambda_T^\vee$ does not include $F_\perp$'s with $(F_\perp)^2 > 0$ because the Noether--Lefschetz divisor is empty for them.
Besides the terms of $F_\perp$'s with $(F_\perp)^2 < 0$, however, $\Phi_{\mathrm{pre}}$ is defined to include an extra term $e_0 q^0 D_{NL(0)}/\Gamma_T$;
we do not provide a description of the divisor
``$D_{NL(0)}/\Gamma_T$'' (see \cite{MP} for details), but all of its necessary
properties are provided later on.
Note that $\Phi_{\mathrm{pre}}$ in this definition does not have a term\footnote{
\label{fn:Mod_0_subtlety}
This leads to the subtle fact that $\Phi$ lies in $\Mod_0$ rather than $\Mod$ in \eqref{eq:Phi'-modform-IIA}.}
proportional to $e_{\gamma}q^0$ for a non-zero isotropic $\gamma \in G_T$.
Note also that $\Phi_{\mathrm{pre}}$ is defined for $\Lambda_T$ without referring to a K3-fibration.

Finally, given a K3-fibration $\iota_\pi$,
we obtain a $\C[G_T]$-valued function
by pairing $\Phi_{\rm pre}$ with the $\iota_\pi$-image inside $D(\Lambda_T)/\Gamma_T$:\footnote{
To be precise, this procedure is well-defined only for smooth fibrations.
In the case of a Calabi--Yau three-fold $X$ with finitely many nodal singular K3-fibres (these degenerations are classified as Type I), one has to take a double cover of $X$ and resolve the conifold singularities, so that a smooth fibration is obtained.
One can apply the procedure to this fibration and divide the modular form by 2. We denote as $\Phi$ the modular form defined in this way.
See \cite{GS-DT, D4D2D0} for details.
}

\begin{align}
 \Phi = \sum_{\gamma \in G_T} \Phi_{\gamma}(\tau) e_{\gamma}
     := \Phi_{\rm pre} \cdot \iota_\pi (\P^1) =   \sum_{\gamma \in G_T}
    \sum_{[F_\perp] \in (\gamma + \Lambda_T)/\Gamma_T}
          NL_{[F_\perp], \gamma} \; q^{-(F_\perp)^2/2} \; e_{\gamma}.
\end{align}
Define $NL_{\nu,\gamma} = \sum_{[F_\perp],-(F_\perp)^2/2 = \nu} NL_{[F_\perp],\gamma}$;
it is the Fourier coefficients of $\Phi_\gamma = \sum_{\nu} NL_{\nu,\gamma} q^\nu$.
Ref. \cite{MP} arrives at a statement (by using earlier math results
in \cite{Bor2}, but not relying on the duality with Heterotic string) that
\begin{align}
 \Phi
    \in {\rm Mod}_0\left( 11-\frac{\rho}{2} , \rho_{\Lambda_S} \right).
  \label{eq:Phi'-modform-IIA}
\end{align}
See footnote \ref{fn:Mod_0_subtlety} for why $\Phi$ is in $\Mod_0$,
not in $\Mod$.

The coefficient $NL_{\nu,\gamma}$ with $\gamma = 0$ and $\nu = 0$ in $\Phi$
does not describe the number of Noether--Lefschetz points on $\P^1_{\rm IIA}$.
Following the definition of the divisor ``$D_{NL(0)}/\Gamma_T$'' in \cite{MP},
one arrives at
\begin{align}
  NL_{0,0} = {\rm deg}_{\P^1_{\rm IIA}} (R^2\pi_*({\cal O}_X)) = -2,
\end{align}
where we used $h^{0,3}(X)=1$ and $h^{0,2}(X) = 0$ at the last equality.
So, as a consequence\footnote{
If $X = K3 \times T^2$, where $h^{0,2}(X)=1$,
then $R^2\pi_*({\cal O}_X) = {\cal O}_{T^2}$. So, $NL_{0,0} =
{\rm deg}_{T^2}({\cal O}_{T^2}) = 0$. This is consistent with the
Heterotic string description ($n_0 = -2 + n_0^H = 0$ in the ${\cal N}=4$
supersymmetry situation).}
of (\ref{eq:h10=0}), we have the expansion
$\Phi \sim -2 q^0e_0 + (\text{strictly higher power of $q$})$.

\subsubsection{Heterotic--Type IIA Duality and Effective Theory}
\label{sssec:Het-IIA-dual-basic}

When branches of moduli space of Heterotic theories and Type IIA theories
associated with a pair $(\widetilde{\Lambda}_S, \Lambda_T)$  are identified under the duality,
both descriptions should give rise to the same modular form:
\begin{align}
 \left\{ \Phi^{\rm Het}_{\gamma}(q) \right\}_{\gamma \in G_S} =
 \left\{ \Phi^{\rm IIA}_{\gamma}(q) \right\}_{\gamma \in G_T}.
  \label{eq:Phi'--Phi'-dual}
\end{align}
The isomorphism $G_S \cong_{ab} G_T$ as abelian groups
comes from the mutually orthogonal embedding\footnote{More generally, there is such a unique
isomorphism $\phi_M: G_L \cong_{ab} G_{L'}$
that satisfies $\phi_M^*(-,-)_{L'} = -(-,-)_{L}$
for any mutually orthogonal pair
$L$ and $L'$ of primitive sublattices of a unimodular lattice $M$.
In this article, elements of $G_L$ and $G_{L'}$ are identified without
referring to $\phi_M$ (e.g., in (\ref{eq:et-4})).}
$(\widetilde{\Lambda}_S \oplus \Lambda_T) \subset
{\rm II}_{4,20} \subset (\widetilde{\Lambda}_S^\vee \oplus \Lambda_T^\vee)$.

That is because all the Fourier coefficients of $\Phi/\eta^{24}$
determine physical quantities in the ${\cal N}=2$ supersymmetric effective
theory on 3+1-dimensions; the helicity supertrace is defined by
\begin{align}
  \Omega(0,0;(r,w),q_0) :=
    -2 {\rm Tr}_{{\cal H}(0,0;(r,w),q_0)} \left[ (-1)^{2J_3}(J_3)^2 \right]
\end{align}
on the Hilbert space ${\cal H}(0,0;(r,w),q_0)$ of particles on $\R^{3,1}$
with a given pure electric charge under
the $(\rho+2)$ $U(1)$ gauge fields; $r,q_0 \in \Z$ and $w \in \Lambda_S^\vee$.
$J_3$ is the 3-component of the angular momentum of this space-time.
In the Heterotic language \cite{DDMP-0502}
\begin{align}
  \Omega(0,0;(1,w),q_0) = c^{\rm Het}_{[v]}(v^2/2), \qquad
    v = e^0 + w + q_0 e^4 \in \widetilde{\Lambda}_S^\vee,
\end{align}
where $e^0$ and $e^4$ are the generators of the $U[-1]$ factor in \eqref{eq:restriction-2-Het}, and $w\in \Lambda_S^\vee$.
In the Type IIA language, we apply the electromagnetic duality transformation
in the effective theory for the one of the $(\rho+1)$ $U(1)$ gauge fields
originating from the Ramond--Ramond 3-form field, the one associated with
the 2-form $D_s$; then a D4-brane wrapped
on the fibre class along with a 2-form $F \in H^2(K3;\Z)$ and $N_{\geq 0}$
units of anti-D0-brane gives rise to a particle on $\R^{3,1}$ with a pure
electric charge (Mukai vector on the K3 surface)
\begin{align}
v =  e^0 + F_{\parallel} + q_0(F,N) e^4 \in \widetilde{\Lambda}_S^\vee,
  \qquad   q_0(F,N) = 1-N + \frac{F^2}{2}.
  \label{eq:Mukai-vect-r1}
\end{align}
Here, $\widetilde{\Lambda}_S = U[-1]\oplus \Lambda_S$ is regarded
as the Mukai lattice of $\Lambda_S$-polarized K3 surface,
and $e^0$ and $e^4$ generate the factor $U[-1] \cong_{ab}
H^0(K3;\Z) \oplus H^4(K3;\Z) \cong_{ab}\Z e^0 \oplus \Z e^4$.
$F_{\parallel}$ is the projection of $F \in H^2(K3;\Z)$ to
$\imath^*(H^2(X;\Z)) \cong \Lambda_S$, and $\imath:
({\rm a~fibre~K3})\hookrightarrow X$ is an embedding.
The helicity supertrace is \cite{BSV-95, GopVafa, KKV, KKRS} (and \cite{DM-07})
\begin{align}
  \Omega(0,0;(1,F_{\parallel}),q_0) = c_{[v]}^{\rm IIA}(v^2/2) .
\end{align}
The values of $v^2/2$ in the set of states above exhaust\footnote{
The Gopakumar--Vafa invariants of vertical curve classes
$\beta \in [H_2(X;\Z)]^{\rm vert} \cong \Lambda_S^\vee$ are
equal to $c_{[\beta]}(\beta^2/2)$, and are also physical in the effective
field theory because of its appearance in the prepotential
(\ref{eq:prepot-pert-ansatz}, \ref{eq:matching-chi+GV});
see \cite{HM, MP, KMPS}.
Not all the coefficients $c_\gamma(\nu)$ of $\Phi$ correspond to those Gopakumar--Vafa invariants for some $\Lambda_S$
(e.g., $\Lambda_S = \vev{+2}$), however.} all
$\nu \in h_{\rm min}(\gamma) + \Z_{\geq 0}$, so all the Fourier coefficients
of $\Phi^{\rm Het}/\eta^{24}$ and $\Phi^{\rm IIA}/\eta^{24}$ should be
the same.
The Coulomb branch moduli space $D(\widetilde{\Lambda}_S)$
in (\ref{eq:def-Narain-period-dom}) is parametrized by
$t \in \Lambda_S \otimes \C$ with $(t_2,t_2) > 0$, which is interpreted
as Narain moduli parameters in the Heterotic description and as complexified
K\"{a}hler parameters in the Type IIA description:
\begin{align}
  \mho(t) = e^0 + e^4 \frac{(t,t)}{2} + t.
  \label{eq:def-mho}
\end{align}
The central charge of a BPS state with the electric charge
$v \in \widetilde{\Lambda}_S^\vee$ is proportional to
$\left( p_R^\C(v) \propto (\mho(t), v) \right)$.
We will focus only on the $s_2 \gg 1$ and $s_2 \gg \abs{t_2}$ region
of the special geometry throughout this article; that is
the weak coupling region in the Heterotic description, and
the large base $\P^1$ region in the Type IIA description.\footnote{
That is when we expect little contributions to low-energy physics from
the NS5-branes in Heterotic string and D-branes wrapped on
cycles that are mapped surjectively to $\P^1_{\rm IIA}$ in the Type IIA string.
BPS states of such origins are not used in defining the modular form $\Phi$.}

To wrap up, the moduli space of the Heterotic--Type IIA dual vacua with 4D
${\cal N}=2$ supersymmetry has a branch structure, and each branch is labeled by
a pair of lattices $\widetilde{\Lambda}_S$ and $\Lambda_T$, and
a modular form $\Phi \in {\rm Mod}_0(11-\rho/2,\rho_{\Lambda_S})$.
It is known \cite{McGraw} that one can find a $\C$-basis $\{ \phi_i \}$ of
${\rm Mod}_0(11-\rho/2,\rho_{\Lambda_S})$ so that all the Fourier coefficients
of $\phi_{i,\gamma}(\tau)$ of
$\phi_i = \sum_{\gamma \in G_S} e_\gamma \phi_{i,\gamma}(\tau)$ are integers.
So, $\Phi$ must be in the free abelian group within
${\rm Mod}_0(11-\rho/2,\rho_{\Lambda_S})$ whose rank is the same as
the dimension of ${\rm Mod}_0(11-\rho/2,\rho_{\Lambda_S})$.
This free abelian group is denoted by ${\rm Mod}_0^\Z(11-\rho/2,\rho_{\Lambda_S})$.

Without relying on explicit constructions (such as toric complete
intersection), we can therefore hope to use properties of the free abelian
group ${\rm Mod}_0^\Z(11-\rho/2,\rho_{\Lambda_S})$ to derive some properties
of $\Lambda_S$-polarized K3-fibred Calabi--Yau three-folds.

\subsection{Conditions for $n^V_\gamma = 1$}
\label{ssec:gamma_0-level-quantization-condition}

As seen in section \ref{sssec:review-Het}, $n_\gamma^V$ is the number of BPS vector multiplets of fixed charge $v \in \gamma$ subject to $-2 \leq v^2 < 0$. Since any charged massless gauge boson\footnote{Heterotic string non-perturbative
effects modify the infrared dynamics and the moduli space, as
in Seiberg--Witten theory, but they are all known story \cite{KKLMV}.}
should be a part of non-abelian gauge bosons for some compact Lie group, its multiplicity must be one for each possible charge. This implies that $n_\gamma^V \in \{ 0,1 \}$ for any non-zero $\gamma \in G_S$, so that $n_\gamma \geq -2$.
Let us consider what happens when $n_\gamma^V = 1$.

Suppose $n_{\gamma_0}^V = 1$ for a given non-zero $\gamma_0 \in G_S$. Fix $v_0 \in \gamma_0 \subset \widetilde{\Lambda}_S$ such that $-2 \leq (v_0,v_0) < 0$.
Let us consider the points in the Coulomb branch moduli space $D(\widetilde{\Lambda}_S)$ where the states of charge $v_0$ become massless: $p_R(v_0) = 0$. At these points, the massless vector bosons of charge $v_0$ should be a part of non-abelian gauge bosons. In the language of the Heterotic worldsheet theory, this gauge symmetry is described by some current algebra carried by the left-mover.
Therefore we have $SU(2)$ current algebra (that may be a sub-algebra of a larger current algebra) that consists of
\begin{align}
    J^\pm(z) := e^{ \pm i p_L(w_0) \cdot X_L} \mathcal{O}^\pm,
    \quad
    J^3(z) := \frac{i}{-(v_0,v_0)}\; p_L(v_0) \cdot \partial X_L,
\end{align}
where $X_L$ is the internal free chiral bosons in the $c=\rho$ sector, and $\mathcal{O}^\pm$ are some vertex operators of conformal weight $[(\gamma_0,\gamma_0)/2]_{\mathrm{frac}}$ in $\mathcal{H}_{\gamma_0,V}^{(22,0)}$. Note that $J^\pm$ have conformal weight 1 as well as $J^3$, because a charge $v_0$ with $-2 \leq (v_0,v_0) < 0$ is under consideration.
Their OPEs are
\begin{align}
    J^3(z) J^\pm(0) \sim \frac{\pm 1}{z} J^\pm(0),\qquad
    J^3(z) J^3(0) \sim \frac{-1/(w_0,w_0)}{z^2}.
\end{align}
This means that\footnote{
In other words, $\mathcal{O^\pm}$ are from the coset $SU(2)_k/U(1)_k$.
This coset model is known as parafermion theory with $\Z_k$ symmetry.
}
 the level of the current algebra is $k= -2/(v_0,v_0)$.

Let us consider some constraints from unitarity. First, the level $k$ needs to be a positive integer. Second, any state (of $U(1)$-charge $v \in \widetilde{\Lambda}_S^\vee$) should have half-integral spin in terms of this $SU(2)$:
this results in
\begin{align}
     \frac{(v_0,v)}{(v_0,v_0)} \in \frac{1}{2} \Z \qquad \forall v \in \widetilde{\Lambda}_S^\vee,
\end{align}
because that\footnote{
This condition is required for all $v \in \widetilde{\Lambda}_S^\vee$, because
we assumed in section \ref{ssec:Het-II-review}
(footnote \ref{fn:all-charge-exist-assume}) that
there exists at least one state of charge
$v$ for any $v \in \gamma \subset \widetilde{\Lambda}_S^\vee$.
}
 is the charge of $e^{i (p_L(v)\cdot X_L + p_R(v)\cdot X_R)}$ under $J_3$.
%
These mean the following constraints\footnote{
It follows from these constraints that, for any non-zero isotropic $\gamma$
in $G_S$, we get $n^V_{\gamma_0} = 0$. The assumption $n_{\gamma_0}=0$
in (\ref{eq:restriction-3-Het})---automatic in a Type IIA geometric phase
(footnote \ref{fn:Mod_0_subtlety})---further implies $n^H_{\gamma_0}=0$ then.
}
 for $\gamma_0 \in G_S = \widetilde{\Lambda}_S^\vee/\widetilde{\Lambda}_S$:
\begin{align}
    \frac{(\gamma_0,\gamma_0)}{2} \equiv \frac{-1}{k} \bmod{\Z} \quad \text{and} \quad
    k\gamma_0 \equiv 0 \in G_S,
    \quad \text{for some } k \in \Z_{\geq 2}.
    \label{eq:gamma_0-level-quantization-condition}
\end{align}
The possibility $k=1$ does not have to be included here, because
$k=1$ would simply mean $\gamma_0 = 0$
in (\ref{eq:gamma_0-level-quantization-condition}), where we know
that $n_{\gamma_0=0}^V=1$ from the beginning.

Not all $\gamma \in G_S$ satisfy the
conditions (\ref{eq:gamma_0-level-quantization-condition}), but those
that satisfy (\ref{eq:gamma_0-level-quantization-condition}) are not
extremely rare.
Table \ref{tab:quantization-example} shows the list of such $\gamma$'s
for some of the $\rho=1$ lattices, $\Lambda_S=\vev{+2n} \cong_{ab} \Z e$.
\begin{table}[tbp]
    \centering
    \begin{tabular}{c||c|c|c|c|c|c|c}
        $n$     & 1          & 2       & 3       & 4          & 5           & 6 & $\cdots$ \\ \hline
        $(j,k)$ & $\varnothing$ & $(2,2)$ & $(3,4)$ & $\varnothing$ & $(\pm 4,5)$ & $(\pm 4,3),(\pm 6,2)$ & $\cdots$
    \end{tabular}
    \caption{list of $(\gamma_0,k) = ([\frac{j}{2n} e],k)$ satisfying the condition \eqref{eq:gamma_0-level-quantization-condition}
    when $\Lambda_S = \vev{+2n}$}
    \label{tab:quantization-example}
\end{table}
In the case of lattices of the form $\Lambda_S = U \oplus \vev{-2m} =:_{ab}
U \oplus \Z e$, at least $(\gamma_0 = [\frac{1}{2m}e],\, k= 4m)$ and
$(\gamma_0 = [\frac{2}{2m}e],\, k= m)$
satisfy (\ref{eq:gamma_0-level-quantization-condition}).

There are a couple of different behaviors in the Type IIA geometry
$X_{\rm IIA}$ that correspond to appearance of a massless non-abelian
gauge boson in the low-energy effective field theory.
Let $v = r e^0 + q_0 e^4 + F_{\parallel} \in \widetilde{\Lambda}_S^\vee$
be the U(1) charge of such a gauge boson (as in (\ref{eq:Mukai-vect-r1})).

Suppose $rq_0 = 0$. Then the Calabi-Yau $X_{\rm IIA}$ should have
a curve class $F_{\parallel} \in \Lambda_S^\vee \cong [H_2(X_{\rm IIA};\Z)]^{\rm vert}$
realized algebraically over a generic point of the base $\P^1_{\rm IIA}$
($B_\infty$ in \S 3.2 of \cite{MP}); this is because
\begin{align}
c_{\gamma_0}(F_{\parallel}^2/2) =
    NL_{1+ (F_\parallel)^2/2, \gamma_0} < 0, \qquad  -1 < \frac{F_{\parallel}^2}{2} < 0
\end{align}
is possible only when $\iota_\pi(\P^1)$ stays within a Noether--Lefschetz
divisor $(\sum_{F_{\perp}} D_{NL(F_{\perp})})/\Gamma_T$ with the sum ranging over
$F_\perp$'s with $F_{\parallel}^2 + F_{\perp}^2 = -2$.
The algebraic curve is $ \pm F=  \pm (F_{\parallel} + F_{\perp}) \in {\rm II}_{3,19}$,
which must be a $(-2)$ curve. The vector boson on the spacetime $\R^{3,1}$
is massless when this $(-2)$ curve in the K3-fibres collapses to zero volume ($(t_2,F_\parallel) = 0$).
$F_{\perp}$ must be non-zero since we think of a case $\gamma_0 \neq 0 \in G_S$.
Then there must be non-trivial monodromy on $F_{\perp} \in \Lambda_T$ so that
$F$ is in $L_S$, but not in $\Lambda_S$.
$\Lambda_S$ is a proper subset of $L_S$ then.

In the case of $\Lambda_S = U \oplus \vev{-2m}$ and $\gamma_0 = \pm [e/m]$
(the level $k$ is $m$), the following interpretation seems to work:
in terms of lattices, $L_S \cong U \oplus A_1[-1]^{\oplus m}$ and
the lattice $\vev{-2m} \subset \Lambda_S$ is embedded diagonally in
$A_1[-1]^{\oplus m}$; in terms of geometry, each K3 fibre of $X \rightarrow \P^1$
has $m$ points of $A_1$ singularity, and those singular points form
a curve in $X$ that is an $m$-fold cover over the base $\P^1$. The gauge
kinetic term of the vector field is $ms$ in Type IIA reasoning, which agrees nicely with the
the gauge kinetic function $ks$ for gauge fields associated with
the level-$k$ current algebra in Heterotic constructions \cite{Ginsparg:1987ee}.

There should also be geometry / lattice interpretation along the line of
$\Lambda_S \subsetneq L_S$ also for the
case $\Lambda_S = U \oplus \vev{-2m}$ and $\gamma_0 = \pm [e/(2m)]$,
but we have not been able to find a functioning interpretation yet.

Suppose instead that $r q_0 \neq 0$. The massless vector boson is then
a D4--D2--D0 bound state, not just of D2- and D0-branes ($r \neq 0$).
The K\"{a}hler parameter $t$ must be of order unity for the vector boson
to become massless (for moderate choices of $r$, $q_0$, and $F_{\parallel}$),
so the base $\P^1_{\rm IIA}$ may be large ($s_2 \gg 1$), but
the fibre K3 is not safely in the large radius geometric regime.
In any case of $\Lambda_S = \vev{+2n}$ with $\gamma_0 \in G_S$ satisfying
(\ref{eq:gamma_0-level-quantization-condition}), the U(1) charge $w_0$ for
such a massless vector boson should be the one of this category, because
$F_{\parallel}^2/2$ is positive definite in $\Lambda_S = \vev{+2n}$.

\subsection{Examples of $\Phi$}
\label{ssec:examples-of-Phi}

One can list up modular forms
$\Phi \in {\rm Mod}^\Z_0(11-\rho/2,\rho_{\Lambda_S})$ that satisfy the condition $n_0 = -2$
and the lower bounds $n_\gamma \geq -2$ or $n_\gamma \geq 0$ depending on
$\gamma \in G_S$ as seen in
section \ref{ssec:gamma_0-level-quantization-condition}.
The easiest and well-known case is when $\Lambda_S$ is unimodular:
\begin{align}
(\widetilde{\Lambda}_S, \Lambda_T) = & \;
       (U^{\oplus 2}, U^{\oplus 2} \oplus E_8^{\oplus 2}[-1]),  \quad
   (U^{\oplus 2} \oplus E_8[-1], U^{\oplus 2} \oplus E_8[-1]), \\
  &  (U^{\oplus 2} \oplus E_8[-1]^{\oplus 2}, U^{\oplus 2}).  \nonumber
\end{align}
The rank $\rho$ is $2,10,18$ for each case and
$\Phi$ should be a scalar-valued modular form of weight $(22-\rho)/2$
with $n_0 = -2$. So $\Phi = -2E_4E_6$ and $-2E_6$ for the first and
second case, respectively. There is no candidate of $\Phi$ for the
third case; $\widetilde{\Lambda}_S = U^{\oplus 2} \oplus E_8[-1]^{\oplus 2}$
(i.e., zero instantons in $E_8 \times E_8$ in the Heterotic string)
cannot be realized at least in our setup reviewed in
section \ref{ssec:Het-II-review}.

\subsubsection{Cases $\Lambda_S = \vev{+2}$, $\vev{+4}$, and $\vev{+6}$}
\label{sssec:rho=1}

We attempt at assessing how well/poorly the combination of the modular
invariance, integrality of the BPS indices, and their lower bounds explains
possible topological choices of K3-fibrations of Calabi--Yau three-folds
for non-unimodular $\Lambda_S$. In section \ref{sssec:rho=1},
we first work on the cases $\Lambda_S = \vev{+2}$, $\vev{+4}$, and
$\vev{+6}$, where a set of independent generators
of ${\rm Mod}_0(21/2,\rho_{\Lambda_S})$ is known \cite{Kawai, KKRS, MP, HK},
as summarized in appendix \ref{apdx:A21}.
The list of $\Phi$'s determined in this way is compared against a list of
Calabi--Yau three-folds with those
$\Lambda_S$-polarized K3-fibration explicitly constructed in \cite{KKRS}.

Let us illustrate the procedure using the case $\Lambda_S = \vev{+2}$
as an example. To prepare a notation, let $e$ be the generator
of the free abelian group $\vev{+2n} =:_{ab} \Z e$; the formal basis
elements of $\C[G_S]$ [resp. the low-energy BPS indices] are denoted by
$e_\gamma = e_{j/2n}$ [resp. $n_\gamma = n_{j/2n}$] for short, when
$\gamma = (j/2n + \Z)e \in G_S$ for some $j \in \{0,1,\dots, 2n-1\}$.
Now, one can\footnote{Here, $[-, -]_i$ is the Rankin--Cohen bracket.
See appendix \ref{apdx:A1}.}
choose $\phi_{(i)} = [\theta_{\vev{+2}},E_{10-2i}]_i,\,i=0,1$
as a $\C$-basis of the vector space ${\rm Mod}_0(21/2,\rho_{\vev{-2}})$
\cite{KKRS, HK}.
The modular form $\Phi$ is parametrized by
\begin{align}
  \Phi & \; = -2 (\phi_{(0)}+\phi_{(1)}/2) - \frac{n_{1/2}}{4} \phi_{(1)},
    \label{eq:deg2-Phi-paramtrz-main} \\
  & \; = e_0 \left(-2 + (300-56n_{1/2})q + (217200-13680 n_{1/2})q^2
               + \cdots \right) \nonumber \\
  & \; + e_{1/2} \left( n_{1/2} q^{\frac{1}{4}} + (2496 + 360 n_{1/2}) q^{\frac{5}{4}}
      + (665600 + 30969 n_{1/2}) q^{\frac{9}{4}} + \cdots \right);
\end{align}
this linear combination is chosen so that the coefficients of $e_0$ and
$e_{1/2} q^{1/4}$ be $n_0 = -2$ and $n_{1/2}$ respectively.\footnote{
Because $\Phi$ was parametrized by
$\Phi = -2 (\phi_{(0)}+\phi_{(1)}/2) + n/4 \phi_{(1)}$ in \cite{KKRS, HK},
the parameter $n$ should be interpreted as $-n_{1/2}$. }
It follows from the discussion in section \ref{ssec:gamma_0-level-quantization-condition}
that $n_{1/2}$ should be a non-negative integer.
This condition indicates that the Euler number
$\chi = -c_0(0) = -[\Phi_0/\eta^{24}]_{q^0}$ of $X_{\mathrm{IIA}}$ is quantized and bounded from below:
\begin{align}
 \chi(X_{\rm IIA}) = -252 + 56n_{1/2}, \qquad n_{1/2} \in \Z_{\geq 0}.
   \label{eq:chi-deg2}
\end{align}
%
It is also bounded from above, because $h^{2,1}(X_{\rm IIA}) \geq 0$
(and hence $\chi(X_{\rm IIA}) \leq 2h^{1,1} = 4$) in the Type IIA
language.\footnote{In the Heterotic language, $c_0(0) = (252-56n_{1/2})$
is $\dim_\C([{\cal H}_{0,(1/2,1/2)}]_{L_0=1}) - 2 \dim_\C([{\cal H}_{0,(0,0)}]_{L_0=1})
- 4$, where the last term $-4$ is from $\eta^{-2}$ of the first term
in (\ref{eq:ContributionsToPhi}). By definition,
$\dim_\C([{\cal H}_{0,(0,0)}]_{L_0=1}) =: \rho$, and the assumption that
the Heterotic construction in consideration has a Type IIA dual in the
geometric phase implies that
$\dim_\C([{\cal H}_{0,(1/2,1/2)}]_{L_0=1})\geq 2$ because the Type IIA
compactifications has at least one hypermultiplet containing the dilaton.
So, $c_0(0) \geq -2(\rho+1)$.}
So,
\begin{align}
  n_{1/2} \in \{ 0,1,2,3,4\}.
\end{align}
As a result, we see that when $\Lambda_S = \vev{+2}$, there are only finite possibilities for $\Phi,\chi$, and BPS indices $n_{\abs{\gamma}}$.

\bigskip

Similar procedure can be carried out for $\Lambda_S = \vev{+4}$ and $\vev{+6}$.
See appendix \ref{apdx:A21} for details. Here we just quote the results on Euler number in Table \ref{tab:EulerNumber-BPSIndicies-rank1}.
The lattice $\Lambda_S = \vev{+4},\vev{+6}$ also allows only finite possibilities for $\chi,\{n_\gamma\}$ and $\Phi$.

\begin{table}[tbp]
    \centering
    \begin{tabular}{cl}
        $\Lambda_S$ & $\chi(X)$ \\ \hline
        $\vev{+2}$ & $\chi = -252 + 56 n_{1/2}$ \\
        $\vev{+4}$ & $\chi = -168 + 128 n_{1/4} + 14 n_{2/4}$ \\
        $\vev{+6}$ & $\chi = -148 + 108 n_{1/6} + 54 n_{2/6} + 2 n_{3/6}$
    \end{tabular}
    \caption{The Euler number $\chi$ in terms of the low-energy BPS indices $n_{\gamma}$. $n_{2/4}$ and $n_{3/6}$ are not less than $-2$, and all the rest are non-negative.
}
    \label{tab:EulerNumber-BPSIndicies-rank1}
\end{table}

The authors of \cite{KKRS} scanned combinatorial data in the toric complete intersection construction and produced some explicit examples of Calabi-Yau three-folds with a $\vev{+2n}$-polarized regular K3 fibration: see Table 1 in \cite{KKRS}.
All of their examples satisfy the integrality of $n_\gamma$ and the bounds of $n_\gamma$ and $\chi$ as described above, of course, but do not exhaust all possibilities under these conditions.
For example, when $\Lambda_S = \vev{+2}$, their list includes threefolds with $\chi = -252$, $-196$, $-140$, $-84$, which correspond to $n_{1/2}=0,1,2,3$ in the parametrization above,
but it does not include any cases where $\chi = -28$ and $n_{1/2} = 4$. The discussion above indicates that there cannot exist Calabi--Yau three-folds with a $\vev{+2}$-polarized regular K3 fibration (whether toric complete intersection or not) with $\chi < -252$, or $-252 < \chi < -196$, etc.
But, we did not find any theoretical reason to rule out the case $n_{1/2} = 4$ and $\chi= -28$.

Similar stories hold for $\Lambda_S = \vev{+4}$ and $\vev{+6}$; Table 1 in \cite{KKRS} shows geometric realizations for $n_{1/4} = 0, n_{2/4} = -2,0,2,4,6$ and $n_{1/6} = n_{2/6} = 0$, $n_{3/6} = 0, 8, 10, 14, 16, 20$. But there exist other values of integers $(n_{1/4},n_{2/4})$ or $(n_{1/6},n_{2/6},n_{3/6})$
satisfying
the bounds $n_{1/2},n_{1/4},n_{1/6},n_{2/6} \geq 0$ and $n_{2/4},n_{3/6}\geq -2$.

The absence of examples with odd $n_{2/4}$ and $n_{3/6}$ in geometry
constructions is presumably explained as follows.
The BPS index $n_{2/4}$ [resp. $n_{3/6}$] in the case of $\Lambda_S = \vev{+4}$
[resp. $\vev{+6}$] is $NL_{4/8,2/4}$ [resp. $NL_{9/12,3/6}$].
To the Noether--Lefschetz number $NL_{4/8,2/4}$, for example, both
$F_\perp = (2/4) e'$ and $F_\perp = - (2/4) e'$ give rise to separate contributions $NL_{[F_\perp],2/4}$ where $[F_\perp] = F_\perp \Gamma_T$; here, $e'$ is the
generator of $\vev{-4} =:_{ab} \Z e'$ in
$\Lambda_T \cong \vev{-4} \oplus U^{\oplus 2} \oplus E_8[-1]^{\oplus 2}$.
They are separate contributions,
because the two $F_\perp$'s shown above
are not in a common orbit\footnote{
In the argument here, we discuss only the contributions $NL_{[F_\perp],2/4}$
only from the $\Gamma_T$-orbits $[e'/2]$ and $[-e'/2]$. 
Note in v.2: It turns out that the $\Gamma_T$-orbit decomposition 
of $\{ F_\perp \in e'/2 + \Lambda_T \; | \;  - (F_\perp)^2/2 = 1/2 =\nu\}$ 
consists of one more orbit besides $[e'/2]$ and $[-e'/2]$. So, unless we
can prove that $NL_{[F_\perp],2/4}$ from this orbit is even, we cannot  
rule out an odd $n_{2/4}$. 
Also for $\Lambda_S = \vev{+6}$, $\{F_\perp \in e'/2 + \Lambda_T \; | \; 
(F_\perp)^2 = -3/2 \}$ consists of three $\Gamma_T$ orbits. 
}
of $\Gamma_T$. Their Noether--Lefschetz divisors
$D_{NL(F_\perp)}$ in $D(\Lambda_T)/\Gamma_T$ are the same, so the sum
of the two contributions is twice the single contribution\footnote{
The authors do not have confidence to say that the divisor--curve
intersection number $NL_{[F_\perp],2/4}$ is definitely an integer, because
$D(\Lambda_T)/\Gamma_T$ has orbifold singularity associated with
K3-surfaces of a non-trivial group of purely non-symplectic automorphisms.
If $NL_{[F_\perp],2/4} \in \Z$, then $2NL_{[F_\perp],2/4} \in 2\Z$. }
 $NL_{[F_\perp],2/4}$ (cf. \cite{MP}).
This argument is adapted in an obvious way\footnote{
This argument is not applicable to the case $\Lambda_S = \vev{+2}$.
That is because $F_\perp = (1/2)e$ and $F_\perp = -(1/2)e$ are in one
orbit under $\Gamma_T$.  That is consistent with the fact that
Ref. \cite{KKRS} found threefolds for
$\Lambda_S = \vev{+2}$ with odd $n_{1/2}$'s. 
 }
to the case of $\Lambda_S = \vev{+6}$ and $n_{3/6}$.

We could not rule out $n_{1/4},n_{1/6},n_{2/6}$ that are non-zero,\footnote{
$n_{1/4}\geq 2$ and $n_{1/6} \geq 2$ are ruled out because $\chi(X)$ would
be larger than $2h^{1,1}$ then.
} or $n_{3/6}=2,4,6$.
We may have missed some additional physical/mathematical constraints,\footnote{
The procedure explained in section \ref{sec:finerC} and exemplified in appendix \ref{sssec:integrals-U+W} is one of the ways to find a
constraint on $n_{\gamma}$'s. } or
it is possible that some of such Calabi--Yau three-folds may exist, either outside of the scanned range of the combinatorial data in \cite{KKRS},
or in the form that do not allow their realization by  complete intersections
in toric varieties.

\subsubsection{Linear Relations on the Spectrum of a Local Effective Field Theory}
\label{sssec:lin-reltn}

As stated already in section \ref{sssec:review-Het}, the classification
invariant $\Phi$ is in the free abelian group
whose rank is the $\dim_\C ({\rm Mod}_0(11-\rho/2,\rho_{\Lambda_S}))$,
and is also completely
determined by the low-energy BPS indices $\{ n_{|\gamma|} \}_{\pm \gamma \in G_S^<}$
(because the weight $(-1-\rho/2)$ of $\Phi/\eta^{24}$ is strictly negative
for any $\rho \in \{ 1,2,\dots,20 \}$). This implies immediately that there is a
linear relation among $n_\gamma$'s when
$\dim_\C({\rm Mod}_0(11-\rho/2,\rho_{\Lambda_S}))$ is strictly less than
$d^< = |G_S^</\pm|$. As is clear from the Heterotic description,\footnote{
Modular form $\Phi$'s can be defined in a Type IIA compactification on
a Calabi--Yau $X$ that does not necessarily have a K3-fibration;
for a divisor $P$ in $X$, $\Phi$ is the generating function of the helicity supertrace for states originating from a D4-brane wrapped on $P$.
It is then in ${\rm Mod}(11-r/2,\rho_{\Lambda})$, where the lattice $\Lambda$
is the sublattice of $H^2(P)$ corresponding to the image
$\imath_P^*(H^2(X))$ of the embedding $\imath_P: P \hookrightarrow X$;
 $r := {\rm rank}(\Lambda)$. See \cite{MSW-97}.
Here, the weight $(-1-r/2)$ of $\Phi/\eta^{24}$ is always strictly negative,
and the $n_\gamma$'s are the multiplicities of states whose central charge
may vanish at a positive K\"{a}hler parameter, even
in this more general set-up \cite[\S3]{DM-07}. So, the same argument
as in the main text also holds; whenever
$\dim_\C({\rm Mod}(11-r/2,\rho_\Lambda))$ is strictly less than
$| G_\Lambda/\pm|$, there is a linear relation among the low-energy
BPS indices.

For a general divisor $P$ in $X$, however, such a linear relation
is among the multiplicities of states whose U(1) charges are not
necessarily mutually local. So, it cannot be regarded as a
prediction on a spectrum of a Lagrangian-based local effective field theory.
In the set-up discussed in the main text, $P$ is the total fibre class
$D_s$, where $P \cdot P = 0$, and $r = h^{1,1}(X)-1$, not $h^{1,1}(X)$.
This property makes all the states from a D4-brane on $P$ free from magnetic
charge (obvious in the Heterotic description from the start).
}
the low-energy BPS indices $n_\gamma$'s with $\gamma \neq 0$ are the multiplicities of fields with purely electric charge under the $(\rho+2)$ gauge bosons.
So, such a linear relation is that of the spectrum of $\mathcal{N}=2$ supersymmetric Lagrangian-based effective field theory on $\R^{3,1}$
with charge lattice $\Lambda_S$.
It cannot be related to the 4D triangle anomaly because of non-chiral matter representations (possibly to 6D box anomaly if $\Lambda_S$ contains $U$), but it originates from the modular invariance of $\Phi$.

Examples of the lattice $\Lambda_S$ with such a prediction are found
by using the dimension formula
(\ref{eq:RR-thm-ModForm}, \ref{eq:diff-dim-Mod-Mod0}) if $\rho < 18$
(so $11-\rho/2 > 2$). Within rank-1 $\Lambda_S$'s,
the lattice $\vev{+2n} = \vev{+14}$ is the first example, where
$\dim_\C({\rm Mod}_0(21/2,\rho_{\vev{14}})) = 7$, less than $d^{<}=d = 8$.
We found that
\begin{align}
  14n_{1/14} + 8n_{2/14} -13n_{3/14} - 6n_{4/14} - n_{5/14} -6n_{6/14} + n_{7/14} + 28 = 0 ;
\end{align}
see appendix \ref{apdx:A22} for necessary details. In the series of $\rho = 3$ lattices
$\Lambda_S = U \oplus \vev{-2m}$ with $m \in \N$,
the $m=2$ case already has a prediction, because $\dim_\C({\rm Mod}_0(19/2,\rho_{\vev{-4}})) = 2$
and $d^<=d=3$.
\begin{align}
  n_{1/4} = 8 n_{2/4} + 96.
\end{align}
Details are found in appendix \ref{apdx:A21}.
In both of the series $\Lambda_S = \vev{+2n}$ and
$\Lambda_S = U \oplus \vev{-2m}$, we confirmed that
the dimension of ${\rm Mod}_0(11-\rho/2,\rho_{\Lambda_S})$ lies
strictly below $d^<$ and also above zero for large $n$'s and $m$'s,
by evaluation of the dimension
formula (\ref{eq:RR-thm-ModForm}, \ref{eq:diff-dim-Mod-Mod0}).
The swampland surely exists within the space of local effective field theories,\footnote{The lattice $\Lambda_S$ is characterized within the language of local
effective field theory on $\R^{3,1}$; it appears in the
prepotential (\ref{eq:prepot-pert-ansatz}). }
if we restrict our attention to the class of Heterotic--Type IIA dual vacua
reviewed in section \ref{ssec:Het-II-review}.

It is also found that the vector spaces
${\rm Mod}_0(21/2,\rho_{\vev{+2n}})$
and ${\rm Mod}_0(19/2,\rho_{\vev{-2m}})$ continue to have strictly positive
dimensions\footnote{
There are more modular forms of a fixed weight and for $\Gamma(4n)$
for large $n$. So, that is not surprising.} for larger $n$ and $m$
(by numerically
evaluating (\ref{eq:RR-thm-ModForm}, \ref{eq:diff-dim-Mod-Mod0})).
So, the modular invariance of $\Phi$ and the integrality of its
coefficients alone do not rule out existence of Calabi--Yau three-folds with a $\Lambda_S=\vev{+2n}$-polarized K3-fibration for an arbitrary
large $n$, or also with $\Lambda_S = U \oplus \vev{-2m}$ for an arbitrary
large $m$.
%

\subsubsection{Cases with $\rho=20$ and $\rho=19$}

Here, we have a look at the cases with $\rho = 20$ and $19$.
In some of them, we will see that the vector space
${\rm Mod}(11-\rho/2,\rho_{\Lambda_S})$ is empty, and that
there cannot be such a lattice-polarized regular K3-fibration
in a Calabi--Yau three-fold, so studies from both sides agree nicely.

When $\rho = 20$, the lattice $\Lambda_T$ is of rank-2, positive definite, and
even. One can see that the vector space ${\rm Mod}(1,\rho_{\Lambda_S})$
is empty in the following way. For any $\Phi \neq 0$ in this vector space,
$\theta_{\Lambda_T} \cdot \Phi$ must be a scalar-valued weight-2 modular form
starting with $-2 + {\cal O}(q)$. Because there is no such modular form, the vector-valued modular form $\Phi$ should have been zero.

One can also arrive at almost the same conclusion independently by using
geometry available in the Type IIA language. If a Calabi--Yau three-fold
$X_{\rm IIA}$ has a K3-fibration with a generic fibre having $\rho = 20$
Neron--Severi lattice, the fibre K3 surface has a fixed complex structure
over the entire base $\P^1_{\mathrm{IIA}}$, so $X_{\rm IIA}$ must be of the form
$(\rho=20 {\rm ~K3}) \times \P^1_{\rm IIA}$. This is not a Calabi--Yau
three-fold, so there should not be such a K3-fibred Calabi--Yau three-fold.
It should be noted, however, that this second argument does not rule out
non-geometric phase\footnote{
Because we only think of Type IIA compactifications in the
$s_2 \gg 1$ regime, this premise almost implies that the base
$\P^1$ is in the geometric phase.
But we still need to find an appropriate technical language to extend this argument to cover constructions without entirely geometric $X_{\rm IIA}$.}
Type IIA constructions with $\rho=20$, and hence the first argument is stronger.

Similar arguments also rule out a family of $\rho = 19$ cases
\begin{align}
 \widetilde{\Lambda}_S = U^{\oplus 2} \oplus E_8[-1]^{\oplus 2} \oplus \vev{-2n},
    \qquad
 \Lambda_T = U \oplus \vev{+2n},
\end{align}
where $n \in \Z_{>0}$. The first argument for the $\rho =20$ cases can be
repeated by replacing $\theta_{\Lambda_T}$ with $\theta_{\vev{+2n}}$, to see
that the vector space ${\rm Mod}(3/2, \rho_{\Lambda_S})$ is empty.
So there is no suitable $\Phi$ in this case.

The absence of such a $\Phi$ (with $n_0 = -2$) is also understandable in terms of geometry.
The Fourier coefficients of $\Phi$ are the intersection
numbers of the Noether--Lefschetz divisors in $D(\Lambda_T)/\Gamma_T$
(they are points in the $\rho=19$ cases) and the image $\iota_\pi(\mathbb{P}^1_{\mathrm{IIA}})$ of the base $\mathbb{P}^1_{\mathrm{IIA}}$.
The moduli space\footnote{In the case of $\Lambda_T = U \oplus \vev{+2n}$,
${\rm Isom}(\Lambda_T) \cong \Gamma_0(n)_+$ (see \cite{Antoniadis:95-quartic}),
but $\Gamma_T \cong \Gamma_0(n)$.}
$D(\Lambda_T)/\Gamma_T = {\cal H}/\Gamma_0(n)$ contains the
large complex structure limit point
(when ${\cal H}/\Gamma_0(n)$ is compactified);
if $\iota_\pi$ is surjective, then there are points in the base $\P^1_{\rm IIA}$ where the K3-fibration is not
regular.
All the $\rho=19$ K3-fibrations studied in \cite{Doran-17} are in this category.
If $\iota_\pi$ were to be a constant map, then
$X_{\rm IIA}$ would not be a Calabi--Yau three-fold (see the second argument
for the $\rho=20$ case).

This geometry-based argument for the absence of 
Calabi--Yau three-folds with a $\Lambda_S$-polarized regular K3-fibration holds true for all the $\rho=19$ cases,
not necessarily for the $\Lambda_T = U \oplus \vev{+2n}$ cases discussed
above.
The authors do not have a proof yet that ${\rm Mod}_0(3/2, \rho_{\Lambda_S})$
is empty for such general $\Lambda_T$'s with $\rho=19$, however.

The consequence that the dimension of ${\rm Mod}_0(11-\rho/2, \rho_{\Lambda_S})$ is smaller for larger
$\rho$ is understandable intuitively in itself.
A K3 fibration is specified, after all, by specifying a map from the base $\P^1$ to
the period domain $D(\Lambda_T)$; less complicated geometry $D(\Lambda_T)$
allows less variety in the map from $\P^1$ to $D(\Lambda_T)$.

\subsection{Lower Bounds on Euler Numbers}
\label{ssec:bd-EulerN}

We have seen in Table \ref{tab:EulerNumber-BPSIndicies-rank1}
that the Euler number $\chi(X_{\rm IIA})$ of any Calabi--Yau three-fold
$X_{\rm IIA}$ that admits $\Lambda_S$-polarized K3-fibration is given
by a linear sum of the low-energy BPS indices $\{n_{|\gamma|}\}$, for a few
choices of $\Lambda_S$.
In Table \ref{tab:EulerNumber-BPSIndicies-rank1}, all the coefficients
of $\{n_{|\gamma|}\}$ are positive, from which it follows that $\chi(X_{\rm IIA})$
is bounded from below.
Actually, this is true for any choice of $\Lambda_S$, as we see below.

Suppose\footnote{
\label{fn:idea-derive-lin-constr}
There is a more general version of this argument.
A linear relation among Fourier coefficients of arbitrary
$\Phi' \in {\rm Mod}(2k+11-\rho/2,\rho_{\Lambda_S})$
may be obtained by using $\phi \in
{\rm Mod}(2k'+1+\rho/2,\rho_{\Lambda_S}^\vee)$; the combination
$\phi \cdot \Phi'$ must be a scalar-valued ${\rm SL}(2;\Z)$ modular
form of weight $(2k+2k'+12)$.
} $\phi \in {\rm Mod}(3+\rho/2, \rho_{\Lambda_S}^\vee)$. Then
\begin{align}
 \phi \cdot \frac{\Phi}{\eta^{24}} = -2 [\phi]_{q^0} \frac{E_4^2 E_6}{\eta^{24}}
\end{align}
because $\phi \cdot \Phi$ must be a scalar-valued ${\rm SL}(2;\Z)$ modular
form of weight 14 with the leading coefficient
$[\phi \cdot \Phi]_{q^0} = -2 [\phi]_{q^0}$. By comparing the coefficients of
the $q^0$ term on both sides,\footnote{
The same relation is obtained also from $0 = \oint_{\tau \sim i \infty} d\tau
\phi \cdot (\Phi/\eta^{24}) = [\phi \cdot (\Phi/\eta^{24})]_{q^0}$, because
$\phi \cdot \Phi/\eta^{24}$ is of weight 2 (e.g., \cite{Bruinier-bookBor}).}
we obtain one linear relation of
$\{ n_\gamma \}$ and $c_0(0) = -\chi(X_{\rm IIA})$ for one
$\phi \in {\rm Mod}(3+\rho/2,\rho_{\Lambda_S}^\vee)$.




A non-trivial $\phi$ can be constructed by a theta function for a suitable
lattice. The $(1, 7+\rho)$ lattice $\Lambda_S \oplus E_8[-1]$ has a primitive
sublattice\footnote{
\label{fn:how2find-LinLambdsS+E8}
To see this, choose an element $x \in \Lambda_S$ of positive norm $2n$;
there exists such $x$ for some $n > 0$, since $\Lambda_S$ is indefinite.
One can choose a primitive element $y \in E_8[-1]$ of norm $-2n$, and
also an element $z \in E_8[-1]$ such that $(y,z) = 1$. Now, the sublattice
$\operatorname{span}_\Z\{ x+y,\, z\} \subset \Lambda_S \oplus E_8$ is
isometric to $U$. } isometric to $U$.
Since $U$ is unimodular, the orthogonal complement
$L := [U^\perp \subset \Lambda_S \oplus E_8[-1]]$
satisfies
\begin{align}
    \Lambda_S \oplus E_8[-1] \cong U \oplus L.
   \label{eq:rho1-findL-for-chiExpr}
\end{align}
This lattice $L$ has signature $(0, 6+\rho)$ and shares the same discriminant
form $(G_S, (-,-)_{G_S})$ with $\Lambda_S$, so the lattice theta function\footnote{
For example, when $\Lambda_S = U \oplus W$ for some even lattice $W$ of
signature $(0,\rho-2)$, there is always an obvious embedding of $U$
into $\Lambda_S \oplus E_8[-1] = U \oplus W \oplus E_8[-1]$.
$L = W \oplus E_8[-1]$, so $\theta_{L[-1]} = \theta_{W[-1]} E_4$. }
$\theta_{L[-1]}$ is in ${\rm Mod}(3+\rho/2, \rho_{\Lambda_S}^\vee)$.
The linear relation for $\phi = \theta_{L[-1]}$,
\begin{align}
    \sum_{b \in L^\vee} c_{[b]}(b^2/2) = 0,
\end{align}
leads to
%
\begin{align}
  \chi(X) = - c_{0}(0)
    &= \sum_{b \in L^\vee}^{(-2 < b^2 < 0)} n_{[b]} + \sum_{b^2 = -2} c_{[b]}(-1) 
    &= \sum_{b \in L^\vee}^{(-2 < b^2 < 0)} n_{[b]} -2 \cdot \#(\text{roots of } L).
   \label{eq:chi-as-lincmb-leBPSind}
\end{align}
%
The sum over $b \in L^\vee$ is a finite sum since $L[-1]$ is positive definite and $c_\gamma(\nu) = 0$ for $\nu < -1$.
The last equality follows from $c_{\gamma}(-1) = 0$ for $\gamma_{\neq 0} \in G_S$ (see \eqref{eq:restriction-3-Het}).
For example, when $\Lambda_S = U \oplus W$ for an even lattice $W$,
the relation 
for $\phi = \theta_{W[-1]}E_4$
yields
\begin{align}
  \chi(X_{\rm IIA}) = \left( \sum {}_{b \in W^\vee}^{0> b^2 > -2} n_{[b]} \right)
    + (-2) \times | \{ b \in W \; | \; b^2 = -2\} | - 480.
\end{align}
In particular, $\chi = -960$ when $W = E_8[-1]$.

The linear coefficient of $n_{|\gamma|}$ on the right hand side of \eqref{eq:chi-as-lincmb-leBPSind} is positive for any
$|\gamma|_{\neq 0} \in G_S/\pm $,
because we have used a lattice theta function
for $\phi$. 
Then the lower bounds on $n_\gamma$'s lead to a lower bound
on $\chi(X)$:
\begin{align}
    \chi \geq -2 \cdot \#\{ b \in L^\vee ; -2 \leq b^2 < 0 \}.
    \label{eq:chi-bdd-blw-gen}
\end{align}
%
If there are no $\gamma\in G_S$ that satisfy the condition \eqref{eq:gamma_0-level-quantization-condition},
then $n_\gamma \geq 0$ for $\gamma_{\neq 0} \in G_S$, and
\begin{align}
    \chi \geq -2 \cdot \#(\text{roots of } L).
\end{align}

The relation (\ref{eq:chi-as-lincmb-leBPSind}) reproduces those in
Table \ref{tab:EulerNumber-BPSIndicies-rank1} by using
$\phi = \theta_{L[-1]}$ for an $L$ determined as in
footnote \ref{fn:how2find-LinLambdsS+E8}.
We applied the same procedure for some $\Lambda_S = \vev{+2n}$ not covered
in section \ref{ssec:examples-of-Phi}, and obtained the result summarized
in Table \ref{tab:rho=1-chiMin}; calculations that led to
Table \ref{tab:rho=1-chiMin} are found in appendix \ref{apdx:A22}.
Note that it is not necessary to work out a basis of
${\rm Mod}_0(11-\rho/2,\rho_{\Lambda_S})$ (where $\Phi$ lies in) in deriving these results.

\begin{table}[tbp]
\begin{center}
  \begin{tabular}{c||c|c|c|c|c|c|cc|}
    $2n$ & 2 & 4 & 6 & 8 & 10 & 12 & 14 & 14  \\
  \hline
   $\chi \geq $ & $-252$ & $-196$ & $-152$ & $-112$ & $-124$ & $-124$ & $-144$ & $-92$
  \end{tabular}
  \caption{\label{tab:rho=1-chiMin} Lower bounds on $\chi(X)$
for $X$'s that have a regular $\Lambda_S = \vev{+2n}$-polarized
K3-fibration. Two lower bounds are listed for $\vev{2n}=\vev{14}$
because we can construct two inequivalent lattices $L$. }
\end{center}
\end{table}

\bigskip

All the lower bounds of $\chi(X_{\rm IIA})$ for individual $\Lambda_S$ 
in Table \ref{tab:rho=1-chiMin} (and $\chi = -960$ for 
$\Lambda_S = U \oplus E_8[-1]$) are safely above the absolute lower bound 
for all Calabi--Yau three-folds $X_{\rm IIA}$ \cite{HS-C},
\begin{align}
  \chi \gtrsim - \frac{5}{3}e^{4\pi} \sim -5\times 10^5.
   \label{eq:bd-HS}
\end{align}
The bound (\ref{eq:bd-HS}) was derived \cite{HS-C} by exploiting the modular invariance of the fundamental string partition function of Type II compactification (with $X_{\rm IIA}$ as the target space),
while the bound (\ref{eq:chi-bdd-blw-gen}) is due to
the modular property of $\Phi$, 
the generating function of the helicity supertraces of
vertical D4--D2--D0 BPS bound states on $X_{\rm IIA}$.
Recall the latter modular property can be easily derived from that of the new supersymmetric index $Z_{\rm new}$ of the fundamental string in the Heterotic description. (See section \ref{sssec:review-Het}.)


\bigskip

One also finds from the relation (\ref{eq:chi-as-lincmb-leBPSind})
that the low-energy BPS indices also have an upper bound. This is
because
\begin{align}
\chi(X_{\rm IIA}) = 2h^{11} - 2h^{21} \leq 2h^{11} = 2(\rho + 1).
\end{align}
This is a generalization of the same observation made already
in section \ref{sssec:rho=1}.
Note, however, that all of the coefficients $n_\gamma$ do not necessarily appear in the equation  \eqref{eq:chi-as-lincmb-leBPSind};
in the case of $\Lambda_S = U \oplus \vev{-2n} =:_{ab} U \oplus \Z e$, for example,
the linear relation (\ref{eq:chi-as-lincmb-leBPSind}) for
$\phi = \theta_{\vev{+2n}}E_4$ has contributions only from
$\gamma = [\frac{x}{2n}e] \in G_S$ with $0 \leq x \leq \sqrt{4n}$
These are only $\mathcal{O}(\sqrt{n})$ coefficients among the $\mathcal{O}(n)$ low-energy BPS indices.
So, we cannot use this argument to claim that only finite choices of the low-energy BPS indices $\{n_\gamma \}$ correspond to geometric phases
in the Type IIA description.

\section{Finer Classifications}
\label{sec:finerC}

The modular form $\Phi$ (new supersymmetric index (Het)/Noether--Lefschetz number generating function (IIA)) is not enough discrete data for classification of branches of moduli space of the Het--IIA dual vacua.
Let us take the $\Lambda_S =U$ case as an example.
In the Heterotic language, all the ${\rm K3} \times T^2$ compactifications
with the 24 instantons on K3 distributed by $(12-n,12+n)$ to the two weakly
coupled $E_8$ gauge groups share the same $\Phi = (-2E_4E_6)$ for all
$0 \leq n \leq 2$, but they form three distinct branches of moduli space.
In the Type IIA language, the modular form
$\Phi$ determines the Gopakumar--Vafa invariants of all the vertical curve
classes of an elliptic-K3 fibred Calabi--Yau three-fold $X_{\rm IIA}$, but
the classical trilinear intersection numbers of the divisors are not completely determined\cite{HM}
(for precise statements, see section \ref{sssec:matching-summary}).
$X_{\rm IIA}$ can be any one of the elliptic fibrations over $F_n$ with $n=0,1,2$.
Presence of such multiple branches of moduli space sharing $\Phi$ has
been reported also for the case of
$\Lambda_S = \vev{+2}$ and $\vev{+4}$ \cite{KKRS} (see also \cite{BW-16}).

We introduce invariants of branches of moduli space of the Het--IIA dual vacua
that can distinguish those sharing a common $\Phi$. That is done by
developing observations and ideas that are found in the literatures.
Those invariants do not rely on supergravity approximation or explicit
construction of geometries, but use modular forms.

\subsection{The Idea}
\label{ssec:fineC-idea}

Consider a branch of moduli space of the Het--IIA dual vacua, where we have
special geometry and hypermultiplet moduli space of fixed dimensions
($h^{1,1}(X_{\rm IIA})$- and $(h^{2,1}(X_{\rm IIA})+1)$-dimensions, respectively,
if the branch contains a geometric phase in the Type IIA language).
We call it the original branch.
It often comes with special loci in the hypermultiplet moduli space
where non-abelian gauge symmetry ${\cal R}$ is enhanced in the effective
theory on $\R^{3,1}$; one ventures into other branches of moduli space
by turning on non-zero Coulomb vacuum expectation values in ${\cal R}$.
Modular forms denoted by $\Psi$ and $\underline{\Phi}$ are assigned to such
a symmetry-enhanced branch (see below for details); the idea is to use the
set of such modular forms as an invariant of the original branch.
We will see in this section \ref{sec:finerC} that the set of $\Psi$'s
or the set of $\underline{\Phi}$'s distinguish multiple branches sharing
the same $\widetilde{\Lambda}_S$, $\Lambda_T$, and $\Phi$; moreover,
the modular form $\Psi$ or $\underline{\Phi}$ of even just one
symmetry-enhanced branch attached to the original branch already
improves the classification by the modular form $\Phi$ alone.

\subsubsection{Higgs Cascades and Modular Forms}
\label{sssec:Higgs-cascade}

Let us first assign two modular forms $\underline{\Phi}$ and $\Psi$
for a symmetry-enhanced branch. We will discuss in
section \ref{sssec:matching-summary} the information of the target-space geometry in the original branch that we can extract from such a modular form $\Psi$ or from
the set of $\Psi$'s associated with all the symmetry-enhanced branches.

We restrict our attention to the case where
${\cal R}$ is one of ADE types
and its non-abelian gauge bosons are given by
left-moving level $k=1$ current algebra in the Heterotic language.\footnote{
\label{fn:level-1-restriction}
It is possible for a gauge group with level $k > 1$ to enhance,
although we do not use it as a probe in this article.
}
The lattice ${\cal R}[-1]$ is chosen within $\Lambda_T$; now we introduce
lattices
\begin{align}
 \underline{\Lambda}_T := \left[ {\cal R}[-1]^\perp \subset \Lambda_T \right], \qquad
 \underline{\widetilde{\Lambda}}_S := \left[ \underline{\Lambda}_T^\perp \subset
    {\rm II}_{4,20} \right]
\end{align}
for the symmetry-enhanced branch. The lattice
$\underline{\widetilde{\Lambda}}_S$ is $\widetilde{\Lambda}_S \oplus
\mathcal{R}[-1]$ or its extension.

It is assumed here that the symmetry-enhanced branch is also realized
without NS5-branes and the likes in the Heterotic description, or without
a degeneration of K3 fibre classified as Type II or III in the
Type IIA description. A related discussion is found at the end of
section \ref{sssec:deg2-calc}.

Under this assumption, there must be a modular form
\begin{align}
\underline{\Phi} \in {\rm Mod}_0(11-\underline{\rho}/2, \rho_{\underline{\Lambda}_S});
  \label{eq:Phipr-Renhanced}
\end{align}
it describes the BPS indices
of the Heterotic description and the Noether--Lefschetz numbers of the
K3-fibre in the Type IIA description in the symmetry-enhanced branch,
just like $\Phi$ does for the original branch.
Here $\underline{\rho} := \rho + {\rm rank}({\cal R})$, and
$\rho_{\underline{\Lambda}_S}$ the representation of ${\rm Mp}(2;\Z)$
associated with the lattice $\underline{\Lambda}_S$.

The modular form $\underline{\Phi}$ of the symmetry-enhanced branch
should be related to $\Phi$ of the original branch in the following way.
At the entrance of the symmetry-enhanced branch (so the Coulomb branch moduli
still stays within the subset $D(\widetilde{\Lambda}_S)$ of
$D(\underline{\widetilde{\Lambda}}_S)$), the non-abelian symmetry ${\cal R}$
remains unbroken. Since this vacuum belongs to both of the original and
symmetry-enhanced branches, the new supersymmetric index $Z_{\rm new}$ of both branches should be equal at this point:
\begin{align}
  \sum_{ \underline\gamma \in \underline{G}_S } \theta_{\underline{\widetilde{\Lambda}}_S[-1] + \underline\gamma}
    \underline{\Phi}_{\underline\gamma}
  =
  \sum_{ \gamma \in G_S } \theta_{\widetilde{\Lambda}_S[-1] + \gamma}
    \Phi_{\gamma}.
\end{align}
Here $\underline{G}_S = \underline{\widetilde{\Lambda}}_S^\vee/\underline{\widetilde{\Lambda}}_S$.
Define $G_0 := \underline{\widetilde{\Lambda}}_S^\vee/(\widetilde{\Lambda}_S \oplus \mathcal{R}[-1]) \subset G_S \times G_{{\cal R}}$
and $[\gamma,\delta] := (\gamma,\delta) + \underline{\widetilde{\Lambda}}_S \in \underline{G}_S$ for  $(\gamma,\delta) \in G_0$.
Then we can rewrite the left hand side of the above as
\begin{align}
    \sum_{(\gamma,\delta) \in G_0} \theta_{\widetilde{\Lambda}_S[-1] + \gamma} \theta_{\mathcal{R} + \delta} \underline{\Phi}_{[\gamma,\delta]}.
\end{align}
Therefore, the modular form $\underline{\Phi}$ of the symmetry-enhanced branch
should reproduce $\Phi$ of the original branch through
\begin{align}
    \Phi_\gamma = \sum_{(\gamma,\delta) \in G_0} \theta_{\mathcal{R} + \delta} \underline{\Phi}_{[\gamma,\delta]}.
   \label{eq:Fpr-FprRenhanced-relatn}
\end{align}
In the case of $\widetilde{\underline{\Lambda}}_S =
\widetilde{\Lambda}_S \oplus \mathcal{R}$, this simplifies to
 \cite{CurioLust, Stieberger-98-02, CDDHKW}
\begin{align}
  \Phi_\gamma = \sum_{\delta \in G_{\cal R}}
     \theta_{{\cal R} + \delta} \underline{\Phi}_{(\gamma,\delta)}.
   \label{eq:Fpr-FprRenhanced-relatn-EZ}
\end{align}

The modular form $\Psi$ is introduced in association with 1-loop\footnote{
\label{fn:Het1loop-in-eff-th-language}
The corrections to special geometry that are regarded as
1-loop contributions in the Heterotic string language are those in the Type IIA language that neither
diverge (tree in Het) nor vanish (non-perturbative in Het) in the large
base limit.} threshold correction $\Delta_{{\cal R}}$ to the coupling constant
of the enhanced non-abelian gauge group ${\cal R}$,
which is given by \cite{AGN-1, KL-95}
\begin{align}
  \Delta_{\cal R} \delta^{IJ}
    &=
    \int_{{\cal H}/{\rm SL}_2\Z} \frac{d\tau_1d\tau_2}{\tau_2} \;
      \parren{ {\cal B}^{IJ} - b_{\mathcal{R}} \delta^{IJ} },
      \label{eq:gauge-th-cr}
      \\
  {\cal B}^{IJ}
    &= \frac{-i}{\eta^2} {\rm Tr}_{\text{R-sector}}^{(c,\tilde{c})=(22,9)} \left[
      e^{\pi i F_R} F_R\; q^{L_0-\frac{c}{24}}
         \bar{q}^{\tilde{L}_0 - \frac{\tilde{c}}{24}}
       \left( \frac{Q^I Q^J}{2} - \frac{\delta^{IJ}}{8\pi \tau_2} \right) \right],
    \nonumber
\end{align}
where the Heterotic string is used as a language.
Here, $I, J \in \{ 1, \dots, {\rm rank}({\cal R})\}$ label
${\rm rank}({\cal R})$ left-moving free bosons\footnote{normalized
so $X_L^I(z)X_L^J(w) \sim - \delta^{IJ} \ln (z-w)$} $X_L^I$, and $Q^I$ is
the zero-mode momentum in the expansion $X_L^I(z) = x^I + Q^I \ln z + {\rm oscillators}$.
$Q^I$ works as the Cartan charge operator of $\mathcal{R}$.
$b_{\mathcal{R}} = (\sum_{\rm hyper}2T_{\mathrm{rep}})-2T_{\cal R}$ is the 1-loop beta
function of the probe gauge group ${\cal R}$.
Contracting the indices $I,J$,
we arrive at \cite{CDDHKW}
\begin{align}
 \Delta_{\cal R} & \; = \int_{{\cal H}/{\rm SL}_2\Z}
   \frac{d\tau_1d\tau_2}{\tau_2} \; ({\cal B}_{\cal R} - b_{\cal R} ),
      \label{eq:gauge-th-cr-2} \\
  {\cal B}_{\cal R} & \; = \frac{1}{24}
    \sum_{\gamma \in G_S} \theta_{\widetilde{\Lambda}_S[-1]+\gamma}
      \frac{\Phi_\gamma \hat{E}_2 - \Psi_\gamma}{\eta^{24}},  \\
  \Psi_\gamma & \; = - \frac{24}{{\rm rank}({\cal R})} \sum_{(\gamma,\delta) \in G_0}
     (\partial^S \theta_{{\cal R}+\delta}) \underline{\Phi}_{[\gamma, \delta]} .
   \label{eq:F-FprRenhanced-relatn}
\end{align}
Here $\partial^S$ is the Ramanujan--Serre derivative
(see appendix \ref{apdx:A1}).
Immediately from (\ref{eq:Phipr-Renhanced}, \ref{eq:F-FprRenhanced-relatn}),
\begin{align}
  \Psi \in {\rm Mod}_0(13-\rho/2,\rho_{\Lambda_S}).
   \label{eq:Psi-in-vctSpace}
\end{align}
Obviously the modular form $\Psi/\eta^{24}$ (for not necessarily unimodular
$\Lambda_S$) is the generalization of
$\Psi/\eta^{24} \in {\rm span}_\C \{ E_4^3, E_6^2 \} / \eta^{24}$
for $\Lambda_S = U$ and $\Psi/\eta^{24} = (-2 E_4^2)/\eta^{24}$
for $\Lambda_S = U \oplus E_8[-1]$ (e.g. \cite{HM, CurioLust, Stieberger-98-02, CDDHKW}).

The modular form $\Psi$ of the symmetry-enhanced branch captures
only a part of information in $\underline{\Phi}$, because $\Psi$ can
be determined from $\underline{\Phi}$ as in (\ref{eq:F-FprRenhanced-relatn}).
In fact, when there is a chain of symmetry enhancements
${\cal R}_1 \subsetneq {\cal R}_2 \subsetneq \cdots $ accompanied by a chain of tunings in the hypermultiplet moduli space, the chain
of the invariants
$(\underline{\widetilde{\Lambda}}_S, \underline{\Lambda}_T, \underline{\Phi})_{\mathcal{R} = \mathcal{R}_i}$
all reproduce one common modular form $\Psi$
through (\ref{eq:F-FprRenhanced-relatn}). This is because the corrections
to the gauge coupling constants remain unchanged by continuous change
in the hypermultiplet vacuum expectation values in ${\cal N}=2$ supersymmetric gauge theories
on $\R^{3,1}$, an observation implicit already in \cite{HM}.
For this reason, the modular form $\underline{\Phi}$ is assigned to
each symmetry-enhanced branch, but $\Psi$ to a chain of symmetry-enhanced
branches attached to the original branch (such a chain is called a
Higgs cascade).

We will see in section \ref{sssec:matching-summary} that the modular form
$\Psi$ for one Higgs cascade attached to the original branch---an arbitrarily
chosen cascade is fine---specifies the diffeomorphism class of $X_{\rm IIA}$
of the original branch (of the individual geometric phase chambers of the
original branch, to be more precise); the modular form $\Phi$ alone
does not have enough information for this purpose in general. Furthermore,
the set of $\Psi$'s for the set of Higgs cascades attached to the original
branch
can also be used as a classification invariant of the original branch.
This viewpoint is sometimes useful for distinguishing two different branches
of moduli space with the same diffeomorphism class of $\Lambda_S$-polarized
K3-fibred Calabi--Yau three-folds. The modular forms $\underline{\Phi}$
are more useful in capturing the network of symmetry-enhanced branches.

\subsubsection{The Space of the Modular Form $\Psi$'s}
\label{sssec:space-of-Psi}

For a given branch of the Het--IIA dual moduli space characterized by
$(\widetilde{\Lambda}_S, \Lambda_T, \Phi)$, the modular form $\Psi$
of a Higgs cascade of the original branch is not a completely arbitrary
element of the vector space (\ref{eq:Psi-in-vctSpace}).
We will derive a few constraints on $\Psi$ in the following.

\paragraph{General Constraints}

First, recall the definition:
\begin{align}
    \frac{\theta_{\widetilde{\Lambda}_S[-1]} \cdot (\Phi E_2 - \Psi)}{\eta^{24}} \delta^{IJ}
    = 12 \cdot \frac{-i}{\eta^2} {\rm Tr}_{\text{R-sector}}^{(c,\tilde{c})=(22,9)} \left[
      e^{\pi i F_R} F_R\; q^{L_0-\frac{c}{24}}
         \bar{q}^{\tilde{L}_0 - \frac{\tilde{c}}{24}}
         Q^I Q^J \right]
\end{align}
for $1 \leq I,J \leq \rank(\mathcal{R})$.
Choose a basis of the left-moving free bosons $X_L^I$ so that
roots of a fixed subalgebra $\mathfrak{su}(2) \subset {\cal R}$ have charges
only in $I=1$; $Q^I = \pm \sqrt{2} \delta^I_1$. Now set $I=J=1$ in the above equation.
Since all the states in the Hilbert space with a definite charge
under ${\cal R}$ have $Q^{I=1}Q^{J=1} \in \frac{1}{2}\Z$,
contributions from states with a charge $(w,Q^1)$ under $\widetilde{\Lambda_S}
\oplus \mathfrak{su}(2)$ and those with a charge $(w,-Q^1)$ add up to
be an integer.
This implies that
\begin{align}
 d_\gamma(\nu) \in 12 \Z, \qquad
    \frac{\Phi_{\gamma} E_2 - \Psi_{\gamma}}{\eta^{24}} =:
     \sum_{\nu \in h_{\mathrm{min}}(\gamma) + \Z} d_{\gamma}(\nu) q^\nu.
   \label{eq:ds-in-(12)}
\end{align}
As an immediate consequence, all the Fourier coefficients $c^\Psi_\gamma(\nu)$ in
\begin{align}
 \frac{\Psi}{\eta^{24}} = \sum_{\gamma \in G_S} e_\gamma \;
     \sum_{\nu \in \Q} c_\gamma^\Psi(\nu) q^\nu
\end{align}
are all integers.

In other words, this comes just from the properties of lattice theta functions of simple Lie algebra:
Using the relations (\ref{eq:F-FprRenhanced-relatn}, \ref{eq:Fpr-FprRenhanced-relatn}), we see that
\begin{align}
  \frac{\Phi_{\gamma} E_2 - \Psi_{\gamma}}{\eta^{24}}
    = 12\; \frac{2}{\rank(\mathcal{R})} \sum_{(\gamma,\delta) \in G_0}
      q \pdiff{\theta_{\mathcal{R}+\delta}}{q}  \cdot
      \frac{\underline{\Phi}_{[\gamma,\delta]}}{\eta^{24}}.
\end{align}
%
Defining $a_\delta^{({\cal R})}(\nu)$ by
 \begin{align}
    \sum_{\nu \in \delta^2/2 + \Z} a_\delta(\nu) q^\nu
    :=
    \frac{2}{\rank(\mathcal{R})} q \pdiff{}{q} \theta_{\mathcal{R}+\delta},
\end{align}
we have
\begin{align}
  d_\gamma(\nu) = 12 \sum_{(\gamma, \delta) \in G_0} \sum_{\nu' + \nu'' = \nu}
   a_\delta(\nu') \; \underline{c}_{[\gamma,\delta]}(\nu'').
  \label{eq:d-is-a*cbar}
\end{align}
The integrality of $a_\delta(\nu)$ can be seen in essentially the same way as the discussion above (see also appendix \ref{apdx:A23}). So we have $d_\gamma(\nu) \in 12\Z$, because the BPS indices $\underline{c}_{[\gamma, \delta]}(\nu)$ are also integers.

As we see later, consistency in the low-energy effective field theory
on $\R^{3,1}$ implies that $d_0(0) \in 24 \Z_{\geq 0}$. We have not tried much
to think whether this condition can be derived directly from consistency of string theory.

Not all the $d_\gamma(\nu)$'s are arbitrary integers divisible by 12 (or 24).
From the fact $a^{({\cal R})}_0(0)=0$ and 
that $\nu_\delta := \min\{ x^2/2 \mid x \in \delta \subset \mathcal{R}^\vee \}$ is strictly positive for $\delta \neq 0$, it follows immediately that
%
\begin{align}
  d_\gamma(-1) = 0, \qquad \qquad \forall \gamma \in G_S, \quad
      {\rm s.t.} \quad (\gamma,\gamma)/2 = 0 \in \Q/\Z.
  \label{eq:d=0-for-isotropics}
\end{align}
We can say a little more.
Define $m_\gamma := d_\gamma([\gamma^2/2]_{\mathrm{frac}}-1)$ for $(\Phi E_2 - \Psi)$,
like we defined $n_\gamma$ for $\Phi$:
\begin{align}
    m_\gamma = 12 \sum_{(\gamma,\delta) \in G_0}^{( [(\gamma,\gamma)/2]_{\rm fr}>\nu_\delta)}
           a_\delta(\nu_\delta) \, \underline{n}_{[\gamma,\delta]}.
\end{align}
Here only $\delta \in G_{\mathcal{R}}$ such that $[(\gamma,\gamma)/2]_{\rm frac} > \nu_\delta$ can contribute to the sum.
In particular,
\begin{align}
    m_\gamma = 0 \qquad \qquad \forall \gamma \in G_S, \quad {\rm s.t.}
   \quad [\gamma^2/2]_{\mathrm{frac}} \leq 1/4,
      \label{eq:low-ms=0}
\end{align}
regardless of ${\cal R}$, because $(\delta = e_1, \mathcal{R} = A_1)$
gives the smallest\footnote{
This is because $A_1$ is a Lie subalgebra of any $\mathcal{R}$.
} $\nu_\delta$ among all possibilities for $(\delta,\mathcal{R})$.
This supersedes the condition (\ref{eq:d=0-for-isotropics}).
See appendix \ref{apdx:A23} for a list of values of $\nu_\delta$ for
various ${\cal R}$'s.

Ref. \cite{HM} sets a constraint that Heterotic string tachyon states
should have zero contribution to the gauge threshold correction
$\Delta_{{\cal R}}$ (because they are not charged under ${\cal R}$) in relating $\Psi$ to $\Phi$, and this reasoning was enough to determine
$\Psi$ completely in the case of $\Lambda_S = U \oplus E_8[-1]$.
The relation \eqref{eq:d=0-for-isotropics} and \eqref{eq:low-ms=0} are
kinds of generalizations of this argument. On the other hand,
$m_\gamma$ for $\gamma$ satisfying $[\gamma^2/2]_{\mathrm{frac}} \geq 1/4$ can
be non-trivial in many cases. An easiest example is for
$\Lambda_S = U+A_1[-1]$ with an enhancement of symmetry ${\cal R}$
so that $A_1 + {\cal R}$ within contained in one weakly coupled $E_8$ of
the Heterotic string theory.

Let us introduce a space denoted by ${\rm Mod}_0^{\Phi}(13-\rho/2,\rho_{\Lambda_S})$ for a given $\Phi$, which is the set of modular forms $\Psi$ satisfying
$d_\gamma(\nu) \in 12\Z$, $d_0(0) \in 24\Z_{\geq 0}$, and (\ref{eq:low-ms=0}).
When the modular form $\Psi$ is for a Higgs cascade attached to the
original branch with $\Phi$, then $\Psi$ must be in this set.
For example, in the case of $\Lambda_S = U$ (where $\Phi = -2 E_4E_6$),
the space ${\rm Mod}_0^\Phi(12,\rho_U)$ consists of
$\Psi = -E_4^3-E_6^2 + (288-d(0)) \eta^{24}$ with $d(0) \in 24\Z_{\geq 0}$.

\paragraph{Field-theory Argument for $d_0(0) \in 24\Z_{\geq 0}$:}
It follows that $d_0(0) \in 24 \Z$, not just in $12 \Z$,
from its relation 
to the 1-loop beta function $b_{\mathcal{R}}$ through $b_{{\cal R}} = d_0(0)/24$ (e.g. \cite{HM});
this relation itself is obvious also from the expression
(\ref{eq:d-is-a*cbar}):
\begin{align}
  \frac{d_0(0)}{24} = - a_0^{({\cal R})}(1)
   + \sum_{(0,\delta)_{\neq 0} \in G_0}^{(0 < \nu_\delta < 1)}
      \frac{a_\delta^{({\cal R})}(\nu_\delta)}{2}
             \underline{n}_{[0,\delta]}(-\nu_\delta) \rightarrow
    b_{{\cal R}} = -2T_{{\cal R}} + \sum_{{\rm halfhyp}} T_\delta,
  \label{eq:beta_R=d00/24}
\end{align}
where we made replacements $a_0^{({\cal R})}(1) \rightarrow 2T_{{\cal R}}$ ands
$a_\delta^{({\cal R})}(\nu_\delta) \rightarrow 2T_\delta$ for $\nu_\delta < 1$ (see appendix \ref{apdx:A23}), and
interpreted $\underline{n}_{[0, \delta]} = \underline{n}_{[0,-\delta]}$
as the (effective) number of half-hyper multiplets in the corresponding representation.
Now, we see that the fact $d_0(0) \in 12\Z$ implies $b_{\mathcal{R}} \in \Z/2$.
But $b_{\mathcal{R}}$ can be in $1/2 + \Z$ only when $\mathcal{R} = A_1$ and there are odd number of hypers in the fundamental representation, which is not allowed because it would cause the SU(2) global anomaly. So $b_{\mathcal{R}} \in \Z$,
and $d_0(0) \in 24 \Z$.

In addition, $b_{\mathcal{R}}$ should be non-negative because there must be plenty of matter fields to Higgs the gauge symmetry $\mathcal{R}$ completely; the Higgsing brings the symmetry-enhanced branch back to the original branch.
In the case of $\Lambda_S = U$ with the probe symmetry set in one of the two weakly coupled $E_8$'s of the Heterotic string, for example, it is known
that there are $I:= 10+12^{-2}d(0)$ instantons in the rest of the $E_8$.
%
%
The condition that $d(0) \geq 0$ corresponds to the fact that
at least $I = 10$ instantons on K3 are necessary to break the $E_8$ symmetry completely.

As a side remark,
one notices (see appendix \ref{apdx:A23})
that the coefficients $a_\delta^{({\cal R})}(\nu)$ for smaller
values of $\nu$ are divisible by $2$ in $D_{r \geq 4}$, by $6=2T_{{\bf 27}}$
in ${\cal R} = E_6$, $12 = 2T_{{\bf 56}}$ in ${\cal R}=E_7$,
and by $60=2T_{{\bf 248}}$ in ${\cal R}=E_8$,
although the authors do not know if this property persists
for arbitrary large values of $\nu$.
So, a modular form $\Psi$ is such that $d_\gamma(\nu)$'s for small values 
of $\nu$
are divisible not just by 12, but by 24 [resp. $12 \times 6$, $12 \times 12$,
or $12 \times 60$] if the Higgs cascade to which $\Psi$ is assigned
has an enhanced symmetry ${\cal R}$ as large as\footnote{
Remember that we pose here the question how large an enhanced symmetry can be within Heterotic compactifications without 5-branes, or within Type IIA compactifications where the K3-fibration remains regular.}
$D_{r\geq 4}$ [resp. $E_6$, $E_7$, or $E_8$].

\paragraph{Extra Degrees of Freedom}

The set ${\rm Mod}_0^\Phi(13-\rho/2,\rho_{\Lambda_S})$ is parametrized by a finite number of $d_\gamma(\nu)$'s in $12\Z$.
Those with $\nu < 1$---the $m_\gamma$'s---are enough in the case of $\rho > 2$,
and those with $\nu < 2$ are enough if $\rho = 1,2$,
because $\Psi/\eta^{24}$ [resp. $\Psi/\eta^{48}$] has a negative weight when $\rho > 2$ [resp. $\rho =1,2$].
Those $d_\gamma(\nu)$'s (or equivalently the Fourier coefficients
$c_\gamma^\Psi(\nu)$'s) may be subject to some linear constraints,
just like we discussed for $\Phi$ in section \ref{sssec:lin-reltn}.

{\bf In the case $\rho=2$,}
the remaining freedom in the space ${\rm Mod}_0^\Phi(0,\rho_{\Lambda_S})$
not specified by the $m_\gamma$'s is in the free abelian group
${\rm Mod}^\Z(k=0, \rho_{\Lambda_S})$.
This abelian group is equivalent to that of a ($\tau$-independent\footnote{
Consider lifting $\phi \in {\rm Mod}(k=0, \rho_{\Lambda_S})$ to a modular
curve $\overline{{\cal H}/\Gamma(N)}$ where $\Gamma(N)$ is in the kernel
of the representation $\rho_{\Lambda_S}$. The lift $\phi$ should be $\C$-valued
functions on the compact curve, so it must be $\tau$-independent.})
vector $\phi = \{ \phi_\gamma = \Delta d_\gamma(0) \} \in 12\Z[G_S/\pm]$
invariant under $\rho_{\Lambda_S}(g)$ for any $g \in {\rm Mp}(2;\Z)$.

It is enough to make sure that $\{ \phi_\gamma \}$ is invariant under $\rho_{\Lambda_S}(T)$ and $\rho_{\Lambda_S}(S)$.
The invariance under
$\rho_{\Lambda_S}(T)$ implies that $\phi_\gamma$ can be non-zero only if $(\gamma,\gamma)/2 = 0 \in \Q/\Z$.
From the invariance under the $\rho_{\Lambda_S}(S)$,
it follows that a non-zero $\phi$ is possible
only when $G_S$ contains a non-zero isotropic element $\gamma_{\neq 0} \in G_S$
or $G_S = \{0 \}$ (so $\Lambda_S = U$). For most of rank-2 $\Lambda_S$'s,
therefore, there is no extra degree of freedom for $\Psi/\eta^{24}$.

Suppose that $G_S$ contains an isotropic\footnote{
A subgroup $H$ of
a discriminant group $G$ is {\it isotropic}, if the restriction
of the discriminant quadratic form on $H$ is trivial.} subgroup $H$
 and that $|H| = \sqrt{|G_S|}$. Whenever there is such a subgroup $H$,
the vector $\phi = \{ \phi_\gamma = 1 \; {\rm if} \; \gamma \in H, \;
\phi_\gamma = 0 \; {\rm otherwise} \}$ is invariant under $\rho_{\Lambda_S}(S)$.
So, there is one independent extra degree of freedom in the form of
$\Delta \Psi = x \phi \eta^{24}$,
with a parameter $x$.
For example, in the case of $\Lambda_S = U[N]$ for some integer $N > 1$, there are
two isotropic subgroups $H \cong \Z_N$ in $G_S = \Z_N \times \Z_N$, so
there are at least two extra degrees of freedom for $\Psi/\eta^{24}$ besides
$n_{|\gamma|}$'s and $m_\gamma$'s. We have neither been able to prove that
all the invariant vectors of a Weil representation are written as linear
combinations of vectors associated with isotropic subgroups $H$ with
$|H| = \sqrt{|G_S|}$, nor to find a counter example.

{\bf In the case of $\rho=1$,} namely, $\Lambda_S = \vev{+2n}$ for some
$n \in \N_{>0}$, the $2(n+1)$ integers $\{ \Delta m_\gamma \}$ and
$\{ \Delta d_\gamma({}_{0 \leq} \nu_{<1})  \}$ must be enough to parametrize
$\Psi$ for a given $\Phi$. They are often redundant, however.\footnote{
because the dimension of ${\rm Mod}_0(25/2,\rho_{\vev{+2n}})$ does not
grow as fast as $\sim (2n)$.} We do not have a general theory
about how many linear constraints exist within 
them
without relying on a case-by-case analysis.
About $\{ \Delta d_\gamma({}_{0 \leq} \nu_{ < 1})\}$,
at least we know that
there is one degree freedom not captured by $\{ m_\gamma \}$; we stay within
the set ${\rm Mod}_0^\Phi(25/2,\rho_{\vev{+2n}})$ under a change by
\begin{align}
 \frac{\Delta \Psi}{\eta^{24}} \propto  - (\Delta d_0(0)) \theta_{\vev{2n}}.
       \label{eq:extra-F-rho=1-univ}
\end{align}

With a case-by-case analysis, it is possible to find out which of those
$d_\gamma(\nu < 1)$'s are linearly independent, when an explicit basis of
the vector space ${\rm Mod}(13-\rho/2,\rho_{\vev{+2n}})$ is available.
An alternative
is to find a basis of ${\rm Mod}(k',\rho_{\vev{-2n}})$ with
$k'\equiv \rho/2 + 1$ mod 2, and find linear constraints
on $\Delta d_\gamma(\nu_{<1})$ as in the discussion in
footnote \ref{fn:idea-derive-lin-constr}.
A basis of ${\rm Mod}(k', \rho_{\vev{-2n}})$ can be worked out by using
the vector space of $\vev{+2n}$-polarized Jacobi forms of weight $(k'+1/2)$
(see \cite{DabMurZagier} or appendix \ref{apdx:A1}).
For any $\phi \in {\rm Mod}(7/2, \rho_{\vev{-2}})$ and
$\phi' \in {\rm Mod}(9/2, \rho_{\vev{-2}})$, for example,
the coefficients $a,b$ and $a',b'$ in
\begin{align}
  \phi \cdot \frac{\Psi}{\eta^{24}} = \frac{a E_4^4 + b E_4 E_6^2}{\eta^{24}},
    \qquad
  \phi' \cdot \frac{\Psi}{\eta^{24}} = \frac{a' E_4^3 E_6 + b'E_6^3}{\eta^{24}}
\end{align}
are determined by $\phi$ and a small number of
$\Delta d_\gamma(\nu_{<1})$'s by comparing the coefficients of
the $q^{-1}$ and $q^0$ terms; comparison of the coefficients of the
$q^1$ term yields a linear constraint on $\{ d_\gamma(\nu_{<1}) \}$ and
$d_\gamma(\nu=1)$ for isotropic $\gamma$'s.

\subsubsection{Modular Forms and Topological Invariants}
  \label{sssec:matching-summary}

Let $X$ [resp. $\underline{X}$] be a Calabi--Yau three-fold with
a regular $\Lambda_S$-polarized [resp. $\underline{\Lambda}_S$-polarized]
K3 fibration, and suppose that $\underline{X}$ with some cycles collapsed
is regarded as a limit of complex structure of $X$ in a way a complex
codimension-2 singularity of type ${\cal R}$ emerges along a curve
$C_{\cal R}$; $\lim_{\rm cpx~str}X = \lim_{\rm Kahler}\underline{X}$.
The modular forms $\Phi$ and $\underline{\Phi}$
assigned to $X$ and $\underline{X}$, respectively, determine such information as
Noether--Lefschetz numbers of $X$ and $\underline{X}$, but there are
also some topological invariants of $X$ and $C_{\cal R}$ that can be
determined from them \cite{HM}.

We start off with quickly reviewing
the matching relation between the data $(\Phi,\Psi)$ and the low-energy
effective theory, and proceed to discuss how we can use such modular
forms for classification of regular $\Lambda_S$-polarized K3-fibrations.


\paragraph{A Quick Summary of the Matching}

The low-energy effective theory on $\R^{3,1}$ has a prepotential ${\cal F}$,
the gravitational coupling $F_1$, and the gauge kinetic function $f_{{\cal R}}$
of the enhanced symmetry ${\cal R}$, when the Type IIA string is
compactified on $ \lim_{\rm cpx str}X = \lim_{\rm Kahler}\underline{X}$.
Those functions of the effective theory in the $s_2 \gg 1$
limit\footnote{We consider only the cases where the curve $C_{\cal R}$
covers the base $\P^1$ just once. (See footnote \ref{fn:level-1-restriction}.)}
\begin{align}
  {\cal F}^{\rm pert} = \frac{s}{2} (t,t) + f^{(1)}(t), \quad
  F_1^{\rm pert}, \quad
  f^{\rm pert}_{\cal R} = s + 4\pi i h^{(1)}(t)
\end{align}
are determined\footnote{See footnote \ref{fn:Het1loop-in-eff-th-language}.}
from the microscopic data $\Delta_{{\cal R}}$ and $\Delta_{\rm grav}$
through the relation\footnote{
$\hat{K} = - \ln (t_2, t_2) + \const$ is the (Heterotic string) tree-level
K\"{a}hler potential of the non-dilaton vector-multiplet scalars.}
\cite{KL-95, deWit, KapLusThs}
\begin{align}
  & 4\pi {\rm Re}(F_1^{\rm pert}) = 24 s_2 + \frac{1}{4\pi}
      \parren{24 V_{\mathrm{GS}} + \Delta_{\rm grav} - b_{\rm grav}\hat{K}},
      \label{eq:grav-matching}
      \\
  & 4\pi {\rm Re}(h^{(1)}) = \frac{1}{4\pi}
      \parren{ V_{\mathrm{GS}} + \Delta_{{\cal R}} - b_{{\cal R}} \hat{K} },
      \label{eq:gauge-matching}
      \\
  & V_{\mathrm{GS}} = \frac{4\pi}{(t_2,t_2)}
  \Im[ (1-it_2^a \partial_{t^a}) f^{(1)} ].
  \label{eq:Vgs-and-prept}
\end{align}

Those low-energy functions are of the following form,\footnote{
In an ${\cal N}=2$ field theory on $\R^{3,1}$, the prepotential itself is not
physical (e.g., \cite{deWit, AFGNT}); different choices of $(2n_V+2)$-tuple
of symplectic
sections $(X^I,F_I)$ related by a ${\rm Sp}(2n_V+2;\Z)$ duality transformation
may have different prepotentials (a prepotential does not exist for some
frames). The prepotential here is for a frame where a D2-brane wrapped on
any real 2-dimensional cycle (in the Type IIA language) is treated as an
electrically charged particle in $\R^{3,1}$.
cf section \ref{sssec:Het-IIA-dual-basic}.}
because we already assume a geometric phase Type IIA compactification:
\begin{align}
  \mathcal{F}^{\rm pert} &=
    \frac{1}{2}s(t,t)_{\Lambda_S}
      + \frac{d_{abc}}{3!}t^at^bt^c
      - \frac{\zeta(3)}{(2\pi i)^3} \frac{\chi}{2}
      + \frac{1}{(2\pi i)^3} \sum_{\beta_{\mathrm{eff}}}
        n_{\beta_{\mathrm{eff}}}^0 {\rm Li}_3(e^{2\pi i (\beta_{\mathrm{eff}}, t)}),
       \label{eq:prepot-pert-ansatz} \\
  4\pi i F_1^{\rm pert} &=
    24s + (c_2)_a t^a
    - \frac{2}{2\pi i} \sum_{\beta_{\mathrm{eff}}}
     ( n_{\beta_{\mathrm{eff}}}^0 + 12 n_{\beta_{\rm eff}}^1 )
              {\rm Li}_1(e^{2\pi i (\beta_{\mathrm{eff}}, t)}),
        \label{eq:F1-ansatz} \\
 s+ 4\pi i h^{(1)} &=
   s+ d'_a t^a
    - \sum_{\beta_{\mathrm{eff}}}
      \frac{A_{\beta_{\mathrm{eff}}}}{2\pi i } {\rm Li}_1(e^{2\pi i (\beta_{\mathrm{eff}}, t)})
      .    \label{eq:gauge-kin-ansatz}
\end{align}
Here, a component description $\{ t^{a=1,\dots, \rho} \}$ is given to
$t \in \Lambda_S \otimes \C$ by choosing an integral basis
$\{D_s,D_{a=1,\dots,\rho}\}$ of $H^2(X;\Z)$ consistent with the filtration
structure\footnote{
\label{fn:tot-fibr-class}
For a Calabi--Yau three-fold, the structure of K3-fibration
$\pi : X_{\mathrm{IIA}} \to \P^1_{\mathrm{IIA}}$ is in one-to-one with a divisor class $D_s$ of $X_{\mathrm{IIA}}$  satisfying $D_s^2=0$ and $\int_{X_{\mathrm{IIA}}} D_s \cdot c_2(TX_{\mathrm{IIA}}) = 24$.
The divisor class characterized in this way is the topological class of the K3 fibre over a generic point in the base $\P^1$ \cite{Aspinwall:1995vk}.
Choice of a divisor $D_a$ has ambiguity $D_a \to D_a + \delta n'_a D_s$ with $\delta n'_a \in \Z$.
In terms of the K\"ahler parameters, this corresponds to $s \to s - \delta n'_a t^a$
and $t^a$ unchanged.
} in (\ref{eq:filtr-coH2}); the divisors $\{D_{a=1,\dots, \rho}\}$ modulo $\Z D_s$
may be regarded as a basis of $\Lambda_S$. The complexified K\"{a}hler class of
$X$ is $t_{CY} = sD_s + t^a D_a = sD_s + t$ when $e^{2\pi i s}$ corrections are
ignored (as we will everywhere in this article).

The sums of exponential terms run over effective vertical curve classes
$\beta_{\rm eff}$, because we retain only the terms that remain non-zero
in the large base ($s_2 \gg 1$) region of the moduli space.
$n_{\beta_{\rm eff}}^r$ is a Gopakumar--Vafa invariant of $X_{\mathrm{IIA}}$. $A_{\beta_{\mathrm{eff}}}$ is related
$r=0$ Gopakumar--Vafa invariants of $\underline{X}_{\mathrm{IIA}}$.
It is well-known that the matching relations (\ref{eq:grav-matching}, \ref{eq:gauge-matching}, \ref{eq:Vgs-and-prept}) determine those parameters in terms
of the coefficients of $\Phi$ and $\Psi$ as (e.g. \cite{HM})
\begin{align}
  \chi = - c_0(0), \quad
  n^0_{w} = c_{[w]}(w^2/2), \qquad
  A_{w} = \frac{d_{[w]}(w^2/2)}{12}, \qquad
  n^1_{w} = \frac{\tilde{c}_{[w]}(w^2/2)-c_{[w]}(w^2/2)}{12}
  \label{eq:matching-chi+GV}
\end{align}
for any $w \in [H_2(X;\Z)]^{\rm vert} \subset \Lambda_S^\vee$ that is effective;
$\tilde{c}_\gamma(\nu)$ is the Fourier coefficient
$[E_2 \Phi_\gamma/\eta^{24}]_{q^\nu}$.
We have nothing to add or discuss about them in this article, however.

The non-exponential part of the low-energy functions capture topological invariants of
$X_{\mathrm{IIA}}$ and the curve $C_{\cal R}$
in $\lim_{\rm cpx~str} X = \lim_{\rm Kahler}\underline{X}$;
$\chi = \chi(X)$ is the Euler number, and
\begin{align}
    d_{abc} t^a t^b t^c &= \int_{X_{\mathrm{IIA}}} t \wedge t \wedge t,
    \qquad t = t^a D_a,    \\
    24s + (c_2)_a t^a & = \int_{X_{\mathrm{IIA}}} c_2(X_{\rm IIA}) \wedge t_{CY} =  24s + t^a \int_X c_2(TX) D_a,   \label{eq:2nd-Chern} \\
    s + d'_a t^a &= \vev{t_{CY}, C_{\cal R}} = s + \vev{t, C_{\cal R}}.
\end{align}
The coefficients $d'_a$ can also be regarded as trilinear intersection numbers in $\underline{X}$ among $D_a$ and a pair of exceptional divisors that emerge
after resolving the singularity of type ${\cal R}$.

Those invariants are determined by\footnote{
Here we have used $\tilde{s} = s+d'_at^a$ for convenience.
Since $d'_a$ is an integer, using $(\tilde{s},t^a)$ instead of $(s,t^a)$ corresponds to the integral basis change $D_a \to D_a -d'_a D_s$.
} the matching conditions (\ref{eq:grav-matching}, \ref{eq:gauge-matching}, \ref{eq:Vgs-and-prept}) as\footnote{
The 1-loop threshold $\Delta_{\mathrm{grav}}$ and $\Delta_{\mathrm{\mathcal{R}}}$ in \eqref{eq:grav-matching}, \eqref{eq:gauge-matching}---see also \eqref{eq:Delta-grav-gauge-as-Borch-integral}---are expanded as in \eqref{eq:unfold-2to1}; the last two lines of \eqref{eq:unfold-2to1} are used to determine the exponential part of the low energy functions and $\chi = -c_0(0)$.
The non-exponential part are determined by the first term in \eqref{eq:unfold-2to1}, or equivalently \eqref{eq:apdx:def-of-P_1P_3}.
}
\begin{align}
  {\cal F}_{\rm cub} := \frac{1}{2}s (t,t) + \frac{d_{abc}}{3!} t^at^bt^c
   & = \frac{1}{2}\tilde{s} (t,t)
    + \frac{1}{3!} P_3(t),
    \label{eq:matching-rslt-cubic-hol}  \\
 4\pi i (F_1)_{\rm nonexp} = 24 s + (c_2)_a t^a & =
       24 \tilde{s} + P_1(t),
      \label{eq:matching-rslt-F1-hol} \\
 s + 4\pi i (h^{(1)})_{\rm nonexp} = s + d'_a t^a & =: \tilde{s},
    \label{eq:matching-rslt-h1-hol}
\end{align}
where $P_3(t)$ and $P_1(t)$ are polynomials of $t$ given by the integrals over the fundamental region $SL(2;\Z)\backslash\UH$:
\begin{align}
    P_3(t_2) & := \frac{-t_2^2}{32\pi} \frac{\abs{t_2}}{\sqrt{2}}
    \int \frac{d\tau_1 d\tau_2}{\tau_2^{3/2}} \;
    \overline{\theta_{\Lambda_S}{(\tau,\bar\tau;t_2)}}\; \frac{\Phi(\tau)\hat{E}_2(\tau,\bar\tau)-\Psi(\tau)}{\eta^{24}},
    \label{eq:def-of-P_3}
    \\
    P_1(t_2) & := \frac{1}{4\pi} \frac{\abs{t_2}}{\sqrt{2}}
    \int \frac{d\tau_1 d\tau_2}{\tau_2^{3/2}} \;
     \overline{\theta_{\Lambda_S}(\tau,\bar\tau;t_2)}\; \frac{\Psi(\tau)}{\eta^{24}}.
     \label{eq:def-of-P_1}
\end{align}
See appendix \ref{sec:unfolding} for details of the integrals;
an evaluation method for the case $\Lambda_S$ has a non-trivial null element 
is reviewed in appendix \ref{ssec:if-nullV-exists}.
Any $\Lambda_S$ with $\rho \geq 5$ is known to have such a non-zero null 
element, and the same is also true for any lattice $\Lambda_S = U \oplus W'$ 
for some even lattice $W'$ of signature $(0,\rho-2)$.
Appendix \ref{ssec:embd-trick} explains how to reduce a case of $\Lambda_S$
without such an element to cases with such an element.

\paragraph{Discussion 1}
For a given $X$, choose any Higgs cascade attached to the branch of moduli
space\footnote{
Mathematically, the relation (\ref{eq:F-FprRenhanced-relatn}) can be
regarded as the definition of the modular form $\Psi$ of a Higgs cascade;
$\underline{\Phi}$ is defined as in \cite{MP}
(and reviewed in section \ref{sssec:review-IIA}).}
 of the Type IIA compactification of $X$. We see in the following that
the pair of modular forms $\Phi$ and $\Psi$ contains complete information
in specifying the diffeomorphism class of $X$.

Let us recall Wall's theorem \cite{Wall}, which states that the set of
diffeomorphism
classes of real six-dimensional, simply-connected, spin, and oriented
manifolds with torsion-free cohomology groups and a given set of Betti numbers
$b_2$ and $b_3$ are in one-to-one with the set
\begin{align}
  \left\{ (\mu, p_1) \; | \; \mu\in {\rm Hom}^{\rm Sym}
      ( H^2 \times H^2 \times H^2,\Z),  \; \;
     p_1 \in {\rm Hom}(H^2, \Z), \; \; (a), \; (b) \right\} / \sim,
\end{align}
where $H^2 \cong \Z^{\oplus b_2}$ and
\begin{itemize}
\item [(a)] $\mu(x,x,y) + \mu(x,y,y) \equiv 0$ mod 2 for any
  $x,y \in H^2$,
\item [(b)] $4\mu(x,x,x) - p_1(x) \equiv 0$ mod 24 for any
   $x \in H^2$;
\end{itemize}
the equivalence relation is such that $(\mu, p_1) \sim (\mu', p'_1)$ if and only if
there is an isomorphism $\phi: H^2 \rightarrow H^2$ so that
$(\mu',p'_1) = (\mu, p_1) \cdot \phi$. For a manifold $X$,
the trilinear symmetric form $\mu$ is the wedge product of $H^2(X;\Z)$,
and $p_1$ the linear form $\int_X p_1(TX) \wedge x$ for $x \in H^2(X;\Z)$.

Its subset of interest in this article is those where $H^2$ contains an element $D_s$ of the property described in footnote \ref{fn:tot-fibr-class}.
It is given by
\begin{align}
  {\rm Diff}_{\Lambda_S} :=
  \{ (d, c_2^{\Lambda_S}, \chi) \; | \;
  & d \in {\rm Hom}^{\rm Sym}(\Lambda_S \times \Lambda_S \times \Lambda_S, \Z), \; \;
  c_2^{\Lambda_S} \in {\rm Hom}(\Lambda_S,\Z), \; \chi \in \Z 
         \nonumber \\
   & 
  {\rm subject~to~}(a', \; b')
  \}
   / \sim_{\Lambda_S},
\end{align}
where $\Lambda_S$ is an even lattice of signature $(1,\rho-1)$ with $\rho = b_2 - 1$ and
\begin{itemize}
\item [(a')] $d_{aab} + d_{abb} \equiv 0$ mod 2 for any
   $a,b\in \{1,\dots, \rho\}$,
\item [(b')] $4d_{aaa} +2 (c_2)_a \equiv 0$ mod 24 for any
   $a \in \{1,\dots,\rho\}$;
\end{itemize}
$d_{abc}$ and $(c_2)_a$ for $a,b,c=1,\dots, \rho$ are the component description
of $d$ and $c_2^{\Lambda_S}$ for some basis $\{ D_{a=1,\dots,\rho} \}$ of $\Lambda_S$;
the equivalence relation is\footnote{Note that an element $D_s \in H^2$ with the property
in footnote \ref{fn:tot-fibr-class} is mapped by $\phi:H^2\cong H^2$
for the relation $(\mu, p_1)\sim (\mu',p'_1)$ to an element $\phi(D_s)\in H^2$
that also has the same property. } given by setting $(d, c_2^{\Lambda_S})
\sim_{\Lambda_S} (d', c_2^{\Lambda_S  \;'})$ if and only if they become identical
for some combination of isometries of $\Lambda_S$ and the basis changes
in footnote \ref{fn:tot-fibr-class}.

The modular form $\Phi$ of $X$ determines the combinations
%
\begin{align}
    P_3(t_2) - \frac{(t_2,t_2)}{8} P_1(t_2)
    = \frac{-\abs{t_2}^3}{32\pi\sqrt{2}}
    \int \frac{d\tau_1 d\tau_2}{\tau_2^{3/2}} \;
    \overline{\theta_{\Lambda_S}}\; \frac{\Phi\hat{E}_2}{\eta^{24}},
    \label{eq:P_3-P_1-Phi}
\end{align}
and hence the combinations
\begin{align}
  d'_{abc} := d_{abc} - \frac{[(c_2)_a C^{\Lambda_S}_{bc} + {\rm cycl.}]}{24},
  \quad\text{where}\quad
  C^{\Lambda_S}_{ab} := (D_a,D_b)_{\Lambda_S}.
    \label{eq:inv-from-Phi}
\end{align}
They remain invariant under the shifts
$D_a \rightarrow D_a + (\delta n'_a) D_s$ with $\delta n'_a \in \Q$ for
a basis $\{D_s, D_{a=1,\dots, \rho}\} \in H^2 \otimes \Q$. So the modular form
$\Phi$ of a Calabi--Yau three-fold $X$ determines an element of
\begin{align}
 {\rm Diff}_{\Lambda_S \; (\Q)} := \left\{ d'_{abc}:=
    (d_{abc}-[(c_2)_aC^{\Lambda_S}_{bc}+{\rm cycl.}]/24) \; | \;
    (a'), \; (b') \right\} / {\rm Isom}(\Lambda_S) \times
   \left\{ \chi \in \Z \right\}.
   \label{eq:def-coarse-classfy-set}
\end{align}

There may be a pair of three-folds $X$ and $X'$ sharing the same
modular form $\Phi$ that are not diffeomorphic to each other.
They must have the same combination
$d_{abc}-[(c_2)_aC^{\Lambda_S}_{bc} + \cdots]/24$, but $(d_{abc}, (c_2)_a)$ of
$X$ may be converted to that of $X'$ only by allowing the shifts with $(\delta n'_a+\Z) \neq 0 \in \Q/\Z$.
\begin{align}
  {\rm Diff}_{\Lambda_S}^{d'} := \left\{ (d_{abc}\in \Z, (c_2)_a \in \Z) \; |
      \; (a'), \; (b'), \; {\rm fixed~} d'_{abc}, \right\}/ \sim_{\Lambda_S}.
\end{align}
With just the modular form $\Psi$ of one arbitrary chosen Higgs cascade
of $X$ (along with $\Phi$), however, the
dictionary (\ref{eq:matching-rslt-cubic-hol},
\ref{eq:matching-rslt-F1-hol}, \ref{eq:matching-rslt-h1-hol})
determines $(d_{abc}, (c_2)_a)$ precisely with the relation
$\sim_{\Lambda_S}$,
because the integrality of $d'_a$ allows only the shifts $s\rightarrow
s - \delta n'_a t^a$ with $\delta n'_a \in \Z$.

To summarize, the modular forms $\Phi$ and $\Psi$ may be seen as
information of the spectrum of BPS states of string theory, or that
of Noether--Lefschetz numbers and curve counting invariants, but
they also carry full information of the diffeomorphism class of the
original manifold $X$. The way to extract the information has already been described.
\begin{align}
\xymatrix{
 & \{ X {\rm 's} \}_{\Lambda_S} \ar[ld]  \ar@{=>}[ldd] \ar[rd] \ar[rdd] \\
[{\rm Mod}_0^\Z(11-\frac{\rho}{2},\rho_{\Lambda_S})]^{\S 2} \supset
  [{\rm Mod}_0^\Z(11-\frac{\rho}{2},\rho_{\Lambda_S})]^{\rm r.mfd} \ar@{=>}[rr] &
 & {\rm Diff}_{\Lambda_S(\Q)} & {\rm Diff}_{\Lambda_S} \ar[l] \\
{\rm Mod}_0^\Phi(13-\frac{\rho}{2},\rho_{\Lambda_S}) \supset
  [{\rm Mod}_0^\Phi(13-\frac{\rho}{2},\rho_{\Lambda_S})]^{\rm r.mfd} \ar@{=>}[rr] &
 & {\rm Diff}_{\Lambda_S}^{d'} \ar[ur]
}
  \label{eq:comm-diagram}
\end{align}
The following discussions explain why some of the maps are drawn in
double lines.

For a given three-fold $X$, there may be multiple Higgs cascades attached
to the original branch of the moduli space, and hence multiple modular form
$\Psi$'s. The arrow from $\{ X\mbox{'s} \}$ to
${\rm Mod}_0^\Phi(12-\rho/2, \rho_{\Lambda_S})$ in (\ref{eq:comm-diagram})
is shown in a double line because of that.
Those $\Psi$'s should yield the same element in
${\rm Diff}_{\Lambda_S}^{d'}$. It follows that the difference $\Delta \Psi$
must be such that the resulting $\Delta P_1(t)$ is of the form
$24(\Delta d'_a) t^a$ with $(\Delta d'_a) \in \Z$.

One may also ask which subset of the diffeomorphism classes
${\rm Diff}_{\Lambda_S}$ of real six-dimensional manifolds are realized
under the restriction that $X$ is a Calabi--Yau three-fold (with $\Lambda_S$-polarized regular K3-fibrations).
Because we do not know well the set of such Calabi--Yau three-folds,
$\{ X \mbox{'s}\}$, one may think of applying the procedure of
assigning $(d'_{abc}, \chi)$ to an abstract general element
$\Phi \in {\rm Mod}_0^\Z(11-\rho/2,\rho_{\Lambda_S})$.
First, it is not true that the resulting $d'_{abc}$ can be always interpreted
as \eqref{eq:inv-from-Phi} for some $d_{abc} \in \Z$ and $(c_2)_a \in \Z$, satisfying (a') and (b'),
if we just require
that\footnote{
The subset of the $\Phi$'s that meet these requirements is denoted by
${\rm Mod}_0^\Z(11-\frac{\rho}{2},\rho_{\Lambda_S})]^{\S 2}$
in \eqref{eq:comm-diagram}.
}
$\Phi \in {\rm Mod}_0^\Z(11-\rho/2,\rho_{\Lambda_S})$ is subject
to the inequalities we derived in section \ref{sec:coarseC}; see
an example in appendix \ref{sssec:integrals-U+W}.
So, the subset of the $\Phi$'s whose $d'_{abc}$ backed by integer
$(d_{abc}, (c_2)_a)$ subject to (a',b') is denoted by
$[{\rm Mod}_0^\Z(11-\rho/2,\rho_{\Lambda_S})]^{\rm r.mfd}$.
Similarly, not a general element of
$\Psi \in {\rm Mod}_0^\Phi(13-\rho/2,\rho_{\Lambda_S})$ yields an element
of ${\rm Diff}_{\Lambda_S}^{d'}$ (see section \ref{ssec:ex-sect3}),
so those that fall into
${\rm Diff}_{\Lambda_S}^{d'}$ forms a subset denoted by
$[{\rm Mod}_0^\Phi(13-\rho/2,\rho_{\Lambda_S})]^{\rm r.mfd}$. We have the map
\begin{align}
 {\rm diff}_{\rm coarse}: [{\rm Mod}_0^\Z(11-\rho/2, \rho_{\Lambda_S})]^{\rm r.mfd}
   & \; \longrightarrow {\rm Diff}_{\Lambda_S \; (\Q)}, \\
 {\rm diff}_{\rm fine}: [{\rm Mod}_0^\Phi(13-\rho/2,\rho_{\Lambda_S})]^{\rm r.mfd}
   & \; \longrightarrow {\rm Diff}_{\Lambda_S}^{d'};
\end{align}
see (\ref{eq:comm-diagram}). The set of diffeomorphism classes represented
by Calabi--Yau three-folds must be\footnote{
A cautionary remark is that the diffeomorphism class of $X$
may not be contained in the image of ${\rm diff}_{\rm fine}$, if
$X$ does not have any symmetry-enhanced branch of $\underline{X}$
with a regular K3-fibration. } within the image of the map
$({\rm diff}_{\rm coarse}, {\rm diff}_{\rm fine})$ in ${\rm Diff}_{\Lambda_S}$.

The map $({\rm diff}_{\rm coarse}, {\rm diff}_{\rm fine})$ can be worked out
by dealing with purely mathematical objects. In setting up the relation
between the modular forms and diffeomorphism classes, however,
we have combined two physics observations under the Heterotic--Type IIA
string duality; the parameters $d_{abc}$ and $(c_2)_a$ in the low-energy
effective theory is determined i) by the topology
of the target space $X$ in a Type IIA string compactification, and
ii) also by 1-loop integrals in the Heterotic string where the integrands
are modular forms.
The ${\cal R}$-independence of $\Psi$
in a given Higgs cascade may well be proved purely in math,
although physics reasonings are enough; the claim that
the difference among the $\Psi$'s from different Higgs cascades
of a given original branch disappears
in the image of ${\rm diff}_{\mathrm{fine}}$
also relies on physics reasonings.

\paragraph{Discussion 2}
For a given $X$ with its modular form $\Phi$, one may specify a
curve class $C \in H_2(X;\Z)$ and ask whether complex structure
of $X$ can be tuned to have singularity of some type ${\cal R}$ along $C$
(and its resolution is still a regular K3-fibration).
In general, there is no guarantee that a modular form $\Psi$
exists in ${\rm Mod}_0^\Phi(13-\rho/2, \rho_{\Lambda_S})$
such that the resulting $P_1(t)$ and $P_3(t)$, hence
the right hand side of (\ref{eq:matching-rslt-cubic-hol}, \ref{eq:matching-rslt-F1-hol}, \ref{eq:matching-rslt-h1-hol}), reproduce all the input data $d_{abc}$, $(c_2)_a$ (from $X$) and $d'_a$ (from $C$) on the left hand side.
If there is no such modular form $\Psi$, we learn that complex structure of $X$ cannot be tuned in that way.
See section \ref{sssec:u-calc} for an example.

\paragraph{Discussion 3}
Suppose that a pair of Calabi--Yau three-folds $X$ and $X'$ have a
diffeomorphism between them, but not a holomorphic one-to-one map.
The Type IIA string
compactifications over $X$ and $X'$ form two different branches of
moduli space then. Such a pair of branches of moduli space cannot be distinguished by the invariants
${\rm diff}_{\rm coarse}(\Phi)$ and
${\rm diff}_{\rm fine}(\Psi)$.

Instead of finding the modular form $\Psi$ for one Higgs cascade of
the branch of $X$ and specify ${\rm diff}_{\rm fine}(\Psi)$, we can specify
the subset of $[{\rm Mod}_0^\Phi(13-\rho/2,\rho_{\Lambda_S})]^{\rm r.mfd}$
of all the $\Psi$'s of the Higgs cascades attached to the original
branch of $X$.
In some cases, this subset of $\Psi$'s is different for $X$ and $X'$, so that it works as an invariant for distinguishing such a pair of Calabi--Yau three-folds.
An example of such a case is discussed in section \ref{sssec:u-calc}.

That idea of extracting an invariant of a branch of moduli space
is faithful to the way we analyze the moduli space by using the low-energy
effective field theory. In practice, however, it is not easy to work out
all the possible ways to tune complex structure of a manifold to obtain
singularity. A close alternative to the idea of using
the set of $\Psi$'s of all the Higgs cascades is i) to think of all the
holomorphic curves in $H_2(X;\Z)$, ii) apply the reasoning in Discussion 2
to eliminate some of those curves, and finally, iii) to extract the set of
$\Psi$'s for those remaining curves. The latter set of $\Psi$'s contains the former set of $\Psi$'s.
Those two set of $\Psi$'s may be called as
the set of $\Psi$'s for curve classes (the latter)
and
the set of $\Psi$'s for Higgs cascades (the former).

The latter idea detects difference in the K\"{a}hler cone,
or in the cone of curves. Let $f:X \rightarrow X'$ be a diffeomorphism;
if $C \in H_2(X;\Z)$ is in the cone of curves of $X$, but $f_*(C)$ is not
in that of $X'$, then the modular form $\Psi$ for $C$ may be in the set
of $\Psi$'s for $X$, but not in the the set for $X'$. In the example of
section \ref{sssec:deg2-calc}, we discuss this latter set of $\Psi$'s.

\paragraph{Discussion 4}
The positive cone\footnote{Just one piece of the two connected components 
of the cone $(t_2,t_2)>0$ is (more than) enough in parametrizing the Coulomb 
branch moduli space 
$D(\widetilde{\Lambda}_S)/{\rm Ker}[
{\rm Isom}(\widetilde{\Lambda}_S) \rightarrow {\rm Isom}(G_S,q_s)]$. } 
of $\Lambda_S \otimes \R$ may contain
multiple chambers separated by walls orthogonal to some elements in
$\Lambda_S^\vee$; some of those chambers correspond to three-folds with
different topology \cite{top-change}. The modular form $\Phi$ remains
the same on both sides of the wall, but the integral
in (\ref{eq:P_3-P_1-Phi}) does not necessarily yield the same
polynomial on both sides of the walls (see appendix \ref{sssec:wall-xing}), 
and hence not necessarily the same $d'_{abc}$;
the arrow from
$[{\rm Mod}_0^\Z(11-\rho/2,\rho_{\Lambda_S})]^{\rm r.mfd}$
to
${\rm Diff}_{\Lambda_S\; (\Q)}$
in (\ref{eq:comm-diagram}) is shown
in a double line because of that.

There are not many things we can say with confidence about whether a symmetry-enhanced branch available on one side of the wall
continues to exist on the other side of the wall. We think it is
likely, however, if $\lim_{\rm cpx.str}X$ has singularity of type $\mathcal{R}$
along a curve $C \subset X$, then a flop transition on $X$
along a curve disjoint from $C$ yields a three-fold $X'$ that
continues to have a symmetry-enhanced phase with the singularity along $C$.
Even in such cases, the map ${\rm diff}_{\rm fine}$ depends
on the choice of a chamber, because the integrals $P_1$ and $P_3$ have
singularity along the walls.

Such invariants as ${\rm diff}_{\rm coarse}(\Phi)$ and
${\rm diff}_{\rm fine}(\Psi)$ are not assigned to branches of
Coulomb-and-hyper moduli space, but for branches of individual
chambers-and-hyper moduli space.

Because the integrals (polynomials) $P_3$ for $\Phi \in 
[{\rm Mod}_0^\Z(11-\rho/2;\rho_{\Lambda_S})]^{\S2}$ and $P_1$ for 
$\Psi \in {\rm Mod}_0^\Phi(13-\rho/2;\rho_{\Lambda_S})$ vary from one 
chamber to another, their subsets 
$[{\rm Mod}_0^\Z(11-\rho/2;\rho_{\Lambda_S})]^{\rm r.mfd}$ and 
$[{\rm Mod}_0^\Phi(13-\rho/2;\rho_{\Lambda_S})]^{\rm r.mfd}$ may also vary.\footnote{
This paragraph was added in v2.} 
In fact, we have confirmed that those subsets remain the same for 
all the chambers in the positive cone; for a given choice of 
$\Phi \in [{\rm Mod}_0^\Z(11-\rho/2;\rho_{\Lambda_S})]^{\S2}$ and 
$\Psi \in {\rm Mod}_0^\Phi(13-\rho/2;\rho_{\Lambda_S})$, suppose 
that $d_{abc}$ and $(c_2)_a$ are all integral and the conditions 
(a') and (b') are both satisfied when $P_3$ and $P_1$ in one chamber is 
used; the wall crossing formula (\ref{eq:Delta-P3}, \ref{eq:Delta-P1})
can be used to prove that $d_{abc}$ and $(c_2)_a$ are still integral 
and the conditions (a', b') are still satisfied for $P_3$ and $P_1$ 
in other chambers. 
The $\mathrm{diff}_{\mathrm{coarse}}$-image of 
$[{\rm Mod}_0^\Z(11-\rho/2;\rho_{\Lambda_S})]^{\rm r.mfd}$ and 
the $\mathrm{diff}_{\mathrm{fine}}$-image of 
$[{\rm Mod}_0^\Phi(13-\rho/2;\rho_{\Lambda_S})]^{\rm r.mfd}$ 
can be different sets for different chambers when seen in the data of 
$d_{abc}$ and $(c_2)_a$'s; when the images for all the chambers in the positive 
cone are joined, the symmetry ${\rm Isom}(\Lambda_S)$ is restored, and 
we can take a quotient $\sim_{\Lambda_S}$.

\paragraph{Discussion 5}
For Heterotic string compactifications reviewed in
section \ref{sssec:review-Het}, the invariants
such as the set of $\Psi$'s and the set of $\underline{\Phi}$'s are
assigned for (individual chambers of) the branches of moduli space.\footnote{
Their Type IIA dual do not necessarily have a geometric phase, so we may think of $\Phi$ in
${\rm Mod}^\Z(11-\rho/2,\rho_{\Lambda_S})$
rather than in
$[{\rm Mod}_0^\Z(11-\rho/2,\rho_{\Lambda_S})]^{\rm r.mfd}$.
}
Those invariants are given in terms of the CFT of the fundamental
string, and are well-defined, without relying on Heterotic supergravity approximation, or a geometric phase.

Those invariants beyond $(\Lambda_S, \Lambda_T, \Phi)$ generalize
the integer $n$ of the $(12+n,12-n)$ instanton number distribution
in the case of $\Lambda_S = U$, and detect difference among branches
of Heterotic string moduli space already present at perturbative
level (with corrections of order $(e^{2\pi i s})$ ignored);
see also \cite{KKRS, BW-16}.

\paragraph{Remark:}
This is a small side remark before closing this section \ref{ssec:fineC-idea}.
The modular form $\underline{\Phi}$'s for a chain of symmetry
enhancement ${\cal R}_1\subset {\cal R}_2 \subset \cdots $ also
contain how many hypermultiplet moduli need to be tuned to have the
enhanced symmetries in the chain:
\begin{align}
  \Delta h^{2,1}_{{\cal R}} & \; := h^{2,1}(X) - h^{2,1}(\underline{X}^{({\cal R})})
  =  h^{1,1}(X)-h^{1,1}(\underline{X}^{({\cal R})})
   - (\chi(X)-\chi(\underline{X}^{({\cal R})}))/2,   \nonumber \\
  & \; = - {\rm rank}({\cal R}) + [c_0(0) - \underline{c}^{({\cal R})}_0(0)]/2.
\end{align}
This $\Delta h^{2,1}_{\cal R}$ was used in \cite{BW-16} to distinguish four different branches
$X = M_{\vev{+2}}^{(n=2,1,0,-1)}$
of moduli space that share the same $\Lambda_S = \vev{+2}$
and $\Phi$ (those that are discussed in
section \ref{sssec:deg2-calc}). But it is enough to have
$\underline{\Phi}^{({\cal R})}$ without $\Delta h^{2,1}_{\cal R}$ as an invariant
of branches of the moduli space.

\subsection{Examples}
 \label{ssec:ex-sect3}

Let us see in simple examples how the invariants $\Phi$, $\Psi$ and ``the set of possible $\Psi$'s" work in distinguishing different branches of the moduli space.

\subsubsection{$\Lambda_S = U$}
\label{sssec:u-calc}

Let us begin with a traditional example, $\Lambda_S = U$.
The maps $({\rm diff}_{\rm coarse},{\rm diff}_{\rm fine})$
are worked out first.

We know that $\Phi = -2E_4E_6$ and a general element of
${\rm Mod}_0^\Phi(12,\rho_{U})$ is $\Psi = -(E_4^3 + E_6^2) +
(288-24b_{\cal R}) \eta^{24}$, where we use
$b_{\cal R} = d(0)/24 \in \Z_{\geq 0}$ as a parameter. It is known
that the integrals $P_3(t)$ and $P_1(t)$ for those $\Phi$ and $\Psi$
are given by \cite{DKL-2, HM, CurioLust, Stieberger-98-02}
\begin{align}
 \frac{1}{3!} P_3(t) & \; = \frac{2}{3!} \rho^3 + \frac{n'-2}{2} \rho^2u
     + \frac{n'}{2}\rho u^2, \\
 P_1(t) & \; = -4\rho + 12(2+n')(\rho+u),
\end{align}
where $n':=2-12^{-2}d(0) = 2- b_{\cal R}/6$.
Here, the component description of
$t \in \Lambda_S \tensor \C = U \tensor \C$
is that of (\ref{eq:Kahler-cone-parametrize-nullE}) associated with an obvious null element $z \in U$.
The expressions above are for the chamber
$0 \leq \rho_2 \leq u_2$;
those for the other chamber $0 \leq u_2 \leq \rho_2$
are obtained by exchanging $\rho$ and $u$.
As we will take a quotient by ${\rm Isom}(U)$,
which includes the $\rho \leftrightarrow u$ exchange,
it is enough to focus on the $0 \leq \rho_2 \leq u_2$ chamber in the following.

One can see that
\begin{align}
  [{\rm Mod}_0^\Z(10,\rho_U)]^{\rm r.mfd} & \; = [{\rm Mod}_0^\Z(10,\rho_{U})]^{\S 2} =
   \{ \Phi = -2E_4E_6 \}, \\
  [{\rm Mod}_0^\Phi(12,\rho_U)]^{\rm r.mfd} & \; =
  \{ b_{\cal R} = 0,6,\dots \} \subsetneq \{ b_{\cal R} \in \Z_{\geq 0} \}
  = {\rm Mod}_0^\Phi(12,\rho_U),
\end{align}
after working out details by using the expressions of $P_3$ and $P_1$ above.
The restriction on the value of $b_{\cal R}$ is from the integrality of
$d_{abc}$'s; once the condition $n' \in \Z$ is imposed, then  (a') and (b')
are automatically satisfied in this $\rho_{\Lambda_S}=U$ case.
Corresponding to the shift
$\tilde{s} \rightarrow \tilde{s} + (\delta n'_a)t^a$ with $\delta n'_a \in \Z$
is $\Delta n' = 2$, $\Delta b_{\cal R}=-12$, which mods out
$[{\rm Mod}_0^\Phi(12,\rho_U)]^{\rm r.mfd}$ in passing to
${\rm Diff}_{\Lambda_S}^{d'}$ by the map ${\rm diff}_{\rm fine}$.

The image of the map $({\rm diff}_{\rm coarse}, {\rm diff}_{\rm fine})$ must be in
\begin{align}
 {\rm Diff}_{\Lambda_S=U} &=
  \left\{ (d_{\rho\rho\rho}, d_{uuu}, N_\rho, N_u) \in \Z^{\oplus 4} \right\}/( \rho \leftrightarrow u ) \times \{ \nu = 0,1 \} \times \{ \chi \in \Z \}, \\
 (c_2)_\rho &= -2 d_{\rho\rho\rho} + 12 N_\rho + 24\delta n'_\rho, \quad
  d_{\rho \rho u} = \nu + 2\delta n'_\rho, \\
 (c_2)_u &= -2 d_{uuu} + 12 N_u + 24\delta n'_u, \quad
  d_{\rho u u} = \nu + 2 \delta n'_u.
\end{align}
Both $\nu = 0, 1$ of ${\rm Diff}_{\Lambda_S}^{d'}$ are
realized by the images of even $n'$ and odd $n'$. Only just one element of
${\rm Diff}_{\Lambda_S \; (\Q)} = \Z^{\oplus 4}/(\rho \leftrightarrow u) \times
\{ \chi \in \Z \}$ is
in the image of ${\rm diff}_{\rm coarse}$, however. It is the element
represented by
$d_{\rho\rho\rho} = 2$, $d_{uuu}=0$, $N_\rho = 4+\nu$, $N_u = 4+\nu$, and
$\chi = -480$.
The modular property of $\Phi$'s behind the scene indicates that the
diffeomorphism classes realized in the form of Calabi--Yau three-folds
are significantly less.\footnote{
In the case of $\Lambda_S=U$, this is not
surprising, because a simple argument \cite{MV-1, MV-2} shows that
$X^{(n)}$ with $n=0,1,2$ (explained shortly in the main text) are
all the possibilities.
}

As is well-known, there are Calabi--Yau three-folds for both of even $n'$
and odd $n'$. Think of a Weierstrass-model elliptic fibration over the
Hirzebruch surface $F_n$ that is Calabi--Yau, and denote it by
$X^{(n)}$; we denote by $D_7$ the zero-section divisor of the Weierstrass-model $X^{(n)}$.
The base surface $F_n$ is a $\P^1$-fibration over $\P^1_{\rm IIA}$,
where $D_f$ is the $\P^1$-fibre class, and the two sections denoted
by $D_+$ and $D_-$ have self-intersection $+n$ and $-n$, respectively.
The pull-back of the divisors $D_f$, $D_+$, and $D_-$ to $X^{(n)}$ are
denoted by $D_s$, $D_3$, and $D_4$, respectively. $D_3 \sim D_4 + n D_s$.
Some of the triple intersection numbers are
\begin{align}
&  D_7 \cdot \left( \begin{array}{cc}
    D_3 \cdot D_3 & D_3 \cdot D_s \\
    D_s \cdot D_3 & D_s \cdot D_s \end{array} \right) =
  \left( \begin{array}{cc} +n & 1 \\ 1 & 0 \end{array} \right), \\
&  D_s \cdot \left( \begin{array}{cc}
      (D_7+D_3)^2 & (D_7 + D_3)\cdot D_3 \\ D_3 \cdot (D_7+D_3) & D_3^2
    \end{array} \right) =
    \left( \begin{array}{cc} & 1 \\ 1 & \end{array} \right),
\end{align}
which mean that we can use $\{ D_s, (D_7+D_3), D_3\}$ as a basis of
$H^2(X^{(n)};\Z)$ in a way that $\{ (D_7 + D_3)_{+ \Z D_s}, (D_3)_{+\Z D_s} \}$
becomes a basis of $\Lambda_S = U$. We use the parametrization
$t_{CY} = s D_s + \rho(D_7+D_3) + u D_3$.
\begin{align}
   \int_X t_{CY}^3   & \; = s\rho u +
    \frac{2}{3!} \rho^3 + \frac{n-2}{2}\rho^2u + \frac{n}{2}\rho u^2,
    \label{eq:triple-U} \\
 \int_X c_2(TX) \wedge t_{CY} & \; =
     24 s -4\rho + 12(2+n)(\rho+u).
    \label{eq:c2-U}
\end{align}

Now, think of a symmetry-enhanced limit $\lim_{\rm cpx~str}X^{(n)}$ of
this $X^{(n)}$ so a singularity of type ${\cal R}$ emerges in the fibre of
$D_- \subset F_n$. Then $(f_{{\cal R}})_{\rm nonexp} = s = \tilde{s}$.
The modular form $\Psi$ for this Higgs cascade must be the one for
$n' = 2-12^{-2}d(0) = n$, because $\tilde{s}\rho u + P_3(t;n'=n)/3!$ and
$24\tilde{s} + P_1(t;n'=n)$ reproduces the topological invariants
(\ref{eq:triple-U}, \ref{eq:c2-U}) of $X^{(n)}$. In particular,
the set ${\rm Diff}_{\Lambda_S}^{d'} \simeq \Z/2\Z$ is realized indeed by
the diffeomorphism classes of $X^{(n=0)} \sim X^{(n=2)}$ and $X^{(n=1)}$.

It is possible to find a broader class of symmetry-enhanced limits of $X^{(n)}$
by using F-theory. Let $C' \in H^2(F_n;\Z)$ be represented by an
irreducible curve (we call it an irreducible curve class). Now,
choose $f$ and $g$ of the Weierstrass model $y^2 = x^3 + fx + g$ so
\begin{align}
 f = -3 h^2 + \sigma a + {\cal O}(\sigma^2), \qquad
 g = 2 h^3 - h \sigma a + {\cal O}(\sigma^2),
  \label{eq:ell-on-Fn-A1Lim}
\end{align}
where $h \in \Gamma(F_n; {\cal O}(-2K_{F_n}))$,
$a \in \Gamma(F_n; {\cal O}_{F_n}(-4K_{F_n}-C'))$, and
$\sigma \in \Gamma(F_n;{\cal O}_{F_n}(C'))$ \cite{6authors};
in this limit, the three-fold $X^{(n)}$ has a singularity of type
${\cal R}=A_1$ along a curve $C$ in the fibre of the $\sigma = 0$ curve
in $F_n$. Think of $C'$ of the form\footnote{
We have discussed the invariant $\Psi$ for the Higgs cascade with $C' = D_-$ ($m=0$);
the choice $C' = D_+$ ($m=n$) corresponds to placing the probe gauge group ${\cal R}$ in the other weakly coupled $E_8$ in the Heterotic language,
but there are more varieties ($m$) in the symmetry-enhancement limits.
}
$C' \sim D_- + m D_f$
labeled by $m \in \Z$, so that $C' \cdot D_f = 1$ in $F_n$, and
$C \cdot D_s = +1$ in $X^{(n)}$.
For this type of symmetry-enhanced limit (Higgs cascade),
we have\footnote{
In this construction
of $\lim_{\rm cpx~str}X^{(n)}$, the curve $C$ of $A_1$ singularity is along
$(x,y)=(c,0)$ in the elliptic fibre, so it does not touch the zero section
divisor $D_7$. So $D_7 \cdot C=0$.
}
  $(f_{\cal R})_{\rm nonexp} = s + m(\rho+u) = \tilde{s}$.
Therefore, we find that the modular form $\Psi$ of this Higgs cascade
must be that of $n'=n-2m$ ($b_{{\cal R}}/6 = 2m+2-n$),
which we find by requiring that the topological invariants
(\ref{eq:triple-U}, \ref{eq:c2-U}) of $X^{(n)}$ must be reproduced by
$\tilde{s}\rho u + P_3(t)$ and $24\tilde{s} + P_1(t)$,
respectively.

The class $C'$ is represented by a curve when $m \geq 0$;
the curve is not irreducible for the choice $m = 1$ in the case of $n=2$,
however, because the divisor $D_- + D_f$ has $D_-$ as a base locus then.
There is also an upper bound, $m \leq 8+4n$; when the divisor class $-4K_{F_n} - C'$ is not effective, $X^{(n)}$ has singular  fibre over any point on $F_n$ because $f= -3h^2$ and $g=2h^3$.
The effectiveness of $-4K_{F_n} - C'$ is translated into the upper bound.
So, we have found a class of Higgs cascades
attached to the branch of the moduli space of $X^{(n)}$ whose invariants are
\begin{alignat}{2}
  X^{(n=2)}: \quad & \{ \Psi_{b_{\cal R}/6 = 2m+2-2} \; | \;
    m = 0, 2,3,\dots, 16 &&\} \subset [{\rm Mod}_0^\Phi(12,\rho_U)]^{\rm r.mfd},  \\
  X^{(n=0)}: \quad & \{ \Psi_{b_{\cal R}/6 = 2m+2-0} \; | \;
         m = 0,1,\dots, 8 &&\} \subset [{\rm Mod}_0^\Phi(12,\rho_U)]^{\rm r.mfd}, \\
  X^{(n=1)}: \quad & \{ \Psi_{b_{\cal R}/6 = 2m+2-1} \; | \;
         m = 0,1, \dots, 12 &&\} \subset [{\rm Mod}_0^\Phi(12,\rho_U)]^{\rm r.mfd}.
\end{alignat}
The difference among $\Psi$'s for one given $X^{(n)}$ is precisely
of the form we expected in Discussion 1.
The set of $\Psi$'s of $X^{(n=2)}$ and $X^{(n=0)}$ are not the same, however,
reflecting the fact that this pair of three-folds have a diffeomorphism
but not a holomorphic one-to-one map between them, and the K\"{a}hler
cones are not identical when
$H^2(X^{(2)};\R)$ and $H^2(X^{(0)};\R)$
are identified by using the diffeomorphism between them.

One will wonder if there are other symmetry-enhancement limits of $X^{(n)}$.
At least we can rule out cases where singularity of type ${\cal R}$ emerges
along a curve $C$ satisfying $C \cdot D_s = 1$ and $C \cdot D_7 \neq 0$
(Discussion 2). To see this, suppose that there is such a limit. Then
$(f_{{\cal R}})_{\rm nonexp} = s + m_\rho \rho + m_u u = \tilde{s}$ with
$m_\rho \neq m_u$; no choice
of $\Psi$ from $[{\rm Mod}_0^\Phi(12,\rho_U)]^{\rm r.mfd}$ for the polynomials
$P_3(t)$ and $P_1(t)$ can reproduce ${\cal F}_{\rm cub}$ and $F_1$ appropriate
for $X^{(n)}$, and hence the assumption must be wrong.\footnote{
This argument still allows a limit of $X^{(n)}$ whose singularity resolution
$\underline{X}^{(n)}$ does not have a {\it regular} K3-fibration over
$\P^1_{\rm IIA}$.} We do not have an argument to rule out the possibility that there are
other limits of $X^{(n)}$ for symmetry-enhancements with $C \cdot D_s =
C \cdot D_7 = 0$ that cannot be obtained in the
form (\ref{eq:ell-on-Fn-A1Lim}).

\subsubsection{$\Lambda_S = \vev{+2}$}
\label{sssec:deg2-calc}

Consider the case $\Lambda_S = \vev{+2} \cong_{ab} \Z e$ now, where
the modular form $\Phi$ is of the form (\ref{eq:deg2-Phi-paramtrz-main}).
The modular form
$\Psi \in {\rm Mod}_0^\Phi(25/2, \rho_{\vev{+2}})$ is parametrized
by $m_0 = d_0(-1)$, $m_{1/2} = d_{1/2}(-3/4)$, and one more,
because $\dim_\C {\rm Mod}_0(25/2,\rho_{\vev{+2}}) = 3$.
We can use the $d_0(0) = 24 b_{{\cal R}}$ as the third parameter.\footnote{
$d_{1/2}(1/4)$ must be linearly dependent with the other three.}
Just one among them, $d_0(0) \in 24\Z_{\geq 0}$, is the free parameter,
while $m_0 = m_{1/2} = 0$ because of (\ref{eq:low-ms=0}).
So, a general element $\Psi$ in ${\rm Mod}_0^\Phi(25/2,\rho_{\vev{+2}})$
is given by\footnote{
The modular form $J$ of \cite{Kawai} corresponds to
$\Psi = - J \eta^{24}$ with $n_{1/2}=0$ and $d_0(0) = 300$. }
\begin{align}
\Psi &= \frac{E_6}{E_4} \Phi - (d_0(0)+1440) \theta_{\vev{+2}} \eta^{24}
   \label{eq:deg2-Psi-paramtrz-main} \\
 &=
 e_0 \left[ -2 + ( 348 - d_0(0) - 56 n_{1/2} ) q
     + ( - 280656 + 22 d_0(0) + 27984 n_{1/2} ) q^2 + \cdots \right] \nonumber \\
  & \quad
 + e_{1/2} \left[ n_{1/2} q^{\frac{1}{4}} + ( -384 -2d_0(0) - 384n_{1/2} )q^{\frac{5}{4}}
     \right. \nonumber \\
 & \qquad \qquad \qquad \qquad \qquad \qquad \qquad \qquad \left.
    + (-1122304 + 48d_0(0) - 77103n_{1/2}) q^{\frac{9}{4}} \cdots \right].
   \nonumber
\end{align}

We computed the integrals $P_3$ and $P_1$ in (\ref{eq:def-of-P_3},
\ref{eq:def-of-P_1}) for those $\Phi$ in (\ref{eq:deg2-Phi-paramtrz-main})
and $\Psi$ in (\ref{eq:deg2-Psi-paramtrz-main}) parametrized by
$n_{1/2}, b_{\cal R} \in \Z_{\geq 0}$; details\footnote{
One can employ the ``embedding trick" to apply the lattice unfolding method.
} are left to appendix \ref{ssec:embd-trick}, and only the result
is shown here:
\begin{align}
  \frac{1}{3!}P_3 (t) = \frac{(4-b_{\cal R}-n_{1/2})}{3!} (t^{a=1})^3, \qquad
  P_1(t) = (52-4b_{\cal R}-10n_{1/2}) \; (t^{a=1}),
\end{align}
where $t \in \Lambda_S \otimes \C$ is parametrized by
$t = (t^{a=1})e$ with $t^{a=1} \in \C$. So, the dictionary
(\ref{eq:matching-rslt-cubic-hol}, \ref{eq:matching-rslt-F1-hol},
\ref{eq:matching-rslt-h1-hol}) yields
\begin{align}
   d_{111} + 6 \delta n'_a = 4-b_{\cal R} -n_{1/2}, \qquad
   (c_2)_1 + 24 \delta n'_1 = (52 - 4b_{\cal R}-10n_{1/2})
  \label{eq:deg2-dict-tpinv-mdfpara}
\end{align}
for some $\delta n'_1 \in \Z$. By comparing this with
\begin{align}
 & {\rm Diff}_{\Lambda_S = \vev{+2}} = \left\{ N \in \Z \right\} \times
       \left\{ \nu \in \{ 0,1,2,\dots,5\} \right\}, \\
 & d_{111} = \nu + 6 (\delta n'_1), \qquad
  (c_2)_1 = 12 N -2\nu + 24 (\delta n'_1), 
\end{align}
we find that
\begin{align}
  [{\rm Mod}_0^\Z(21/2,\rho_{\vev{+2}})]^{\rm r.mfd} & \;
   = [{\rm Mod}_0^\Z(21/2,\rho_{\vev{+2}})]^{\S2} = \{ n_{1/2} = 0,1,2,3,4\}, \\
  [{\rm Mod}_0^\Phi(25/2,\rho_{\vev{+2}})]^{\rm r.mfd} & \;
   = \{ b_{\cal R} = 0,2,4,\dots \}
   \subsetneq \{ b_{\cal R} = 0,1,2,\dots \}
     = {\rm Mod}_0^\Phi(25/2,\rho_{\vev{+2}}).
\end{align}
Within the set ${\rm Diff}_{\Lambda_S \; (\Q)} = \{ (2/3) (d_{111}-(c_2)/4) =
(\nu-2N) \in \Z \} \times \{ \chi \in \Z\}$, the image of
${\rm diff}_{\rm coarse}$ consists of
$\{(\nu-2N, \chi) = (n_{1/2}-6, -252+56n_{1/2}) \; | \; n_{1/2} =0,1,2,3,4\}$.
For a given $d'_{111}$,
all the three elements of
${\rm Diff}_{\Lambda_S}^{d'} \simeq \Z/3\Z =
\{ \Delta \nu/2 = \Delta N = 0_{+3\Z},1_{+3\Z},2_{+3\Z} \}$ are realized by
the image of the map ${\rm diff}_{\rm fine}$ of
$\{ b_{\cal R} = 0,2,4, \dots \} / \{ \Delta b_{\cal R} = 6\}$.

Some diffeomorphism classes of three-folds with a $\vev{+2}$-polarized regular
K3-fibration are constructed by using toric technique, and are listed up
in \cite[Table 1]{KKRS}.\footnote{
 Here we cited only the case $\Lambda_S = \vev{+2}$. They treat also the
case $\Lambda_S = \vev{+4}$ and $\vev{+6}$.}
Those in the list must have the invariants $\Phi$ and
$\Psi + \{ \Delta \Psi \}$ specified in Table \ref{tab:KKRS-intepret-deg2}.
Such choices as $(n_{1/2}, b_{\cal R}+6\Z) = (1,4_{+6\Z})$, $(2, 4_{+6\Z})$,
$(3,2_{+6\Z})$, $(3, 4_{+6\Z})$, and all of $(4,2\Z/6\Z)$ are in the
images of the map $({\rm diff}_{\rm coarse}, {\rm diff}_{\rm fine})$ but are
not found in the table; we have not made an effort to search in a
larger Calabi--Yau topology database.
\begin{table}[tbp]
\begin{center}
    \begin{tabular}{|ccc|cc|} \hline
        \multicolumn{3}{|c|}{\cite[Table 1]{KKRS}} &
        \multicolumn{2}{|c|}{ discrete parameters } \\ \hline
        $\chi$ & $d_{111}$ & $(c_2)_1$ & $n_{1/2}$ & $b_{\mathcal{R}}$ modulo $6\Z$ \\ \hline
        $-252$ & 4 & 52 & 0 & 0 \\
               & 2 & 44 &   & 2 \\
               & 0 & 36 &   & 4 \\
        $-196$ & 3 & 42 & 1 & 0 \\
               & 1 & 34 &   & 2 \\
        $-140$ & 2 & 32 & 2 & 0 \\
               & 0 & 24 &   & 2 \\
        $-84$  & 1 & 22 & 3 & 0 \\
        \hline
    \end{tabular}
\caption{\label{tab:KKRS-intepret-deg2}Topological invariants of
Calabi--Yau three-folds with a regular $\vev{+2}$-polarized K3-fibration
quoted from \cite{KKRS}.  Corresponding parameters $(n_{1/2}, b_{\cal R}+6\Z)$
of the modular forms $\Phi$ and $\Psi$ are shown on the right.  }
\end{center}
\end{table}

There is at least one pair of three-folds in the same diffeomorphism class
but a holomorphic one-to-one map between them may or may not exist, also
in the $\Lambda_S =\vev{+2}$ case. The three-folds denoted by $M_{\vev{+2}}^{(n)}$
with $n=2,1,0,-1$ in \cite{BW-16} are in the diffeomorphism classes
with $n_{1/2}=0$ for all of them, and $b_{{\cal R}} + 6\Z = 0_{+6\Z}$, $2_{+6\Z}$,
$4_{+6\Z}$, and $0_{+6\Z}$, respectively; the pair $M_{\vev{+2}}^{(n=2)}$ and
$M_{\vev{+2}}^{(n=-1)}$ are in the same class. Unlike the diffeomorphic
pair $X^{(n=2)}$ and $X^{(n=0)}$ in the $\Lambda_S=U$ case, however,
the K\"{a}hler cones of $M_{\vev{+2}}^{(n=2)}$ and $M_{\vev{+2}}^{(n=-1)}$
are identical.  To see this, let $\{ D_s^{(n)}, D_{a=1}^{(n)} \}$
be the basis of $H^2(M_{\vev{+2}}^{(n)};\Z)$ characterized by
$(D_{a=1}^{(n)})^3 = 2n$, and $\{ \Sigma_B^{(n)}, \Sigma_F^{(n)}\}$
its dual basis of $H_2(M_{\vev{+2}}^{(n)};\Z)$. Toric techniques are
used to find that $\Sigma_B^{(n)}$ and $\Sigma_F^{(n)}$ generate
the cone of curves for $n = 0,1,2$, while the generators
should be $(\Sigma_B^{(n)} + n \Sigma_F^{(n)})$ and $\Sigma_F^{(n)}$
for $n = 0,-1$. It is not hard to find that the diffeomorphism
$f: M_{\vev{+2}}^{(n=2)} \rightarrow M_{\vev{+2}}^{(n=-1)}$ maps
$\Sigma_F^{(2)}$ to $\Sigma_F^{(-1)}$ and $\Sigma_B^{(2)}$ to
$\Sigma_B^{(-1)}-\Sigma_F^{(-1)}$, so the cone of curves are
identical indeed.

For the three-folds $M_{\vev{+2}}^{(n)}$ with $n=2,1,0,-1$,
the Discussion 2 cannot rule out
a possibility for any $C$ in the cone of curves that complex structure of
$M_{\vev{+2}}^{(n)}$ can be tuned to have singularity of type ${\cal R}$
along $C$. For a curve class $C = \Sigma_B^{(n)} + m \Sigma_F^{(n)}$
(with $m \in \Z_{\geq 0}$ for $n \geq 0$), we have $f_{\cal R} = s + m (t^{a=1})$.
The dictionary (\ref{eq:deg2-dict-tpinv-mdfpara}) with $n_{1/2}=0$
and $-\delta n'_{a=1} = m$ reads $2n = 4-b_{\cal R} +6m$. So, the
modular form $\Psi$ is that of $b_{\cal R} = 6m +4-2n$, if it is possible
to tune complex structure and the corresponding Higgs cascade exists
for this class $C = \Sigma_B^{(n)} + m \Sigma_F^{(n)}$.
The set of $\Psi$'s for curve classes
is
\begin{align}
  M_{\vev{+2}}^{(n)}: \qquad & 
     \left\{ \Psi_{b_{\cal R} = 6m + 4-2n} \; | \; m \geq 0 \right\}
     \subset [{\rm Mod}_0^\Phi(25/2,\rho_{\vev{+2}})]^{\rm r.mfd}.
\end{align}
The three-folds $M_{\vev{+2}}^{(n)}$ with different $n_{+3\Z}$ are also distinguished
in this way. For further attempt at finding difference between
$M_{\vev{+2}}^{(2)}$ and $M_{\vev{+2}}^{(-1)}$, see discussion at the end
of this section.

\paragraph{A Look at ${\cal R}$-dependence}

The analysis up to this point relied on $\Psi$ of a Higgs cascade as a whole,
so it is independent of the choice of the symmetry $\mathcal{R}$.
Now let us use $\underline{\Phi}$ and look into the information
which type of singularity may develop in a given manifold.

Suppose that a three-fold $X$ is in the diffeomorphism class
characterized by $n_{1/2}$ and $b_{{\cal R}} + 6\Z$.
Unless $(b_{\cal R} + 6\Z) = 0 \in \Z/6\Z$, any Higgs cascade attached
to the branch of the Type IIA compactification on $X$
does not lead to an enhancement of singularity of type ${\cal R}=E_6$
(or higher) whose resolution $\underline{X}$ has a regular K3-fibration.

As a test for whether an enhancement of singularity of a given type ${\cal R}$ is possible,
one may ask whether an appropriate modular form $\underline{\Phi}^{({\cal R})}$ can be found.
For example, take $\underline{\Lambda}_S = \vev{+2} \oplus A_3[-1]$.
Then the modular form $\underline{\Phi}$ of the hypothetical branch of
Type IIA on $\underline{X}$ is in the form of\footnote{
$\NRS$ is a shorthand: for modular form $F$ of weight $k$, $ \NRS F := (\partial^S F)/(-\frac{k}{12}) = (\frac{1}{2\pi i} \frac{\partial}{\partial \tau} - \frac{k}{12} E_2) F / (-\frac{k}{12}).$
}
\begin{align}
    \underline{\Phi}
    &= \frac{-68 + 6 \underline{n}_1 + \underline{n}_2 + \underline{n}_3}{72}
          E_6 \theta_{D_5} \tensor \theta_{\vev{+2}}
     + \frac{-52 - \underline{n}_2 - \underline{n}_3}{72} E_4
           \NRS \theta_{D_5} \tensor \theta_{\vev{+2}}
    \nonumber \\ & \quad
     + \frac{8 - 4 \underline{n}_1 - 3 \underline{n}_2 + 6 \underline{n}_3}{48}
          E_4 \theta_{D_5} \tensor \NRS \theta_{\vev{+2}}
     + \frac{-8 + \underline{n}_2 - 2 \underline{n}_3}{16}
          (\NRS)^2 \theta_{D_5} \tensor \NRS \theta_{\vev{+2}}
\end{align}
parametrized\footnote{
Remembering that $\SU(4) \times \SO(10)$ fits into $E_8$, it is a
reasonable idea to try to construct a basis by using
$\theta_{\vev{+2}} \otimes \theta_{D_5}$, derivatives, and the Eisenstein series $E_4$ and $E_6$ (see appendix \ref{apdx:A1}).
The dimension formula (\ref{eq:RR-thm-ModForm}, \ref{eq:diff-dim-Mod-Mod0})  indicates that the vector space
is of 4-dimensions for the weight and type for this $\underline{\Phi}$.
}
by $\underline{n}_{1,2,3} \in \Z$ and $\underline{n}_0 = -2$.
This parametrization is for an obvious reason:
\begin{align}
 \underline{\Phi} &=
    e_0 \tensor e_0 \parren{ -2+(324-56 \underline{n}_1
                          - 8 \underline{n}_2-6 \underline{n}_3)q+ \cdots}
    \nonumber \\ & \quad
    + e_0 \tensor e_1 \parren{ \underline{n}_1 q^{1/4}
         + \cdots }
    + (e_1+e_3) \tensor e_0 \parren{ \underline{n}_2 q^{5/8}
        + \cdots }
    \nonumber \\ & \quad
    + (e_1+e_3) \tensor e_1 \parren{ (16\underline{n}_1-2 \underline{n}_2+8 \underline{n}_3+96) q^{7/8}
        +\cdots }
    \nonumber \\ & \quad
    + e_2 \tensor e_0 \parren{ \underline{n}_3 q^{1/2}
        + \cdots }
    + e_2 \tensor e_1 \parren{ (8 + 10 \underline{n}_1 + 4 \underline{n}_2 - 6 \underline{n}_3) q^{3/4}
        +\cdots }.
\end{align}
By the discussion in section \ref{ssec:gamma_0-level-quantization-condition},
the coefficients $\underline{n}_{1,2,3} \in \Z$ need to satisfy
\begin{align}
    \underline{n}_1,\, \underline{n}_2 \geq 0,\quad \text{and} \quad
    \underline{n}_3,\,
    16\underline{n}_1-2 \underline{n}_2+8 \underline{n}_3+96,\,
    8 + 10 \underline{n}_1 + 4 \underline{n}_2 - 6 \underline{n}_3 \geq -2.
    \label{eq:3.4-n123-constraints}
\end{align}
Because $\Phi$ and $\Psi$ is given by $\Phi = \theta_{A_3}\cdot \underline{\Phi}$ and $\Psi = \NRS \theta_{A_3}\cdot \underline{\Phi}$, we have
\begin{align}
    n_{1/2} = \underline{n}_1,
    \qquad
    b_{\mathcal{R}} = -8 + \underline{n}_2 + \underline{n}_3.
\end{align}
For $n_{1/2}=0$ and $b_{\cal R} = 0,2,4$, for example,
there are solutions $(\underline{n}_1,\underline{n}_2,\underline{n}_3)$
to \eqref{eq:3.4-n123-constraints}.
Therefore, we do not have to rule out a possibility
that there exists tuning of complex structure of $X$
so it develops singularity of type $A_3$
and its
resolution $\underline{X}$ has a regular K3-fibration.
Indeed, in the case of $X = M_{\vev{+2}}^{(n)}$, such an enhancement can be realized as we see below.\footnote{
Of course, there is no guarantee in general that there exists an actual enhancement $\underline{X}$ when we can construct a candidate for modular form $\underline{\Phi}$.
}

\bigskip

The Calabi--Yau three-folds $M_{\vev{+2}}^{(n=2,1,0,-1)}$ are given
by a hypersurface surface equation
\begin{align}
   X_1^2 & \; 
     + F^{(6)}(X_{2,3,4}) = 0. \\
   F^{(6)} & \; = a_0 X_4^6 + b_0 X_4^4(X_2X_3) + c_0 X_4^2 (X_2X_3)^2
     + d_0 (X_2X_3)^3   \nonumber \\
  & \quad + (a_1 X_4^5 X_2 + a_2 X_4^4 X_2^2 + \cdots + a_6 X_2^6) \nonumber \\
  & \quad + (b_1 X_4^3 X_2^2 + b_2 X_4^2 X_2^3 + b_3 X_4 X_2^4 + b_4 X_2^5)X_3
       + (c_1 X_4 X_2^3 + c'_2 X_2^4)X_3^2 \nonumber \\
  & \quad + (a'_1 X_4^5 X_3 + a'_2 X_4^4 X_3^2 + \cdots + a'_6 X_3^6) \nonumber \\
  & \quad + (b'_1 X_4^3 X_3^2 + b'_2 X_4^2 X_3^3 + b'_3 X_4 X_3^4 + b'_4 X_3^5)X_2
       + (c'_1 X_4 X_3^3 + c'_2 X_3^4)X_2^2.
   \label{eq:def-3fd-M2n}
\end{align}
Homogeneous coordinates $X_1$ and $X_{2,3,4}$ of the toric ambient space are
subject to the $\C^\times$ action $X_1 \rightarrow X_1 \lambda^3$ and
$X_{2,3,4} \rightarrow X_{2,3,4} \lambda$ ($\lambda \in \C^\times$) for projectivization.
The coefficients $a_{i=1,\dots, 6}$, $b_{i=1,\dots, 4}$, $c_{i=1,2}$,
$a'_{i=1,\dots, 6}$, $b'_{i=1,\dots, 4}$, $c'_{i=1,2}$, and $a_0$, $b_0$, $c_0$ and
$d_0$ are regarded as sections of appropriate
line bundles of the base $\P^1$; those line bundles should have
the degree specified in Table \ref{tab:coeff-line-bdle-deg} for construction
of $M_{\vev{+2}}^{(n)}$. More details are found in \cite{BW-16}.
\begin{table}[tbp]
\begin{center}
  \begin{tabular}{c||c|c|c|c}
  & $a_i, a'_i$ & $b_i, b'_i$ & $c_i, c'_i$ & 
   $d_0$ \\
 \hline
 deg & $12-2i+(i-4)(2-n)$ & $8-2i + (i-2)(2-n)$ & $4-2i + i(2-n)$ & $2(2-n)$
  \end{tabular}
  \caption{\label{tab:coeff-line-bdle-deg}The ``coefficients'' $a_i$, $b_i$,
etc. in (\ref{eq:def-3fd-M2n}) are sections of ${\cal O}_{\P^1}({\rm deg})$,
with the degree ``deg'' specified in this table. }
\end{center}
\end{table}

Singularity of type ${\cal R}=E_7$ develops in $M_{\vev{+2}}^{(n)}$ for any one
of $n=2,1,0,-1$, when all of the sections $a_{2,\dots, 6}$, $b_{2,\dots, 4}$,
and $c_{1,2}$ are set to zero. The singularity is along the
curve\footnote{We used inhomogeneous coordinates $x_1 = X_1/X_2^3$,
$x_4 = X_4/X_2$, and $x_3 = X_3/X_2$.} $(x_1,x_4,x_3)=(0,0,0)$, which is
in the class $C_{\cal R} = \Sigma_B^{(n)}$; it is of type $E_7$ because
$x_1^2 + b_1 x_4^3 x_3 + d_0 x_3^3 \simeq 0$ is in the direction of the K3 fibre.

It is only in the case of $n=2$, however, that the
K3-fibration in the resolution $\underline{M}$ of the singularity 
remains regular; for $n \neq 2$, the coefficient $d_0$ is a section of a line
bundle of the base $\P^1$ of positive degree. The section $d_0$ vanishes
at some points in the base, and
the three-fold $\underline{M}$ has
reducible fibres at those points.
So, this failure in finding a tuning of complex structure for
$M_{\vev{+2}}^{(n=1,0)}$ is consistent with
$b_{\cal R} + 6\Z = 2_{+ 6\Z}, 4_{+ 6\Z}$ that does not allow interpretation
as a 1-loop beta function of ${\cal R}=E_7$ gauge group.
On the other hand, singularity of type ${\cal R} = A_3$ enhances
along $C_{\cal R} = \Sigma_B^{(n)}$ by setting $a_{3,\dots, 6}$ and $b_{3,4}$
to zero, and we can see by using toric data
(just like in \cite{BW-16}) that their $\underline{M}$'s have a regular K3-fibration, as announced earlier.

Given the diffeomorphism $f: M_{\vev{+2}}^{(2)} \rightarrow M_{\vev{+2}}^{(-1)}$,
the symmetry-enhancement branch of $M_{\vev{+2}}^{(2)}$ with a singularity along
$C_{\cal R} = \Sigma_B^{(2)}$ should be compared with the symmetry-enhancement
branch of $M_{\vev{+2}}^{(-1)}$ with a singularity along
$\Sigma_B^{(-1)} - \Sigma_F^{(-1)}$.
For a singularity of type ${\cal R}=E_7$
for the latter, we can tune $a'_{4,\dots,0}$, $b'_1$, $b_0$, $a_1$, $b_1$,
and $a_2$ to zero, for example, because the hypersurface equation is
$\xi_1^2 + b'_2 \xi_3^3 \xi_2 + a_3 \xi_2^3 \simeq 0$ near the curve\footnote{
Now, the inhomogeneous coordinates are
$\xi_1 = X_1/X_4^3$, $\xi_{2,3} = X_{2,3}/X_4$. } $\xi_1 = \xi_3 = \xi_2=0$.
So, this Higgs cascade for $M_{\vev{+2}}^{(-1)}$ and $C_{\cal R} =
\Sigma_B^{(-1)}-\Sigma_F^{(-1)}$  must have the same modular form $\Psi$
as the Higgs cascade for $M_{\vev{+2}}^{(2)}$ and $C_{\cal R}=\Sigma_B^{(2)}$.
Those two symmetry-enhancement branches are still different, because
$a_3$ is a section of ${\cal O}(3)$ on $\P^1_{\rm IIA}$ in the former,
whereas $d_0$ that of ${\cal O}(0)$ on $\P^1_{\rm IIA}$ in the latter,
and the three-fold $\underline{M}$ for the former does not have a regular
K3-fibration. 
One possibility is that the branches of IIA/$M_{\vev{+2}}^{(2)}$ and 
IIA/$M_{\vev{+2}}^{(-1)}$ are still identical one branch of moduli space, 
where we have just yet to find tuning of complex structure of 
$M_{\vev{+2}}^{(-1)}$ so there is an $E_7$ singularity along 
$\Sigma_B^{(-1)}-\Sigma_F^{(-1)}$ whose resolution remains to have a regular 
K3-fibration (put differently, there may be multiple Higgs cascades 
sharing one common $\Psi$).  
The other possibility is that they are two physically distinct branches 
of moduli space of Heterotic--IIA dual vacua, and that the invariants 
$\Lambda_S$, $\Lambda_T$, $\Phi$, and the set of $\Psi$'s are not enough 
to distinguish the two branches. 

\section{Open Questions}

Practical questions remain:
how small are the subspaces $[{\rm Mod}_0^\Z(11-\rho/2,\rho_{\Lambda_S})]^{\rm r.mfd}$
and $[{\rm Mod}_0^\Phi(13-\rho/2,\rho_{\Lambda_S})]^{\rm r.mfd}$ within
$[{\rm Mod}_0^\Z(11-\rho/2,\rho_{\Lambda_S})]^{\S2}$ and
${\rm Mod}_0^\Phi(13-\rho/2,\rho_{\Lambda_S})$, respectively.
We worked on this question for $\Lambda_S = U$, $\vev{+2}$,
and $U \oplus \vev{-2}$ in this article, but not for a general
$(\Lambda_S, \Lambda_T)$ that fits into ${\rm II}_{3,19}$.
For example, it is possible (at least in theory) to study whether the subspaces
remain non-empty for the series $\Lambda_S = \vev{+2n}$ with large $n$.

The image of the map $({\rm diff}_{\rm coarse},{\rm diff}_{\rm fine})$
restricts possible diffeomorphism classes of real six-dimensional manifolds
that can be realized by Calabi--Yau three-folds with $\Lambda_S$-polarized
regular K3-fibrations. This method has been applied only for $\Lambda_S = U$ and $\vev{+2}$;
we found that the image of ${\rm diff}_{\rm coarse}$ is much smaller than the set ${\rm Diff}_{\Lambda_S\; (\Q)}$ for both $\Lambda_S$'s,
and the image of ${\rm diff}_{\rm fine}$ is all of ${\rm Diff}_{\Lambda_S}^{d'}$
for $(d',\chi)$ in the image of ${\rm diff}_{\rm coarse}$.
One can find out whether that remains to be true for various different
$\Lambda_S$'s, by working out the images of
$[{\rm Mod}_0^\Z(11-\rho/2,\rho_{\Lambda_S})]^{\rm r.mfd}$ and
$[{\rm Mod}_0^\Phi(13-\rho/2,\rho_{\Lambda_S})]^{\rm r.mfd}$.

A few theoretical questions can also be put down.
There are two possibilities for a pair of modular forms
$\Phi \in [{\rm Mod}_0^\Z(11-\rho/2,\rho_{\Lambda_S})]^{\S2}$
and
$\Psi \in {\rm Mod}_0^\Phi(13-\rho/2,\rho_{\Lambda_S})$
that are not in the subsets $[{\rm Mod}_0^\Z(11-\rho/2,\rho_{\Lambda_S})]^{\rm r.mfd}$
and $[{\rm Mod}_0^\Phi(13-\rho/2,\rho_{\Lambda_S})]^{\rm r.mfd}$.
One is that there are more theoretical constraints of string theory that
we failed to capture in sections \ref{sec:coarseC} and \ref{sec:finerC}
and such a $(\Phi, \Psi)$ is in conflict with those constraints.
The other is that such a pair $(\Phi, \Psi)$ is for a branch of
moduli space whose Type IIA description does not involve a geometric phase.
It remains to be an open question how to determine the boundary
between those two possibilities in the space of $(\Phi,\Psi)$.

We have already seen that the image of the map
$({\rm diff}_{\rm coarse}, {\rm diff}_{\rm fine})$ is small compared with the set ${\rm Diff}_{\Lambda_S}$ for some $\Lambda_S$'s.
But not all the diffeomorphism classes
of real six-dimensional manifold in the image are guaranteed to be
realized as a Calabi--Yau three-fold. By taking advantage of large
database of topology of Calabi--Yau three-folds, one may try to get
the feeling how much fraction of the diffeomorphism classes in the image
of $({\rm diff}_{\rm coarse}, {\rm diff}_{\rm fine})$ are indeed guaranteed
to be realized by Calabi--Yau three-folds. Such a study may provide
hints in considering the ``determining the boundary'' issue above.

In pure mathematics literatures, some inequalities on topological
invariants of Calabi--Yau three-folds have been
derived (e.g., \cite[\S2]{Kanazawa}). It is beyond the scope of this article
to study how those inequalities are related to the bounds that
we discussed in this article. Also in physics approach, various
integer parameters are likely not just bounded from below, but also
from above (for maintaining strictly\footnote{
In the case of $\Lambda_S = U$, for example,
we know that the three-fold $X^{(n)}$ with $n > 2$ has 
$\Lambda_S$ strictly larger than $U$
(non-Higgsable phenomenon \cite{MV-2}).
This phenomenon leaves its trace in the fact that
the set $\{ b_{\mathcal{R}} = 2m+2-n \;|\; m \in \Z_{\geq 0} \}$
contains $b_{\mathcal{R}} < 0$ (hence non-Higgsable) when $n > 2$.
} a given lattice $\Lambda_S$).
But we have not given enough thoughts on how this intuition is related
or unrelated to the bounds and classifications discussed in this article.

A few ideas are also available in improving the effort of introducing invariants for classification of the branches of moduli space of the Het--IIA dual vacua.
We just introduced the idea of using the set of
$\Psi$'s of all the Higgs cascade as an invariant of a branch
(than just using its image by ${\rm diff}_{\rm fine}$) in this article;
more knowledge in the cone of curves (and tuning of complex structure
to have certain singularity along a curve class) would make it possible
to compute the set of $\Psi$'s for general $\Lambda_S$'s, not just for
$\Lambda_S = U$. Also, a part of the idea (using $\Delta h^{2,1}_{\cal R}$)
for invariants in the case $\Lambda_S = \vev{+2}$ in \cite{BW-16} has been
incorporated as a part of $\underline{\Phi}^{({\cal R})}$ for general $\Lambda_S$'s in this article,
but a bit more idea beyond $\Delta h^{2,1}_{\cal R}$
in \cite{BW-16} has not been generalized to other $\Lambda_S$'s, or brought
into the language of world-sheet CFT in this article.

Finally, as a reminder, K3-fibrations of a Calabi--Yau three-fold were assumed to be regular in this article. Classifications of Calabi--Yau
three-folds with a non-regular K3-fibration (and their Heterotic duals)
should be considered separately from this article.
Note also that we set some other technical limitations in section \ref{ssec:Het-II-review}
on the class of Heterotic--IIA dual vacua to study in this article.
Structure of branches of the whole vacua is yet to be figured out.

\section*{Acknowledgements}

The authors thank I. Antoniadis, A. Braun, B. Haghighat, K. Kanno,
A. Klemm, and K.M. Lee for valuable discussions and useful comments.
This work is supported in part by the World Premier International Research
Center Initiative (WPI),
Grant-in-Aid New Area no. 6003 (YE and TW),
the FMSP program (YE),
Grant-in-Aid New Area no. 2303, and the Brain circulation program (TW),
all from MEXT, Japan.


\appendix

\section{Modular Forms}

\subsection{Notations and Basic Facts}
\label{apdx:A1}

In this subsection we explain our notations and some basic facts about modular forms.

\paragraph{Metaplectic Group}
The \textit{metaplectic group} $\mathrm{Mp}(2;\Z)$ is defined by
\begin{align}
    \mathrm{Mp}(2;\Z) = \left\{ (A,f(\tau)) \mymid A = \begin{pmatrix} a & b \\ c & d \end{pmatrix} \in \mathrm{SL}(2;\Z),\; f(\tau)^2 = c\tau + d \right\}.
\end{align}
$f$ is a holomorphic function in the upper half plane $\UH \subset \C$.
$f$ specifies the choice of sign $\pm \sqrt{c\tau+d}$, so $\mathrm{Mp}(2;\Z)$
is a double covering of $\mathrm{SL}(2;\Z)$.
The multiplication of two elements in the group ${\rm Mp}(2;\Z)$ is defined
by
\begin{align}
    (A, f(\tau)) (B, g(\tau)) = (AB, f(B\cdot \tau)g(\tau)).
\end{align}

The group $\mathrm{SL}(2;\Z)$ is generated by two elements
\begin{align}
 S = \left( \begin{array}{cc} & -1 \\ 1 & \end{array} \right), \qquad
 T = \left( \begin{array}{cc} 1 & 1 \\  & 1 \end{array} \right),
\end{align}
which satisfy $S^2 = (ST)^3 = -1$.
Similarly, $\mathrm{Mp}(2;\Z)$ is generated by
\begin{align}
 S = \left( \left( \begin{array}{cc} & -1 \\ 1 & \end{array} \right), \sqrt{\tau} \right),
    \qquad
 T = \left( \left( \begin{array}{cc} 1 & 1 \\  & 1 \end{array} \right), 1 \right),
\end{align}
which satisfy $S^2 = (ST)^3 = Z$, where
\begin{align}
    Z = \left( \left( \begin{array}{cc} -1 & \\ & -1 \end{array} \right), i \right),
     \qquad Z^4 = 1.
\end{align}

\paragraph{Vector-valued modular form}

Let $k^{\pm} \in \frac{1}{2}\Z$, and $\rho : \mathrm{Mp}(2;\Z) \to \mathrm{GL}(V)$ be a representation on a vector space $V$.
A real analytic (but not necessarily holomorphic) function $F : \UH \to V$
is called a {\it (vector-valued) modular form of weight $(k^+,k^-)$ and type}
$\rho$ if $F$ satisfies the modular transformation laws
\begin{align}
    F(A\cdot \tau) = f(\tau)^{\,2k^+} \overline{f(\tau)}{}^{\,2k^-} \rho(A,f) F(\tau),
    \qquad  (A,f) \in \mathrm{Mp}(2;\Z).
\end{align}
A modular form $F(\tau,\bar{\tau})$ is said to be {\it almost holomorphic},
if it is a polynomial of $(1/\tau_2)$ with $\tau$-dependent coefficients
and has finite values at all the cusp points. The vector space of almost
holomorphic modular forms of weight
$(k^+,k^-)$ and type $\rho$ is denoted by ${\rm Mod}((k_+,k_-),\rho)$.
The vector space of truly holomorphic modular forms---no $\bar{\tau}$
dependence---in ${\rm Mod}((k^+,0),\rho)$ is denoted by ${\rm Mod}(k^+,\rho)$.
We mainly consider the case where $\rho$ is a Weil representation
(explained shortly in the following).
A subspace $\Mod_0(k,\rho)$ of ${\rm Mod}(k,\rho)$ is also defined below.

\paragraph{Weil representation}
Let $M$ be an even lattice of signature $(b^+,b^-)$, and
$G_M = M^\vee/M$ the discriminant group.
Define $\C[G_M] := \mathrm{span}_\C \{ e_\gamma \mid \gamma \in G_{M} \} $ where $e_\gamma$ is a formal symbol.
The \textit{Weil representation} $\rho_M : \mathrm{Mp}(2;\Z) \to \mathrm{GL}(\C[G_M])$ is defined by
\begin{align}
  \rho_{M}(T) e_\gamma &= e_\gamma\; \mathbb{E}\parren{ \frac{(\gamma, \gamma)}{2} }, \\
  \rho_{M}(S) e_\gamma &=  \sum_{\delta\in G_M} e_\delta \; \frac{1}{\sqrt{|G_{M}|}} \;
    \mathbb{E}\parren{ -\frac{\sgn(M)}{8} - (\delta,\gamma) },
\end{align}
where $\mathbb{E}(x) = e^{2\pi i x}$. The element $Z$ acts as
$\rho_M(Z)e_\gamma = e_{-\gamma}\mathbb{E}(-\sgn(M)/4)$.
The relation $S^2 = (ST)^3$ holds because of
Milgram's formula:
\begin{align}
 \sum_{\gamma \in G_M} \mathbb{E}\left((\gamma, \gamma)/2 \right) =
  \sqrt{|G_M|} \; \mathbb{E} \left( \sgn(M)/8\right).
\end{align}

If two even lattices $M_1,M_2$ are primitive sublattices of a certain unimodular lattice $L$ and are orthogonal to each other inside $L\tensor\R$, then $\rho_{M_1}$ and $\rho_{M_2}$ are the dual (contragredient) representation of $\mathrm{Mp}(2;\Z)$ of each other. For example $\rho_{\Lambda_S}$ and $\rho_{\Lambda_T}$ in the main text are dual.

Any modular form $F \in {\rm Mod}(k, \rho_M)$ takes value only in the
vector subspace $\C[G_M]^+ := {\rm span}_\C\{ e_\gamma + e_{-\gamma} \; |
\; \gamma \in G_M\} \subset \C[G_M]$, when $k/2 \equiv {\rm sgn}(M)/4$ mod $\Z$.

\paragraph{Subspaces of ${\rm Mod}(k,\rho)$ of interest}
A modular form $\Phi \in \Mod(k,\rho_M)$ has a Fourier expansion
\begin{align}
    \Phi(\tau)
    = \sum_{\gamma \in G_M} e_\gamma \Phi_\gamma
    = \sum_{\gamma \in G_M} e_\gamma \sum_{\nu \in \gamma^2/2 + \Z} x_\gamma(\nu) q^\nu,
    \qquad x_\gamma(\nu) = 0 \text{~~for~~} \nu < 0.
\end{align}
Here $q = e^{2\pi i \tau}$. $\gamma^2/2$ denotes the quadratic form on $G_M$. We define $\Mod_0(k,\rho)$ by imposing the cusp condition on the components $\Phi_\gamma$ for isotropic (but non-zero) $\gamma$'s:
\begin{align}
    \Mod_0(k,\rho) = \{ \Phi \in \Mod(k,\rho) \mid x_\gamma(0) = 0 \text{~if~} \gamma \neq 0 \text{~and~} \gamma^2/2 = 0 \}.
\end{align}

In this article, the Fourier coefficients $x_\gamma(\nu)$ are denoted
by $[\Phi_\gamma]_{q^\nu}$.

\paragraph{Dimensional formula} The dimension of the vector space
${\rm Mod}(k,\rho_M)$ is determined by the following formula in the case of $k >2$
and $k/2 \equiv {\rm sgn}(M)/4$ mod $\Z$ \cite{Bor2}:
\begin{align}
  {\rm dim}_\C \left({\rm Mod}\left(k, \rho_M \right)\right) & \; =
   d + \frac{dk}{12}  \label{eq:RR-thm-ModForm} \\
   - & \; \alpha\left( \mathbb{E}\left(\frac{k}{4}\right)\rho_M(S) \right)
   - \alpha \left( \left(\mathbb{E}\left(\frac{k}{6}\right) \rho_M(ST)\right)^{-1}\right)
   - \alpha(\rho_M(T)),  \nonumber
\end{align}
where $d := \dim_\C \C[G_M]^+$, and $\alpha(X) := \sum_{i=1}^d \beta_j$
when $\beta_j \in [0,1)$ is the complex phase (divided by $2\pi$) of the each eigenvalue of
the representation matrix $X|_{\C[G_M]^+} \sim \mathbb{E} (\diag(\beta_j))$.
The restriction $k > 2$ of this formula is due to the fact that
$\dim_\C {\rm Mod}(k,\rho_M^\vee) = \dim_\C {\rm Mod}(2-k,\rho_M)$
is not necessarily zero for $k \leq 2$.

The dimension of the subspace ${\rm Mod}_0(k,\rho_M)$ is
\cite[Chap. 1.2.3]{Bruinier-bookBor}
\begin{align}
  \dim_\C ({\rm Mod}_0(k,\rho_M))
   =   \dim_\C ({\rm Mod}(k,\rho_M)) -
    \# \{ \pm \gamma  \in G_M /\pm \; | \; \gamma \neq 0, \;
      \gamma^2/2 =0 \}.
  \label{eq:diff-dim-Mod-Mod0}
\end{align}

\paragraph{Siegel theta function}
Let $M$ be an even lattice of signature $(b^+,b^-)$.
Fix a point $v$ in the Grassmannian $Gr(M) = Gr(M \tensor \R; b^+)$, i.e.
a pair of positive/negative definite $b^{\pm}$-dimensional subspaces of
$M \tensor \R$, orthogonal to each other, and define
$v_{\pm}: M \tensor \R \to M \tensor \R$ as the orthogonal projections.
The \textit{Siegel theta function} is defined by
\begin{align}
    \theta_M(\tau,\bar\tau;v) &= \sum_{\gamma\in M^\vee/M} e_{\gamma}
   \sum_{\lambda\in M+\gamma}
   q^{v^2_+(\lambda)/2} \overline{q}^{v^2_-(\lambda)/2},
    \quad q := e^{2\pi i \tau}.
\end{align}
Here $v^2_{\pm}(\lambda) := \abs{ (v_{\pm}(\lambda),v_{\pm}(\lambda)) }$ are both non-negative.
$\theta_M(\tau,\bar\tau;v)$ is a $\C[G_M]$-valued modular form of weight $(b^+/2,b^-/2)$ and type $\rho_M$.

\paragraph{Eisenstein series}
\begin{align}
E_2 =& 1 - 24 \left( q + 3 q^2 + \cdots \right)
    = 1 - 24 \sum_{n=1}^\infty q^n \sigma_1(n)
    = 1 - 24 \sum_{m=1}^\infty \frac{m q^m}{1-q^m}, \\
E_4 =& 1 + 240\left( q + 9 q^2 + \cdots \right)
   = 1 + 240 \sum_{n=1}^\infty q^n \sigma_3(n),
   = 1 + 240 \sum_{m=1}^\infty \frac{m^3 q^m}{1-q^m}, \\
E_6 =& 1 - 504\left( q + 33 q^2 + \cdots \right)
   = 1 - 504 \sum_{n=1} q^n \sigma_5(n)
   = 1 - 504 \sum_{m=1} \frac{m^5 q^m}{1-q^m}.
\end{align}
$E_4(\tau)$ and $E_6(\tau)$ are modular forms of weight 4 and 6, respectively,
for ${\rm SL}(2;\Z)$, but $E_2(\tau)$ is not modular (it is Mock modular).
The space of scalar-valued modular forms can be identified with the polynomial
ring $\C[E_4,E_6]$.

The $q^0$-term vanishes in the combination
\begin{align}
 E_4^3 - E_6^2 = \left(1+3 \cdot 240 q + \cdots \right) - \left(1 - 2 \cdot 504 q - \cdots \right)
   = 1728 q + \cdots = 1728 \eta^{24},
\end{align}
where
\begin{align}
    \eta = q^{1/24} \prod_{n = 1}^\infty (1-q^n).
\end{align}

\paragraph{Ramanujan--Serre derivative and Rankin--Cohen bracket}
For a modular form $F \in {\rm Mod}(k,\rho)$, the \textit{Ramanujan--Serre derivative}
$\partial^S : {\rm Mod}(k,\rho) \to {\rm Mod}(k+2,\rho)$ is defined by
\begin{align}
 \partial^S F :=
  \left(\frac{1}{2\pi i} \frac{\partial}{\partial \tau} - \frac{k}{12} E_2
  \right) F = \eta^{2k} q \partial_q \left(\frac{F}{\eta^{2k}} \right).
\end{align}

Vector-valued modular forms and their Ramanujan--Serre derivatives
can be multiplied to produce yet another vector-valued modular form; the \textit{Rankin--Cohen bracket} $[F,G]_n$ of a pair of modular forms
$F \in {\rm Mod}(w_F,\rho_F)$ and $G \in {\rm Mod}(w_G,\rho_G)$ and
$n \in \N$ is
\begin{align}
 [F, G]_n := \frac{1}{(2\pi i)^n} \sum_{r+s=n} (-1)^r
 \binom{w_F + n-1}{s}
 \binom{w_G + n-1}{r}
   \frac{\partial^r F}{\partial \tau^r} \frac{\partial^s G}{\partial \tau^s}.
\end{align}
For example, $[F,G]_0 = FG$, and
$[F, G]_1 = w_F F q(\partial_qG) - w_G q(\partial_q F) G$.
It is known that
$[F, G]_n \in {\rm Mod}(w_F + w_G + 2n, \rho_F \otimes \rho_G)$.
%
%

\paragraph{Lattice-polarized Jacobi forms and vector-valued modular forms}
For a positive definite even lattice $M$, and $k \in \frac{1}{2} \Z$,
a holomorphic function $\phi: {\cal H} \times (M \otimes_\Z \C)  \ni (\tau, z)
\longmapsto \C$
is said to be a \textit{Jacobi form of weight $k$ and index $M$}, if it is satisfies
\begin{align}
  \phi \left( \frac{a \tau + b}{c\tau + d}, \frac{z}{c\tau + d} \right)
   & \;  = (c\tau + d)^k \mathbb{E}\left( \frac{c \; (z,z)}{2(c\tau + d)} \right)
         \phi(\tau, z), \\
  \phi \left( \tau, z + \mu + \lambda \tau \right) & \; =
    \mathbb{E} \left( - (\lambda,z) - \tau \frac{(\lambda,\lambda)}{2} \right)
        \phi ( \tau, z),
\end{align}
where the bilinear form $(-,-)$ of $M$ has been extended linearly to
$M \otimes_\Z \C$.
The classical definition of a Jacobi
form of weight $k$ and index $m$ is regarded as that of a Jacobi form of weight $k$ and index $M$ with the lattice
$M= \vev{+2m}$.

To any Jacobi form $\phi(\tau,z)$ of weight $k$ and index $M$, one can
assign a vector-valued modular form of weight $(k-1/2)$ and type $\rho_M^\vee$.
That is through
\begin{align}
 \phi(\tau, z) =  \sum_{x \in G_M}
    \left( \sum_{\lambda \in x + M}  q^{\frac{(\lambda,\lambda)}{2}}
       e^{2\pi i (\lambda, z)} \right) \;
     f_x(\tau)  \Longleftrightarrow \sum_{x \in G_M} e_x f_x(\tau),
  \label{eq:1to1-Jacob-vvModForm}
\end{align}
and this is one-to-one. The modular form $\sum_x e_x f_x(\tau)$ is holomorphic
at the cusps if and only if the expansion
$\phi(\tau,z) = \sum_{n \in \Z}\sum_{\lambda \in M^\vee}
c(n,\lambda) q^n e^{2\pi i (\lambda,z)}$ has support
in $n \geq (\lambda,\lambda)/2$. See \cite{EichlerZagier, DabMurZagier}
for more information.

\subsection{Explicit Examples}
\label{apdx:A2}

Some of the modular forms used in the main text are written down
explicitly here.

\subsubsection{Explicit Basis of ${\rm Mod}_0(11-\rho/2,\rho_{\Lambda_S})$
}
\label{apdx:A21}

{\bf The case $\Lambda_S = \vev{+2}$:} (cf \cite{KKRS, MP, HK})
The vector space
${\rm Mod}_0(21/2,\rho_{\vev{+2}})$ is 2-dimensional over $\C$.
One can use $\phi_{(i)} := [\theta_{\vev{+2}}, E_{10-2i}]_i$
with $i=0,1$ as a basis. Their Fourier expansions are
\begin{align}
 \phi_{(0)} & \; =
   e_0 \left( 1-262 q -135960 q^2 + \cdots \right)
 + e_{1/2} \left( 2q^{\frac{1}{4}}  - 528 q^{\frac{5}{4}} - 270862 q^{\frac{9}{4}} + \cdots\right), \\
 \phi_{(1)} & \; =
    e_0\left( 0 +224 q + 54720 q^2 + \cdots \right)
  + e_{1/2} \left( -4q^{\frac{1}{4}} -1440 q^{\frac{5}{4}} -123876 q^{\frac{9}{4}} + \cdots \right),
\end{align}
where $\{e_0,e_{1/2}\}$ is the basis of $\C[G_S]$.

{\bf The Case $\vev{+4}$:} (cf \cite{KKRS, HK})
The vector space ${\rm Mod}_0(21/2, \rho_{\vev{+4}})$ is of 3-dimensions,
and is 
generated by $\phi_{(i)} := [\theta_{\vev{+4}},E_{10-2i}]_i$
with $i=0,1,2$. Their Fourier expansions are
\begin{alignat}
 \phi_{(0)} & \; = 1e_0 &&+q^{1/8}(e_{1/4}+e_{3/4}) + 2 q^{1/2}e_{2/4} - 264 q e_0
      -263 q^{9/8}(e_{1/4}+e_{3/4}) + \cdots, \\
 \phi_{(1)} & \; = && -q^{1/8}(e_{1/4}+e_{3/4})-8q^{1/2}e_{2/4} + 240 q e_0
      -249 q^{\frac{9}{8}}(e_{1/4}+e_{3/4}) +  \cdots , \\
\frac{64}{21} \phi_{(2)} & \; = && + q^{1/8}(e_{1/4}+e_{3/4}) + 32 q^{1/2}e_{2/4} - 576 q e_0 + 1017 q^{\frac{9}{8}} (e_{1/4}+e_{3/4}) + \cdots.
\end{alignat}

The modular form $\Phi$ for a Heterotic--IIA dual vacuum is parametrized
by the low-energy BPS indices $n_0 = -2$, $n_{1/4}$, and $n_{1/2}$. It must be
\begin{align}
 \Phi & \; =  -2 \phi_{(0)} + \frac{(n_{1/2}-32n_{1/4}-60)}{24} \phi_{(1)}
  + \frac{(n_{1/2}-8n_{1/4}-12)}{24} \frac{64}{21}\phi_{(2)}, \\
  & \; = e_0 \left( -2 + (216 -14 n_{1/2}-128 n_{1/4}) q
      + (153900 -57344n_{1/4}-568n_{2/4})q^2 + \cdots    \right)
      \nonumber \\
  & + (e_{1/4}+e_{3/4}) \left(n_{1/4} q^{\frac{1}{8}}
        + (640 - 7 n_{1/4} + 32 n_{1/2}) q^{\frac{9}{8}}
        + (273028 - 272 n_{1/4}+544n_{2/4})q^{\frac{17}{8}}
  \cdots  \right)
      \nonumber \\
  & \quad + e_{2/4} \left( n_{1/2} q^{1/2}
      + (10032 + 4864 n_{1/4} -188 n_{1/2}) q^{\frac{3}{2}} + \cdots \right).
\end{align}
It follows that
\begin{align}
 \chi(X_{\rm IIA}) = 48 - NL_{1,0} = 
   -168+128n_{1/4}+14n_{2/4}.
  \label{eq:chi-deg4}
\end{align}

{\bf The Case $\vev{+6}$:} The vector space ${\rm Mod}_0(21/2,\rho_{\vev{+6}})$
is of 4-dimensions, and is generated by
$\phi_{(i)} = [\theta_{\vev{+6}},E_{10-2i}]_i$ with $i=0,1,2,3$.
The modular form $\Phi$ for a Heterotic--IIA vacuum is parametrized by
the low-energy BPS indices $n_0=-2$, and $n_{\gamma}$ with $\gamma = 1/6, 2/6, 3/6$:
\begin{align}
 \frac{\Phi}{\eta^{24}} = F^{-1} + n_{1/6} F^{-11/12} + n_{2/6} F^{-8/12}
    + n_{3/6} F^{-3/12},
\end{align}
where we can use the following \cite{HK}
\begin{align}
 F^{-1} & \; = \frac{-2}{q}e_0 + 148 e_0 + 336 q^{\frac{1}{12}}(e_{1/6}+e_{5/6})
  + 2730 q^{\frac{4}{12}}(e_{2/4}+e_{4/6}) + 35360 q^{\frac{9}{12}}e_{3/6} + \cdots, \nonumber \\
 F^{-11/12} & \; = \frac{q^{\frac{1}{12}}}{q}(e_{\frac{1}{6}}+e_{\frac{5}{6}}) - 108 e_0-134q^{\frac{1}{12}}(e_{\frac{1}{6}}+e_{\frac{5}{6}})
  + 924 q^{\frac{4}{12}}(e_{\frac{2}{6}}+e_{\frac{4}{6}}) + 20196 q^{\frac{9}{12}} e_{\frac{3}{6}} + \cdots, \nonumber \\
 F^{-8/12} & \; = \frac{q^{\frac{4}{12}}}{q}(e_{\frac{2}{6}}+e_{\frac{4}{6}}) - 54 e_0 + 56 q^{\frac{1}{12}}(e_{\frac{1}{6}}+e_{\frac{5}{6}})
  + 214 q^{\frac{4}{12}}(e_{\frac{2}{6}}+e_{\frac{4}{6}}) - 3024 q^{\frac{9}{12}} e_{\frac{3}{6}} + \cdots, \nonumber \\
 F^{-3/12} & \; = \frac{q^{\frac{9}{12}}}{q} e_{3/6} -2 e_0 + 3q^{\frac{1}{12}}(e_{1/6}+e_{5/6})
  - 6 q^{\frac{4}{12}}(e_{2/6}+e_{4/6}) + 14 q^{\frac{9}{12}} e_{3/6} + \cdots .
\end{align}
It follows that
\begin{align}
 \chi(X_{\rm IIA}) = - c_0(0) = - 148 + (108 n_{1/6}+54 n_{2/6}+2n_{3/6}).
\end{align}
All the details so far in this appendix \ref{apdx:A21} are used
in section \ref{sssec:rho=1}.

{\bf The Case $\Lambda_S = U \oplus \vev{-4}$:} There are two linearly
independent holomorphic Jacobi forms of weight 10 and index 2
(see \cite{DabMurZagier}).
One can work out a basis of the 2-dimensional vector space
${\rm Mod}_0(19/2,\rho_{\vev{-4}})$ by using those holomorphic Jacobi forms.
The modular form
$\Phi$ in this vector space is parametrized by $n_0=-2$ and $n_{2/4}$, as
\begin{align}
 \Phi & \; =
 e_0 \left( -2 +(336-18n_{2/4}) q  + (116340 -16n_{2/4}) q^2 + \cdots \right)
    \nonumber \\
& + (e_{1/4}+e_{3/4}) \left( (96+8n_{2/4}) q^{\frac{7}{8}} +(66976 + 120n_{2/4}) q^{\frac{15}{8}} +(2539488-1368n_{2/4})q^{\frac{23}{8}} + \cdots \right)  \nonumber \\
& + e_{2/4} \left( n_{2/4} q^{\frac{1}{2}} +(10192-120n_{2/4}) q^{\frac{3}{2}}  + (771456 + 900n_{2/4}) q^{\frac{5}{2}} + \cdots \right).
\end{align}
We can read out $n_{1/4}=8n_{2/4}+96$ from $\Phi$ above;
this is used in section \ref{sssec:lin-reltn}.

\subsubsection{Some Lower Bounds on $\chi(X)$ for $\Lambda_S=\vev{+2n}$}
\label{apdx:A22}

\begin{table}[tbp]
    \centering
    \begin{tabular}{cl}
        $\Lambda_S$ & $\chi(X_{\rm IIA})$ \\ \hline
        $\vev{+2}$ & $\chi = -252 + 56 n_{1/2}$ \\
        $\vev{+4}$ & $\chi = -196 + 128 n_{1/4} + 14 (n_{2/4}+2)$ \\
        $\vev{+6}$ & $\chi = -152 + 108 n_{1/6} + 54 n_{2/6} + 2 (n_{3/6}+2)$ \\
        $\vev{+8}$ & $\chi = -112 + 112 n_{1/8} + 56 n_{2/8} + 16 n_{3/8}$ \\
        $\vev{+10}$ & $\chi = -124 + 88 n_{1/10} + 66 n_{2/10} + 24 n_{3/10} + 2  (n_{4/10}+2) + 32 n_{5/10}$\\
        $\vev{+12}$ & $\chi = -124 + 96 n_{1/12} + 60 n_{2/12} + 32 n_{3/12} + 6 (n_{4/12}+2) + 96 n_{5/12} + 10 (n_{6/12}+2)$ \\
        $\vev{+14}$ & $\chi = -144 + 56n_{1/14} + 54n_{2/14} + 54n_{3/14} + 2n_{4/14} + 2n_{5/14} + 54n_{6/12}$\\
        $\vev{+14}$ & $\chi = -92 + 84n_{1/14} + 70n_{2/14} + 28n_{3/14} + 14n_{4/14} + 0n_{5/14} + 42n_{6/12} + 2(n_{7/14}+2)$
    \end{tabular}
    \caption{Euler number $\chi$ in terms of BPS indices $n_{\gamma}$. $n_\gamma$ or $(n_\gamma+2)$ above is non-negative. For example, $\chi \geq -196$ when $\Lambda_S = \vev{+4}$. The last two lines show an example where Weyl-inequivalent sub-lattices of $E_8$ for for $L[-1]$ give different expressions of $\chi$ in terms of $\{n_\gamma\}$.}
    \label{tab:EulerNumber-BPSIndicies-rank1-apdx}
\end{table}

In section \ref{ssec:bd-EulerN}, we derive an expression for $\chi(X)$
for Calabi--Yau three-folds $X$ that have a regular $\Lambda_S$-polarized
K3-fibration, by using $\phi \in {\rm Mod}(3+\rho/2,\rho_{\Lambda_S}^\vee)$ whose
Fourier coefficients at lower powers of $q$'s are all positive.
In this appendix \ref{apdx:A22}, we give details of $\phi = \theta_{L[-1]}$
in some of the $\rho=1$ cases,\footnote{
In the $\rho=1$ cases ($\Lambda_S = \vev{+2n}$), we can
use knowledge on the vector space of holomorphic Jacobi
forms of weight 4 and index $n$ to construct a basis of ${\rm Mod}(7/2,\rho_{\vev{+2n}}^\vee)$.
So, it is also possible to find $\phi$'s with positive Fourier coefficients
from this vector space, instead of finding lattices $L$ and using
$\phi = \theta_{L[-1]}$.}
and the expression for $\chi(X)$ that follows.

In the case of $\Lambda_S = \vev{+2n} =: \Z e$, choose a primitive element
$y' \in E_8$ of norm $2n$ and think of a decomposition 
(\ref{eq:rho1-findL-for-chiExpr}) as in 
footnote \ref{fn:how2find-LinLambdsS+E8}. One can then show that 
the positive definite $L[-1]$ is isometric to the sublattice 
$[(y')^\perp \subset E_8]$. In the $n=1$ case,
$L[-1] \cong E_7$; the relation (\ref{eq:chi-as-lincmb-leBPSind})
for $\phi = \theta_{E_7}$ reproduces $\chi(X) = -252 + 56 n_{1/2}$.
Similarly, the expressions of $\chi(X)$ for $n=2,3$ in
section \ref{ssec:examples-of-Phi}
(Table \ref{tab:EulerNumber-BPSIndicies-rank1}) are also
reproduced from (\ref{eq:chi-as-lincmb-leBPSind}) by choosing
$L[-1]$ as above. Here, we write $n_{x/2n} = n_{\frac{x}{2n} e}$ for
$[(x/2n)e] \in G_{L[-1]}$.

We apply the same procedure to the $n=4,5,6$ cases. It turns out that
we can use the lattice theta functions $\phi = \theta_{L[-1]}$ shown
in the following to obtain an expression for $\chi(X)$:
\begin{align}
& \theta_{L[-1]}^{n=4}(\tau) =
   e_0(1+56q)+(e_{\frac{1}{8}}+e_{\frac{7}{8}})(56q^{15/16})
     +(e_{\frac{2}{8}}+e_{\frac{6}{8}})(28q^{3/4})+(e_{\frac{3}{8}}+e_{\frac{5}{8}})(8q^{7/16})
   \nonumber \\ & \qquad\qquad +e_{\frac{4}{8}}(0q^0+70q)
       + \mathcal{O}(q^{1+\epsilon}), \\
 & \theta_{L[-1]}^{n=5}(\tau) =
    e_0(1+60q)+(e_{\frac{1}{10}}+e_{\frac{9}{10}})(44q^{19/20})
   +(e_{\frac{2}{10}}+e_{\frac{8}{10}})(33q^{4/5})
   +(e_{\frac{3}{10}}+e_{\frac{7}{10}})(12q^{11/20})
    \nonumber\\&\qquad\qquad
    +(e_{\frac{4}{10}}+e_{\frac{6}{10}})(q^{1/5})+e_{\frac{5}{10}}(32q^{3/4})
     + \mathcal{O}(q^{1+\epsilon}),    \\
 & \theta_{L[-1]}^{n=6}(\tau) =
    e_0(1+46q)+(e_{\frac{1}{12}}+e_{\frac{11}{12}})(48q^{23/24})
  +(e_{\frac{2}{12}}+e_{\frac{10}{12}})(30q^{5/6})
  +(e_{\frac{3}{12}}+e_{\frac{9}{12}})(16q^{5/8})
    \nonumber\\&\qquad\qquad
    +(e_{\frac{4}{12}}+e_{\frac{8}{12}})(3q^{1/3})
    +(e_{\frac{5}{12}}+e_{\frac{7}{12}})(48q^{23/24})
    +e_{\frac{6}{12}}(10q^{1/2})    + \mathcal{O}(q^{1+\epsilon}).
\end{align}
The relation (\ref{eq:chi-as-lincmb-leBPSind}) for those
$\phi = \theta_{L[-1]}$ are shown in
Table \ref{tab:EulerNumber-BPSIndicies-rank1}.
For all of $\Lambda_S = \vev{+2n}$ with $n=1,2,3,4,5,6$,
we know from the dimension formula \eqref{eq:RR-thm-ModForm}, \eqref{eq:diff-dim-Mod-Mod0}
that all the $(n+1)$ Fourier
coefficients
$\{ n_{\abs{\gamma}} \}_{ \abs{\gamma} \in G_S/\pm }$
of $\Phi/\eta^{24}$ are linearly
independent for $\Phi \in {\rm Mod}_0(21/2,\rho_{\vev{+2n}})$; so there
cannot be any other linear expressions of $\chi(X) = -c_0(0)$ in terms of those independent $n_{\abs{\gamma}}$'s.
So, it is enough just to find
one $L[-1]$.

In the $n=7$ case, there must be one linear relation among the
8 coefficients $n_{|\gamma|}$ with $| \gamma | \in G_S/\pm$,
because the dimension formula
indicates that the vector space
${\rm Mod}_0(21/2,\rho_{\vev{+14}})$
is 7-dimensional.
So, an expression of the form $\chi(X) = - c_0(0) =
\sum_{|\gamma|} \kappa_{|\gamma|} n_{|\gamma|}$ is not expected to be unique.
Indeed, we can think of two choices of
\begin{align}
  y' = 3 e'_1 + 2 e'_2 + e'_3 \in E_8, \qquad
  y' = 3 e'_1 + (e'_2 + \cdots + e'_6) \in E_8;
\end{align}
here, the lattice $E_8$ is expressed as the abelian group
$\Z^{\oplus 8} \oplus \parren{ (1/2,\dots, 1/2)+\Z^{\oplus 8} }$, and the
intersection form on this is given by setting $(e'_i,e'_j) = + \delta_{ij}$
for the generators $e'_i$ of the $i$-th factor of $\Z^{\oplus 8}$.
The sublattices of $E_8$ isometric to $L[-1]$ are worked out for both 
choices, and it turns out that the corresponding lattice theta functions
\begin{align}
  & \theta_{L[-1]}(\tau) =
     e_0(1+72q)+(e_{\frac{1}{14}}+e_{\frac{13}{14}})(28q^{27/28})
    +(e_{\frac{2}{14}}+e_{\frac{12}{14}})(27q^{6/7})
    +(e_{\frac{3}{14}}+e_{\frac{11}{14}})(27q^{19/28})
     \nonumber\\&\qquad\quad
     +(e_{\frac{4}{14}}+e_{\frac{10}{14}})(q^{3/7})
     +(e_{\frac{5}{14}}+e_{\frac{9}{14}})(q^{3/28})
     +(e_{\frac{6}{14}}+e_{\frac{8}{14}})(27q^{5/7})
     +e_{\frac{7}{14}}(0 q^{1/4})      + \mathcal{O}(q^{1+\epsilon}) \nonumber
\end{align}
and
\begin{align}
    & \theta_{L[-1]}(\tau) =
     e_0(1+44q)+(e_{\frac{1}{14}}+e_{\frac{13}{14}})(42q^{27/28})
    +(e_{\frac{2}{14}}+e_{\frac{12}{14}})(35q^{6/7})
    +(e_{\frac{3}{14}}+e_{\frac{11}{14}})(14q^{19/28})
     \nonumber\\&\quad\quad
     +(e_{\frac{4}{14}}+e_{\frac{10}{14}})(7q^{3/7})
     +(e_{\frac{5}{14}}+e_{\frac{9}{14}})(0q^{3/28})
     +(e_{\frac{6}{14}}+e_{\frac{8}{14}})(21q^{5/7})
     +e_{\frac{7}{14}}(2 q^{1/4})    + \mathcal{O}(q^{1+\epsilon}), \nonumber
\end{align}
are not the same. Two expressions for $\chi$ are obtained from the relation
(\ref{eq:chi-as-lincmb-leBPSind}), and are shown
in Table \ref{tab:EulerNumber-BPSIndicies-rank1-apdx}.

Multiple choices of $\phi \in {\rm Mod}_0(3+\rho/2,\rho_{\Lambda_S}^\vee)$ result
in multiple expressions for $\chi(X)$ in terms of
$\{ n_{\abs{\gamma}} \}_{ \abs{\gamma} \in G_S/\pm }$, as we have seen above
in the $\Lambda_S = \vev{+14}$ case.
When we form a linear combinations of such $\phi$'s so that the leading
Fourier coefficient vanishes (i.e., a cusp form), then we obtain
linear relations among $\{ n_{\abs{\gamma}} \}$ discussed
in section \ref{sssec:lin-reltn}. Multiple expressions for $\chi(X)$ are
consistent because of the linear relations among those low-energy BPS indices.

\subsubsection{The First Derivative of Some Lattice Theta Functions}
\label{apdx:A23}

For any ${\cal R}$ of ADE type, the coefficients
$a^{({\cal R})}_\delta(\nu)$ in
\begin{align}
  \frac{2}{{\rm rank}({\cal R})} q \partial_q \theta_{{\cal R}}
     =: \sum_{\delta \in G_{{\cal R}}} e_\delta a_\delta^{({\cal R})}(\nu) q^\nu
\end{align}
are all integers.
To see this, we use Lemma 6.1 and 6.2 of \cite{Bor1}
for the lattice ${\cal R}$.  We have a formula
\begin{align}
 a^{({\cal R})}_\delta(\nu) = 
    \frac{\sum_{{\bf b} \in {\cal R}^\vee}^{(({\bf b},{\bf b})_{{\cal R}}/2 = \nu)} (({\bf b}, {\bf a})_{{\cal R}})^2}
         {({\bf a},{\bf a})_{{\cal R}}}
\end{align}
for any ${\bf a} \in {\cal R} \otimes \R$. By using any root in ${\cal R}$
as the vector ${\bf a}$, we see that any pair ${\bf b}$ and $-{\bf b}$
in the sum\footnote{${\bf b}=0$ does not form a pair, but $a_0(0)=0$ obviously.
} contributes by
\begin{align}
   \frac{(({\bf b},{\bf a})_{{\cal R}})^2 + ((-{\bf b},{\bf a})_{{\cal R}})^2}
        {({\bf a},{\bf a})_{{\cal R}}} \in \Z.
\end{align}

The values of $a_\delta^{({\cal R})}(\nu)$ for $\nu < 1$ are recorded here.
\begin{align}
    & \mathcal{R}  && \sum_{\delta \in G_{{\cal R}}} \sum_{\nu}^{\nu < 1}
       e_\delta \; a^{({\cal R})}_\delta(\nu) q^\nu \\
    & A_1: && = e_1\; q^{1/4}, \\
    & A_2: && = (e_1 + e_2)\; q^{1/3}, \\
    & A_3: && = (e_1 + e_3)\; q^{3/8} + e_2\; 2 q^{1/2}, \\
    & A_4: && = (e_1 + e_4)\; q^{2/5} + (e_2 + e_3)\; 3q^{3/5}, \\
    & A_5: && = (e_1 + e_5)\; q^{5/12} + (e_2 + e_4)\; 4q^{2/3} + e_3\; 6q^{3/4}, \\
    & A_6: && = (e_1 + e_6)\; q^{3/7} + (e_2 + e_5)\; 5q^{5/7} + (e_3 + e_4)\; 10q^{6/7}, \\
    & A_7: && = (e_1 + e_7)\; q^{7/16} + (e_2 + e_6)\; 6q^{3/4} + (e_3 + e_5)\; 15q^{15/16} 
   \\
    & A_8: && = (e_1 + e_8)\; q^{4/9} + (e_2 + e_7)\; 7q^{7/9} 
   \\
    & A_{r \geq 8}: && = (e_1 + e_r)\; q^{\frac{r}{2(r+1)}} + (e_2 + e_{r-1})\; (r-1) q^{\frac{2(r-1)}{2(r+1)}} 
   \\
    & D_{r=2m}: && = (e_{01} + e_{10})\; 2^{r-3} q^{r/8} + e_{11}\; 2 q^{1/2}
      \qquad {\rm ignore~the~}q^{r/8}{\rm ~term~for~}r \geq 8, \\
    & D_{r=2m+1}: && = (e_{1} + e_{3})\; 2^{r-3} q^{r/8} + e_{2}\; 2 q^{1/2}
      \qquad {\rm ignore~the~}q^{r/8}{\rm ~term~for~}r > 8, \\
    & E_6: && = (e_1 + e_2)\; 6q^{2/3}, \\
    & E_7: && = e_1\; 12q^{3/4}.
\end{align}
All the $a_\delta(\nu)$'s recorded here are equal to $2T_{R}$, where $T_{R}$
is the Dynkin index of the fundamental representation $R$ of the algebra
${\cal R}$ associated with $\delta \in G_{{\cal R}}$.
%
%
%
%

\section{Evaluation of Integrals by Lattice Unfolding}
\label{sec:unfolding}

We need to evaluate the integrals $\Delta_{{\rm grav}}$
in (\ref{eq:def-Delta-grav}, \ref{eq:def-Znew}) and
$\Delta_{{\cal R}}$ in (\ref{eq:gauge-th-cr-2}), just like in
Ref. \cite{DKL-2,HM}. The evaluation method in \cite{DKL-2}
(with extension by \cite{HM}), however, is applicable immediately
only for lattices $\Lambda_S$ of the form $\Lambda_S = U \oplus W$ with
a lattice $W$ of signature $(0,\rho-2)$. Instead, we rely on the evaluation
method presented in \cite{Bor-Grass}; here, we describe the outline
of the evaluation method in \cite{Bor-Grass}, and quote results relevant
to this article (a review is also found in \cite[\S3]{MM}).
The method is applied to cases of our interest in appendices \ref{sssec:integrals-U+W} and \ref{sssec:unfold-deg2};
some of those results are used in the main text. The embedding
trick is presented in a more general form in appendix \ref{ssec:embd-trick} than in the original \cite[\S8]{Bor-Grass};
we use this general form in the calculation in \ref{sssec:unfold-deg2}.

A class of integrals considered in \cite{Bor-Grass} was\footnote{
When the integral shows some divergence, we understand the integral as
regularized by subtracting the integrand by $\mathrm{const} \times
\tau_2^{b^+/2-2}$ (equivalent of integrating in IR degrees of freedom)
\cite{DKL-2} or replacing
$d\tau_2/\tau_2^2 \to d\tau_2/\tau_2^{2+s}$ as in \cite{Bor-Grass}.}
\begin{align}
    I_M(v,F) := \int_{SL(2;\Z)\backslash\UH} \frac{d\tau_1 d\tau_2}{\tau_2^2}
     \tau_2^{b^+/2}\; \overline{\theta_M(\tau,\bar\tau;v)} F(\tau,\bar\tau),
   \label{eq:def-integral-Borch}
\end{align}
where $M$ is an even lattice of signature $(b_+,b_-)$, and $v$ denotes a point in the Grassmannian $Gr(M) := Gr(M \otimes \R;b_+)$.
One point $v \in Gr(M)$ has the same information as isometries
$M \rightarrow \R^{b_+,b_-}$
modulo
$\SO(b_+) \times \SO(b_-)$,
when the lattice $M$
is indefinite.
In the case $M$ is negative definite ($b_+=0$),
$v$ has empty information, and $\overline{\theta_M(\tau,\bar{\tau};v)}$ is
the theta function $\theta_{M[-1]}(\tau)$.
The other factor $F(\tau,\bar\tau)$ in the integrand is a $\C[M^\vee/M]$-valued almost holomorphic modular form of weight $(\frac{b^+-b^-}{2},0)$ and
type $\rho_M$;
%
\begin{align}
    F(\tau,\bar\tau) &= \sum_{\gamma\in M^\vee/M} e_\gamma \sum_{\nu \in\Q,\, \nu \geq \nu_\mathrm{min}} \sum_{k=0}^{k_{\mathrm{max}}}
      c_\gamma^{(k)}(\nu) q^\nu \parren{\frac{-3}{\pi\tau_2}}^k, \quad c_\gamma^{(k)}(\nu) \in \C.
    \label{eq:def-integral-Borch-2}
\end{align}

Those integrals with $M = \widetilde{\Lambda}_S$ are of immediate relevance,
because
\begin{align}
    \Delta_{\mathrm{grav}} = I_{\widetilde{\Lambda}_S} \left(p,
      \frac{\Phi \hat{E}_2}{\eta^{24}}\right),
    \quad
    \Delta_{\mathrm{gauge}} = \frac{1}{24}I_{\widetilde{\Lambda}_S}\left(p,
       \frac{\Phi \hat{E}_2 - \Psi}{\eta^{24}}\right),
  \label{eq:Delta-grav-gauge-as-Borch-integral}
\end{align}
where we used the right-mover and left-mover momenta $p = (p_R,p_L)$
in the Heterotic description to refer to a choice of
$v \in Gr(\widetilde{\Lambda}_S)$. We can just take
$\nu_\mathrm{min} = -1$ and $k_\mathrm{max} = 1$, and both
$\Delta_{\rm grav}$ and $\Delta_{\rm gauge}$ are within the class of
integrals (\ref{eq:def-integral-Borch}, \ref{eq:def-integral-Borch-2})
introduced above.

\bigskip

In evaluating the integral $I_M(v,F)$, Ref. \cite{Bor-Grass} relates it
to an integral in the same class, but with a lattice $M'$ of
$b_+(M')=b_+(M)-1$, $v' \in Gr(M')$, and $F'$
of the form\footnote{
The value of $k_{\rm max}$ does not change under this reduction, but the value of $\nu_{\rm min}$ may not be the same as before if one applies the embedding trick (see appendix \ref{ssec:embd-trick}).
} (\ref{eq:def-integral-Borch-2}) with $M$ replaced by $M'$,
as we review in appendices \ref{ssec:if-nullV-exists}
and \ref{ssec:embd-trick}.
This procedure is called the lattice unfolding method.
At the end, we are left with evaluating integrals of the form
(\ref{eq:def-integral-Borch}, \ref{eq:def-integral-Borch-2}) for a
negative definite lattice $M''$.
Since $\overline{\theta_{M''}}(\tau)$ is holomorphic, the integral $I_{M''}(F'')$ can be regarded as that for 0-dimensional lattice: $I_{M''}(F'') = I_{\{0\}}(\overline{\theta}_{M''}(\tau)F'')$. This type of integrals can be evaluated by simple partial integrals.
When $F(\tau) = \phi(\tau)\hat{E}_2^m$ with $\phi$ some scalar-valued modular form of weight $-2m$, the formula is \cite{LSW-PhysRept, DKL-2}
\begin{align}
    I_{\{0\}}(\phi\hat{E}_2^m) = \frac{\pi}{3(m+1)} [\phi E_2^{m+1}]_{q^0}.
    \label{eq:apdxB:b_+=0-formula}
\end{align}
%

\subsection{Lattice Unfolding Formula}
\label{ssec:if-nullV-exists}

When the lattice $M$ of signature $(b_+,b_-)$ has a non-zero element $z$
of norm $z^2=0$, a lattice $M' := [z^\perp \subset M]/\Z z$ has signature
$(b_+-1,b_--1)$. For $v \in Gr(M)$, $v' \in Gr(M')$ is determined by
the $(b_+-1)$-dimensional vector subspace of the $b_+$-dimensional positive
definite subspace corresponding to $v$, orthogonal to $z$.
Discussions in \cite{Bor-Grass} rewrite $I_M(v;F)$ as a sum of
$I_{M'}(v',F')$ for an appropriately chosen $F'$ and additional terms that
are completely determined in terms of $v$ and $F$.

Since $U[-1]$ has a non-zero null vector, evaluation of
$I_{\widetilde{\Lambda}_S}$ with $\widetilde{\Lambda}_S = U[-1] \oplus \Lambda_S$
can be reduced to that of $I_{\Lambda_S}$ (see appendix \ref{apdx:B11}).
When the lattice $\Lambda_S$ of signature $(1,\rho-1)$ also has a non-zero null 
vector, we can again reduce $I_{\Lambda_S}$ to an integral for a smaller lattice
$[z^\perp \subset \Lambda_S]/\Z z$ of signature $(0,\rho-2)$; see appendix \ref{sssec:when-thereis-nullE}.
Since this lattice
is negative definite, we can apply the formula \eqref{eq:apdxB:b_+=0-formula}.

When the lattice $\Lambda_S$ does not have a non-zero null vector,
we can use the embedding trick (explained in appendix \ref{ssec:embd-trick})
to think of $I_{\Lambda_S}(v',F')$ as
$I_{\widetilde{M}}(\tilde{v},\widetilde{F})$ for some lattice $\widetilde{M}$ and
some modular form $\widetilde{F}$.
Here $\widetilde{M}$ contains $\Lambda_S$ as a sublattice and
$b_+(\widetilde{M})=b_+(\Lambda_S)$, and it can be chosen so that it has
a non-zero null element. In this way, the integrals for all $\Lambda_S$ can
be reduced to integrals with some negative definite lattice.

\subsubsection{From $\widetilde{\Lambda}_S$ to $\Lambda_S$}
\label{apdx:B11}

In the lattice unfolding process from the lattice
$M=\widetilde{\Lambda}_S$ to $M' = \Lambda_S$,
the positive definite $(b_+=2)$-plane $v \in Gr(M)$
is specified by the real and imaginary parts $\mho$ in (\ref{eq:def-mho}).
Let $\{e^0, e^4\}$ be a basis of $U[-1]$ with the intersection form
$(e^0,e^0)=(e^4,e^4)=0$ and $(e^0,e^4)=-1$; they are the generators
of $H^0(K3;\Z)$ and $H^4(K3;\Z)$, as in
section \ref{sssec:Het-IIA-dual-basic}.
When we choose $z=-e^4$, $M'$ is $\Lambda_S$ and the positive definite
$(b_+(M')=1)$-plane $v'$ in $M' \otimes \R$ is the imaginary part of $\mho$,
namely $\R t_2$. Following \cite[Thm. 7.1]{Bor-Grass}, one finds
that\footnote{
Here, $k_{\mathrm{max}} \leq 1$ is assumed.
The value of ``const" in the expression is regularization dependent.
}
\begin{align}
    I_{\widetilde{\Lambda}_S}(v,F) = &\;
    \frac{\abs{t_2}}{\sqrt{2}} I_{\Lambda_S}(v',F)
      \label{eq:unfold-2to1} \\
    & + c_0^{(0)}(0) \sqbra{ - \log(2\pi t_2^2) + \const  } - c_0^{(1)}(0)  \frac{6\zeta(3)}{\pi^2 t_2^2}
     \nonumber \\
    & + \sum_{0\neq w \in \Lambda_S^\vee}
      \sqbra{   2 c_{[w]}^{(0)}(w^2/2) \Li_1(e^{2 \pi i(w \;\widehat{\cdot}\; t)})
              - \frac{24}{t_2^2} c_{[w]}^{(1)}(w^2/2) Li_3(e^{2 \pi i(w \;\widehat{\cdot}\; t)})
    }.   \nonumber
\end{align}
Here $(w \;\widehat{\cdot}\; t) = (w,t_1) + i\abs{(w,t_2)}$ and $Li_3(e^{i z}) = \Li_3(e^{i z}) + \Im(z)\Li_2(e^{i z})$.
The modular form $F$ is used as $F'$ in the first term on the right hand side.
The details of the second and the third line are necessary when working out the matching calculation (\ref{eq:matching-chi+GV}), but are not relevant to the matching (\ref{eq:matching-rslt-cubic-hol},
\ref{eq:matching-rslt-F1-hol}, \ref{eq:matching-rslt-h1-hol}) that we need
to discuss in this article.

The integral $I_{\widetilde{\Lambda}_S}(v,F)$ as a function of
$v \in Gr(\widetilde{\Lambda}_S)$ has singularity only along real
codimension-$(b_+(\widetilde{\Lambda}_S)=2)$ subspace,
whereas $I_{\Lambda_S}$ as a function of $v' \in Gr(\Lambda_S)$ has
singularity (``wall-crossing") along real codimension-$(b_+(\Lambda_S)=1)$
walls;
$I_{\widetilde{\Lambda}_S}(v,F)$ has logarithmic singularity at the locus
$\mho(w)=0$ for some $w \in \widetilde{\Lambda}_S^\vee$, while
$I_{\Lambda_S}(v',F)$ has conical singularity at $(w,t_2)=0$ for some
$w \in \Lambda_S^\vee$. This implies that
there is (partial) cancellation of singularity between
$I_{\Lambda_S}(v',F')$ and the third line of \eqref{eq:unfold-2to1} so the sum of them remains non-singular at the codimension-1 walls
(except the codimension-2 points).\footnote{
The equation \eqref{eq:unfold-2to1} holds even when $t_2$ does not have a large norm (although \cite[Thm. 7.1]{Bor-Grass} proves the corresponding claim  only when the norm $(t_2,t_2)$ is sufficiently large), because both sides of \eqref{eq:unfold-2to1} are well-defined even for small $t_2$ and they remain (real-)analytic inside any single chamber, so they must be the same.
}
See appendix \ref{sssec:wall-xing} for the wall-crossing formula of $I_{\Lambda_S}$. For more information, see \cite[\S6]{Bor-Grass}.

\subsubsection{$I_{\Lambda_S}$ for $\Lambda_S$ with a Null Element}
\label{sssec:when-thereis-nullE}

In the rest of appendix \ref{ssec:if-nullV-exists}, we discuss
evaluation of $I_{\Lambda_S}(v',F)$
for $\Lambda_S$ that has a non-zero null element $z$.
Let $z$ be a non-zero primitive null element of $\Lambda_S$, and
$N$ be the GCD of the values of $(z,\lambda) \in \Z$ for
$\lambda \in \Lambda_S$. We then choose $\tilde{z} \in \Lambda_S$ so
that $(z,\tilde{z})=N$, and a free abelian subgroup $W \subset \Lambda_S$
such that $z \perp W$ and $\Lambda_S \cong_{ab}
\Z z \oplus \Z\tilde{z} \oplus W$;
here $\cong_{ab}$ stands for an isomorphism between abelian groups.
The same $W$ also denotes the lattice $[z^\perp \subset \Lambda_S]/\Z z$.
The complexified K\"{a}hler parameter
$t \in \Lambda_S \otimes \C$ can be parametrized by
\begin{align}
  t = u z + \rho \tilde{z} + a, \qquad u, \rho \in \C, \quad
   a \in W \otimes \C.
  \label{eq:Kahler-cone-parametrize-nullE}
\end{align}

Here is additional preparation.
The pairing $(-,z) : \Lambda_S \to \Z$ induces a homomorphism $(-,z) : G_S \to \Z_N$,
where $\Z_N := \Z/N\Z$.
The kernel of this homomorphism
$\ker (-,z)$ includes the subgroup
$\vev{z/N} \isom \Z_N$ of $G_S$ generated by $z/N + \Lambda_S$.
We have a natural isomorphism $j: \ker(-,z)/\vev{z/N} \isom G_W$
of finite abelian groups, which preserves the quadratic form.
%
%
The subset $j^{-1}(\lambda) \subset {\rm ker}(-, z)$ for $\lambda \in G_W$
is of the form $\delta_0 + \vev{z/N}$ for some $\delta_0$ that depends on
$\lambda$. The linear map
\begin{align}
    \C[G_S] \to \C[G_W],  \qquad
    \left\{
  \begin{array}{ll}
       e_{\delta} \mapsto e_{j(\delta)} & (\delta \in {\rm ker}(-,z)), \\
       e_{\delta} \mapsto 0 & (\delta \not\in {\rm ker}(-,z)),
   \end{array}
         \right.
\end{align}
is compatible with the Weil representation \cite[Thm. 5.3]{Bor-Grass}.
For a modular form $F$ of type $\rho_{\Lambda_S}$, we denote by $F_W$ the modular form of type $\rho_W$ obtained by composing $F : \UH \to \C[G_S]$ and this linear map.

Now
Thm. 7.1, Lemma 7.3, and Thm. 10.2 of Ref. \cite{Bor-Grass} can be
used to rewrite the integrals $I_{\Lambda_S}(v',F)$ in terms of
$u$, $\rho$, $a$, and the coefficients $c^{(k)}_\gamma(\nu)$ of $F$.
We just quote the results for the case $k_{\mathrm{max}} \leq 1$:
when $t_2$ lies in a fundamental
%
%
chamber\footnote{
The positive cone, one of two components of
$\{ t_2 \in \Lambda_S \otimes \R \; |
\; (t_2,t_2)>0 \}$,
is divided into {\it chambers} based on which
subset of $\Pi := \left\{ w \in \Lambda_S^\vee \; | \;
2\nu_{\rm min} \leq w^2 < 0 \right\}$ is characterized as
$\left\{w \in \Pi \; | \; (w,t_2)>0 \right\}$. We call a chamber ${\cal C}$
a {\it fundamental chamber}, if it contains a region $|(z,t_2)| \ll |t_2|$
(equivalently $N|\rho_2| \ll |t_2|$).
There may be multiple fundamental chambers in general. }\raisebox{4pt}{,}\footnote{
This formula is guaranteed to be valid in fundamental chambers.
For evaluating the integral in other chambers, one needs to use the
wall-crossing formula \eqref{eq:wall-crossing-ILambda_S}.
See appendix \ref{sssec:wall-xing} for details.}
$\mathcal{C}_z$,
\begin{align}
    & \frac{\abs{t_2}}{\sqrt{2}} I_{\Lambda_S}(v',F)
    = \frac{t_2^2}{2N\rho_2} I_W(F_W)
    \label{eq:unfolding-formula-LambdaS}
    \\ \nonumber & \qquad
    + 2\pi \sum_{b\in W^\vee} \sum_{\delta\in j^{-1}([b])}
    \parren{
        N \rho_2\, c^{(0)}_\delta(b^2/2) \mathbb{B}_2( \cdots )
      + \frac{2N^3\rho_2^3}{t_2^2}\, c^{(1)}_\delta(b^2/2) \mathbb{B}_4( \cdots )
    },
\end{align}
where the argument of the functions\footnote{
The function $\mathbb{B}_m(x)$ is defined (for $m > 0$) by
\begin{align}
 \mathbb{B}_m(x) = -m! \sum_{n \neq 0} \frac{\mathbb{E}(nx)}{(2\pi i n)^m}
  = - \frac{m!}{(2\pi i)^m}
      \left( {\rm Li}_m(\mathbb{E}(x)) + (-1)^m {\rm Li}_m(\mathbb{E}(-x)) \right).
      \nonumber
\end{align}
It satisfies $\mathbb{B}_m(x+1) = \mathbb{B}_m(x)$ and
$\mathbb{B}_m(-x) = (-1)^m \mathbb{B}_m(x)$ by definition.
Its value for $0 < x < 1$ is given by the usual Bernoulli polynomial
$B_m(x)$; e.g.
    $B_2(x) = x^2 - x + 1/6$ and
    $B_4(x) = x^4 - 2x^3 + x^2 - 1/30$.
}
 $\mathbb{B}_2$ and $\mathbb{B}_4$ are
$(b,a_2/(N\rho_2)) + N^{-1}(\delta,\tilde{z})$.
The first term $I_W(F_W)$ can be evaluated by the
formula \eqref{eq:apdxB:b_+=0-formula}.

When one expands the right hand side, there appear some terms with $\rho_2$ in the denominator, but they should cancel out because of consistency with wall-crossing behavior of $I_{\Lambda_S}$ (see \cite[Thm. 10.3]{Bor-Grass}).\footnote{
Roughly, the claim is proved by observing that different choices of $z$ put different variable (e.g. $u_2$) in the denominator and may cause different type of singularity. But the wall-crossing is just a polynomial so the fractional terms must vanish.
}
Due to this cancellation,
the integral $I_{\Lambda_S}(v',F) \times |t_2|^{1+2k_{\rm max}}$
turns out to be 
a chamber-wise
homogeneous function of the components of $t_2$ of degree $(1+2k_{\rm max})$.

The integrals of our interest are those
in (\ref{eq:Delta-grav-gauge-as-Borch-integral}), and we find it useful
to assign notations for the following combinations (the same
as (\ref{eq:def-of-P_1}, \ref{eq:def-of-P_3})):
\begin{align}
    P_1(t_2) := \frac{1}{4\pi} \frac{\abs{t_2}}{\sqrt{2}}  I_{\Lambda_S}(v',\Psi\eta^{-24}),
    \qquad
    P_3(t_2) := \frac{-t_2^2}{32\pi} \frac{\abs{t_2}}{\sqrt{2}} I_{\Lambda_S}(v',(\Phi\hat{E}_2-\Psi)\eta^{-24});
    \label{eq:apdx:def-of-P_1P_3}
\end{align}
$P_1$ is a chamber-wise polynomial of degree 1 and $P_3$ of degree 3.
They are used to determine some of the functions of the low-energy
effective theory
in (\ref{eq:matching-rslt-cubic-hol}, \ref{eq:matching-rslt-F1-hol}).

\subsubsection{Examples: $\Lambda_S = U \oplus W$}
\label{sssec:integrals-U+W}

The polynomials $P_3$ and $P_1$ have simpler expressions when
$\Lambda_S$ is of the form $U \oplus W$ for some even and negative
definite lattice $W$.
In the fundamental chambers,\footnote{
This expression is for fundamental chambers that contain the region
with $(b,a_2) < \rho_2$ for $b$'s appearing in the sum.
}
they are
\begin{align}
    \frac{1}{3!} P_3(t)
    &=
  \frac{2}{3!} \rho^3
 + \frac{n'-2}{2} \rho^2 u
 + \frac{n'}{2} \rho u^2
 + \frac{n'}{2} (u + \rho)(a,a)
 \nonumber \\ & \quad
 - \frac{u+\rho}{2} \sum_{b \in W^\vee} \frac{d_{[b]}(b^2/2)}{12}(b,a)^2
 + \frac{2u\rho+(a,a)}{4}\sum_{(b,a_2) > 0} \frac{d_{[b]}(b^2/2)}{12}(b,a)
 \nonumber \\ & \quad
 + \frac{1}{3!} \frac{1}{2} \sum_{\substack{b \in W^\vee\\ (b,a_2) > 0}} c_{[b]}(b^2/2) (b,a)^3,
  \\
 P_1(t)  &=
  -4\rho + 12(2+n')(u+\rho)
 - \sum_{\substack{b \in W^\vee\\ (b,a_2) > 0}} c^\Psi_{[b]}(b^2/2) (b,a),
\end{align}
where we have taken care of cancellation referred
below the formula \eqref{eq:unfolding-formula-LambdaS}.
The parameter $n'$ extracts\footnote{
Suppose that
$W \subset U^{\oplus 2} \oplus E_8^{(1)}$,
the probe gauge group $\mathcal{R}$ is in $E_8^{(2)}$,
and there are $I = 12-n'$ instantons on K3
(of the $\mathrm{K3}\times T^2$ internal space)
in $[\mathcal{R}^\perp \subset E_8^{(2)}]$
in the Heterotic language.
Then $\Psi$ should be the one with $d_\gamma(\nu)$'s that are related to $n'$ in this way.
} a combination of $d_\gamma(\nu)$'s through
$(2-n') := 144^{-1} \sum_{b} d_{[b]}(b^2/2)$.

An element $\Phi \in [{\rm Mod}_0^\Z(11-\rho/2,\rho_{\Lambda_S})]^{\S 2}$
and $\Psi \in {\rm Mod}_0^\Phi(13-\rho/2,\rho_{\Lambda_S})$ is
in $[{\rm Mod}_0^\Z(11-\rho/2,\rho_{\Lambda_S})]^{\rm r.mfd}$ and
$[{\rm Mod}_0^\Phi(13-\rho/2,\rho_{\Lambda_S})]^{\rm r.mfd}$, respectively,
if the following conditions are satisfied.
For integrality of the coefficients $d_{abc}$ of the $\rho^2 u/2$ and
$\rho u^2/2$ terms in $P_3$,
\begin{align}
 \left( 2 - \frac{1}{12}\sum_{b \in W^\vee} \frac{d_{[b]}(b^2/2)}{12} \right)
   =: n' \in \Z.
   \label{eq:geom-cond-U+W-1}
\end{align}
The $u\rho a$ term and $aaa/3!$ term have integer coefficients, only if
\begin{align}
  \sum_{b \in W^\vee}^{(b,a_2)>0} \frac{d_{[b]}(b^2/2)}{12} b \in 2 W^\vee
    \label{eq:geom-cond-U+W-2}
\end{align}
and
\begin{align}
  \sum_{b \in W^\vee}^{(b,a_2)>0} c_{[b]}(b^2/2) \; (b,r_1) (b,r_2) (b, r_3) \in 2 \Z
  \qquad \qquad {\rm for~} r_{1,2,3} \in W,
    \label{eq:geom-cond-U+W-3}
\end{align}
respectively. The condition (a') is translated into
\begin{align}
  \sum_{b \in W^\vee}^{(b,a_2)>0} c_{[b]}(b^2/2) \; (b,r_1) (b,r_2) (b,r_1+r_2) \equiv 0
    \quad ({\rm mod~} 4)
     \qquad \qquad {\rm for~} r_{1,2} \in W,
\end{align}
and the condition (b') to
\begin{align}
  \sum_{b \in W^\vee}^{(b,a_2)>0} \left(
       c_b(b^2/2) (b,r)^3 - c_{[b]}^\Psi(b^2/2) (b,r) \right)
    \equiv 0 \quad ({\rm mod~}12)
       \qquad \qquad {\rm for~} r \in W,
\end{align}
or equivalently, to
\begin{align}
    \sum_{b \in W^\vee}^{(b,a_2)>0} c_{[b]}(b^2/2) \; (b,r) ((b,r)+1)((b,r)-1)
    \equiv 0 \quad ({\rm mod~}12)
       \qquad \qquad {\rm for~} r \in W.
\end{align}

The special case $W = \{0\}$ has been treated in
section \ref{sssec:u-calc}. The $W = A_1[-1]$ case yields an example
where $[{\rm Mod}_0^\Z(11-\rho/2,\rho_{\Lambda_S})]^{\rm r.mfd}$
is a proper subset of $[{\rm Mod}_0^\Z(11-\rho/2,\rho_{\Lambda_S})]^{\S2}$.
To see this, note that the modular forms $\Phi$ and $\Psi$ are parametrized
by $n_{1/2} \in \Z_{\geq 0}$ and $m_{1/2} \in 12\Z$ (after using
$\dim_\C({\rm Mod}_0(23/2,\rho_{\vev{-2}})) = 2$ and imposing $n_0=-2$ and
$m_0=0$). The integrality of $n' = 2-[56-n_{1/2}+m_{1/2}]/12$ is translated into
$n_{1/2} \in 8 + 12\Z$, so $[{\rm Mod}_0^\Z(11-\rho/2,\rho_{\Lambda_S})]^{\rm r.mfd}$
is strictly smaller. The other four conditions above follow automatically for $n' \in \Z$. One may also find from this that the image of 
$\mathrm{diff}_{\mathrm{fine}}$ in the set ${\rm Diff}_{\Lambda_S}^{d'}$ is $\Z/2\Z$
for $\Lambda_S = U\oplus W$ with $W=A_1[-1]$. 


\subsection{Wall-crossing Behavior}
 \label{sssec:wall-xing}

We give some comments on the wall-crossing behavior. 
See section 6 of \cite{Bor-Grass} for details. 
$I_{\Lambda_S}(v,F)$ as a function of $v = \R t_2 \in Gr(\Lambda_S)$ shows conical singularities along the walls $(w,t_2) = 0$ for some\footnote{
$w$ that gives singularity satisfies $-2 \leq w^2 < 0$;
the condition $w^2 < 0$ [resp. $w^2/2 \geq \nu_{\mathrm{min}} = -1$] follows from $t_2^2 > 0$ [resp. $c_\lambda(\nu) = 0$ for $\nu < \nu_{\mathrm{min}}$].
} $w$'s in $\Lambda_S^\vee$.
These real-codimension-1 walls separate $Gr(\Lambda_S)$ into many chambers; $I_{\Lambda_S}$ is analytic in each chamber but shows jump from its analytic continuation when $t_2$ crosses a wall.
Let $I_{\Lambda_S}(v,F;\mathcal{C})$ be the analytic continuation of the restriction of $I_{\Lambda_S}(v,F)$ to a chamber $\mathcal{C}$.
The difference for two different chambers $\mathcal{C}_1$ and $\mathcal{C}_2$ is given by
\begin{align}
    & I_{\Lambda_S}(v;\mathcal{C}_1) - I_{\Lambda_S}(v;\mathcal{C}_2)
    \label{eq:wall-crossing-ILambda_S} \\ \nonumber
    & \quad = \sum_{\substack{(w,\mathcal{C}_1) > 0 \\ (w,\mathcal{C}_2) < 0}}
    \left\{
     c_{w}^{(0)}(w^2/2)\;(-8\sqrt{2}\pi)(w,v)
     +c_{w}^{(1)}(w^2/2)\;(-32\sqrt{2}\pi)(w,v)^3
     \right\}.
\end{align}
Here $(w,\mathcal{C}) > 0$ means $(w,t_2) > 0$ for any $t_2 \in \mathcal{C}$.
Note that the difference as shown above is a polynomial of $v = t_2/\abs{t_2}$.
Using this fact, \cite[Thm. 10.3]{Bor-Grass} shows that $I_{\Lambda_S}(v,F)$ gives chamber-wise polynomial of $v$ with degree at most 3 in our case (in fact, without even degree terms).

As a corollary, we obtain wall-crossing formulas for $P_1(t)$ and $P_3(t)$, which are directly relevant to the topological invariants of $X_\mathrm{IIA}$:
\begin{align}
    \frac{1}{3!} P_3(t; \mathcal{C}_1) - \frac{1}{3!} P_3(t; \mathcal{C}_2)
    & = \sum_{w}
    \braces{
     \frac{c_{w}(w^2/2)}{3!} (w, t)^3
    + \frac{(t,t)}{2} \frac{d_{w}(w^2/2)}{12} \; (w, t)
    },   \label{eq:Delta-P3}
    \\
    P_1(t;\mathcal{C}_1) - P_1(t;\mathcal{C}_2)
    & = - \sum_{w} 2c^\Psi_{w}(w^2/2) (w,t).
         \label{eq:Delta-P1}
\end{align}
The summation over $w$ is the same as in \eqref{eq:wall-crossing-ILambda_S}.

\subsection{Embedding Trick}
 \label{ssec:embd-trick}

Even when $\Lambda_S$ has no non-zero null elements, one can evaluate the integral $I_{\Lambda_S}(v,F)$ by embedding $\Lambda_S$ into a larger lattice(s) $\widetilde{M}$ with a non-zero null element so that $I_{\Lambda_S}(v,F)$ is equal to (linear combination of) $I_{\widetilde{M}}(v,G)$ for a suitable modular form(s) $G$. One can then apply the lattice unfolding method to compute it. This method is called the ``embedding trick" in \cite{Bor-Grass}. In this section, we explain the original embedding trick and its slight modifications. We also treat a concrete example (the case $\Lambda_S = \vev{+2}$).

The original embedding trick is as follows \cite[Thm. 8.1]{Bor-Grass}. To begin with, choose a pair of (negative definite) Niemeier lattices $M_1,\,M_2$ with different numbers of roots:
\begin{align}
    r_1 - r_2 \neq 0, \qquad r_i := \text{the number of roots in } M_i.
\end{align}
Then we obtain
\begin{align}
    1 = \parren{ \bar\theta_{M_1}(\tau) - \bar\theta_{M_2}(\tau) } \frac{1}{(r_1-r_2)\eta(\tau)^{24}};
    \label{eq:et-3}
\end{align}
this is because the right hand side is a scalar-valued modular form of weight 0 with the term $q^{-1}$ vanishing and the coefficient of $q^0$ normalized.
By inserting the expression on both sides to the integrand of $I_{\Lambda_S}(v,F)$, we get
\begin{align}
  I_{\Lambda_S}(v,F) =
     I_{\Lambda_S \oplus M_1}(v, G)
  -  I_{\Lambda_S \oplus M_2}(v, G),
\end{align}
where $G = F\,[(r_1-r_2)\eta^{24}]^{-1}$ and $v \in Gr(\Lambda_S)$ is regarded as the same positive definite $(b_+=1)$-dimensional subspace $v \in Gr(\Lambda_S) \subset Gr(\Lambda_S \oplus M_i)$.
Since $\Lambda_S \oplus M_i$ has $U$ as direct summand (see \cite[\S 8]{Bor-Grass}), we can apply the lattice unfolding formula to evaluate the right hand side.

There are a few points to keep in mind. First, $G$ has a pole of higher order at cusps than $F$ (i.e. $\nu_{\mathrm{min}}(G) = \nu_{\mathrm{min}}(F)-1 = -2$). Second, in many cases (e.g., in appendix \ref{sssec:unfold-deg2}), we need to care about wall-crossings of $I_{\Lambda_S\oplus M_i}$ for walls between $v$ and the fundamental region in $Gr(\Lambda_S \oplus M_i)$ where the unfolding formula (\ref{eq:unfolding-formula-LambdaS}) is valid.

This embedding trick adds Niemeier lattice to the original lattice and increases the rank by as many as $\rank M_i = 24$; sometimes it is troublesome to handle such a big lattice and consider all the relevant wall-crossings.
But actually, it suffices to add a smaller lattice: choose sublattices $N_i \subset M_i$ such that $N_1 \cong N_2 \;(\cong: N)$.
Then by decomposing the theta function of $M_i$ as
\begin{align}
    \bar\theta_{M_i} (\tau) = \sum_{\delta \in N_i^\vee/N_i} \bar\theta_{N_i + \delta} \bar\theta_{N_i^\perp + \delta} (\tau),
    \label{eq:et-4}
\end{align}
we can rewrite \eqref{eq:et-3} to
\begin{align}
    1 = \sum_{\delta \in N^\vee/N} \bar\theta_{N+\delta}(\tau) h_\delta(\tau),
    \qquad
    h_\delta(\tau) =
    \frac{\bar\theta_{N_1^\perp+\delta}(\tau)-\bar\theta_{N_2^\perp+\delta}(\tau)}{(r_1-r_2)\eta(\tau)^{24}}.
    \label{eq:et-5}
\end{align}
$\{h_\delta\}$ is a modular form of type $\rho_N$.
Inserting this expression to the integrand of $I_{\Lambda_S}(v,F)$, we get
\begin{align}
    I_{\Lambda_S}(v,F) = I_{\Lambda_S \oplus N}(v,G),
    \qquad
    G_{\gamma,\delta} = F_\gamma h_\delta.
\end{align}
Here $G$ is a modular form of type $\rho_{\Lambda_S \oplus N}$.
Note that also in this case $\nu_{\mathrm{min}}(G)$ is less than $\nu_{\mathrm{min}}(F) = -1$ (but still $\nu_{\mathrm{min}}(G) \geq -2$).

After all, an important point for the embedding trick is to find a decomposition
\begin{align}
    1 = \sum_{\delta \in N^\vee/N} \bar\theta_{N+\delta}(\tau) h_\delta(\tau)
    \label{eq:et-0}
\end{align}
for some lattice $N$ and a modular form $h$, not necessarily associated with
Niemeier lattices.
There are many choices. For example, let\footnote{
We can choose $N$ to be rank-1 for any $\Lambda_S$. This is because
$\Lambda_S$ has necessarily an element of norm $2m$ for some positive
integer $m$; the lattice $\Lambda_S \oplus N$ with $N = \vev{-2m}$ has
a non-zero null element then.
} $N = \vev{-2m}$,
and $\varphi(\tau,z)$ be a weak Jacobi form of weight 0 and index $m$,
satisfying $\varphi(\tau,z = 0) \neq 0$.
We can theta-expand $\varphi$ using a modular form $h$:
\begin{align}
    \varphi(\tau,z) = \sum_{\delta \in \Z_{2m}}
    \bar\theta_{\vev{-2m} + \delta}(\tau,z)\, h_{\delta}(\tau).
    \label{eq:et-1}
\end{align}
Since a modular form of weight 0 and holomorphic at cusps is necessarily just a constant, setting $z = 0$ in the above equation leads to the required decomposition \eqref{eq:et-0} up to some normalization. If $m\geq 2$, there are multiple choices for such $\varphi$.

So, even for a given lattice $\Lambda_S$, there are multiple different ways
to use the embedding trick, $I_{\Lambda_S}(v,F)= \sum_i I_{\widetilde{M}_i}(v,G_i)$. There is no
unique choice for $\widetilde{M} = \Lambda_S \oplus N$; even for a given $N$,
the choice of $G$ is not necessarily unique. One can use just any version
of the embedding trick, so practical calculations of one's interest are easier.

\subsubsection{Example: $\Lambda_S = \vev{+2}$}
\label{sssec:unfold-deg2}

Let us compute $I_{\Lambda_S = \vev{+2}}(F)$ for $F = \Psi \eta^{-24},\, (\Phi \hat{E}_2 - \Psi)\eta^{-24}$ using the embedding trick.\footnote{
When $F$ is holomorphic on the upper half-plane (e.g. $\Psi\eta^{-24}$),
$I_{\vev{+2n}}(F)$ can be also calculated by using \cite[Cor 9.6]{Bor-Grass},
which uses Zagier's modular form and Stokes' theorem. }
They contribute to (\ref{eq:Delta-grav-gauge-as-Borch-integral})
through (\ref{eq:unfold-2to1}); as a reminder, $\Phi$ and $\Psi$ to
be used here (for $\Lambda_S=\vev{+2}$) are those
in \eqref{eq:deg2-Phi-paramtrz-main}, \eqref{eq:deg2-Psi-paramtrz-main}.

As a practical implementation of the embedding trick,
we choose $N = \vev{-2}$, so $\widetilde{M} = \vev{+2} \oplus \vev{-2}$,
and $G_{\widetilde{M}} = \Z_2 \times \Z_2$. We will use the notation
$\Z_2 \cong \{ 0,1\}$ (instead of $\{0,1/2\}$) in this
appendix \ref{sssec:unfold-deg2}.
Let us use the weak Jacobi form $\varphi(\tau,z)$ of weight 0 and index 1 (i.e., index $\vev{+2}$) normalized so that $\varphi(\tau,z=0)=1$,
and determine a vector-valued modular form $h$
through (\ref{eq:1to1-Jacob-vvModForm}, \ref{eq:et-1}).
Now
\begin{align}
  I_{\vev{+2}}(v, \Psi \eta^{-24}) = I_{\widetilde{M}}(v, G), \qquad
  I_{\vev{+2}}(v, (\Phi \hat{E}_2 - \Psi)\eta^{-24}) = I_{\widetilde{M}}(v,\widetilde{G}),
    \label{eq:integrals-interest-deg2-embdT-applied}
\end{align}
where $G = \sum_{\delta,\gamma \in \Z_2} e_{\delta \gamma} h_\delta \Psi_\gamma \eta^{-24}$
and $\widetilde{G} = \sum_{\delta \gamma \in \Z_2} e_{\delta, \gamma} h_\delta
(\Phi_\gamma \hat{E}_2-\Psi_\gamma)\eta^{-24}$. Because
\begin{align}
    h(\tau)
    = \frac{e_0}{12} \parren{ 10 + 108 q^{4/4} + \mathcal{O}(q^{8/4}) } + \frac{e_1}{12} \parren{ \frac{1}{q^{1/4}} - 64 q^{3/4} + \mathcal{O}(q^{7/4})},
    \label{eq:deg2-h}
\end{align}
$\nu_{\mathrm{min}}(G)$ and $\nu_{\mathrm{min}}(\widetilde{G})$ are both $-5/4$,
rather than $-1$.

Let us describe the chamber structure in $\widetilde{M} \otimes \R$.
To start, we introduce a parametrization of the space
$\{ \tilde{t}_2 \in \widetilde{M} \otimes \R\}$ as in
(\ref{eq:Kahler-cone-parametrize-nullE}).
We denote a generator of $\Lambda_S = \vev{+2}$ and $N = \vev{-2}$
by $v_{+2}$ and $v_{-2}$, respectively. In a basis
$(z,\tilde{z}) = (v_{+2} + v_{-2},\,-v_{-2})$, the intersection form of
$\widetilde{M}$ is given by
\begin{align}
    \begin{pmatrix} 0 & 2 \\ 2 & -2 \end{pmatrix}.
\end{align}
So we can use $z$ as the null element for the lattice unfolding method;
now $\tilde{t}_2 = u_2 z + \rho_2 \tilde{z}$.
Next, the walls of interest are of the form $(\tilde{t}_2, \lambda)=0$
for some $\lambda \in \widetilde{M}^\vee$, with $\nu_{\rm min} \leq \lambda^2/2 < 0$.
This condition is equivalent to $0 < m^2-n^2 \leq 5$ for
$\lambda = n v_{+2}/2 + m v_{-2}/2 \in \widetilde{M}^\vee$,
$n, m \in \Z$. So, there are only finite number of solutions $(n,m)$:
\begin{align}
    (n,m) = (0,\pm 1) {\rm ~and~} (0,\pm 2),\, (\pm 1, \pm 2),\, (\pm 2, \pm 3).
\end{align}
See Table \ref{tab:deg2-embded-listofwalls} for the list of those walls
in $\widetilde{M} \otimes \R$.
\begin{table}[tbp]
\begin{center}
    \begin{tabular}{cccc}
        $(n,m)$ & $\lambda^2/2$ & $(\lambda,t_2 = u_2 z + \rho_2 \tilde{z})$ & wall at $u_2/\rho_2 =$ \\ \hline
        $\pm(2,-3)$ & $ -5/4 $ & $\pm(5u_2-3\rho_2)$ & $3/5$ \\
    \hline
        $\pm(1,-2)$ & $ -3/4 $ & $\pm(3u_2-2\rho_2)$ & $2/3$ \\
    \hline
        $\pm(0,1)$ & $-1/4$ & $\mp(u_2-\rho_2)$ & $1$ \\
        $\pm(0,2)$ & $ -1 $ & $\mp 2(u_2-\rho_2)$ &   \\
     \hline
        $\pm(1,2)$ & $ -3/4 $ & $\mp(u_2-2\rho_2)$ & $2$ \\
      \hline
        $\pm(2,3)$ & $ -5/4 $ & $\mp(u_2-3\rho_2)$ & $3$ \\
        \hline
    \end{tabular}
    \caption{\label{tab:deg2-embded-listofwalls}The list of walls where
  $I_{\widetilde{M}}(\R\tilde{t}_2, G)$ and $I_{\widetilde{M}}(\R\tilde{t}_2,\widetilde{G})$
   are singular. They are sorted in the order of their slopes in the
  $\rho_2$--$u_2$ plane. They are all within the positive cone
$(2u_2-\rho_2)\rho_2 > 0$ (and $\rho_2 > 0$) of the lattice $\widetilde{M}$.}
\end{center}
\end{table}


The integrals (\ref{eq:integrals-interest-deg2-embdT-applied}) are evaluated,
first, in the fundamental chamber ${\cal C}_z$ by using the lattice unfolding
formula (\ref{eq:unfolding-formula-LambdaS}). The fundamental
chamber is the region $0 < 3\rho_2 < u_2$.
The result is
\begin{align}
 P_1(\tilde{t}_2; {\cal C}_0) & \; =
   (2u_2-\rho_2)\left(10-b_{{\cal R}}- \frac{7}{2}n_{1/2} \right)
 + \rho_2 \frac{(248-18b_{{\cal R}} -29 n_{1/2})}{6},
    \label{eq:P1-deg2-in-fundCmbr} \\
 P_3(\tilde{t}_2; {\cal C}_0) & \; =
     (2u_2-\rho_2)^2 \rho_2 \frac{(-b_{{\cal R}} -3n_{1/2})}{4} \nonumber \\
  & \quad
    + (2u_2-\rho_2)\rho_2^2 \frac{(20-3b_{\cal R} + 5n_{1/2})}{4}
   + (\rho_2)^3 \frac{1-7n_{1/2}}{3}; \label{eq:P3-deg2-in-fundCmbr}
\end{align}
in using the formula (\ref{eq:unfolding-formula-LambdaS}) for the lattice
$\widetilde{M}$ and the null element $z=(v_{2+}+v_{2-})$, the lattice $W$ is
$\{ 0\}$, and $N=2$; the subgroup ${\rm ker}(-,z)$ is
$\{ 0, (z/2)_{+\widetilde{M}} \} = \{ (0,0), (1,1) \} \subset G_{\widetilde{M}}$,
which is also equal to $j^{-1}(0)$ for $0 \in G_W$;
relevant Fourier coefficients of $G$ and $\widetilde{G}$ are computed
from (\ref{eq:integrals-interest-deg2-embdT-applied}) and
(\ref{eq:deg2-Phi-paramtrz-main}, \ref{eq:deg2-Psi-paramtrz-main}).

The fundamental region ${\cal C}_z$ is separated from the
$\Lambda_S \otimes \R \subset \widetilde{M}\otimes \R$ locus
by the two walls $u_2 = 3\rho_2$ and $u_2 = 2\rho_2$.
The integrals in the chamber $\rho_2 \leq u_2 \leq 2\rho_2$
are obtained by adding the following terms
to (\ref{eq:P1-deg2-in-fundCmbr}, \ref{eq:P3-deg2-in-fundCmbr}),
\begin{align}
 \Delta P_1 & \; = \frac{-2}{6} (u_2-3\rho_2) + \frac{10n}{6}(u_2-2\rho_2), \\
 \Delta P_3 & \; = \frac{2}{12}(u_2-3\rho_2)^3
    - \frac{10n_{1/2}}{12}(u_2-2\rho_2)^3,
\end{align}
because of the wall crossing formula (\ref{eq:wall-crossing-ILambda_S}).
The first and second terms are associated with the walls at
$u_2 = 3\rho_2$ and $u_2 = 2\rho_2$, respectively.
Relevant Fourier coefficients are $[G_{10}]_{q^{-5/4}} = -2/12$,
$[G_{01}]_{q^{-3/4}} = 10n/12$, $[h_1\Phi_0 \eta^{-24}]_{q^{-5/4}} = -2/12$,
and $[h_0\Phi_1\eta^{-24}]_{q^{-3/4}} = 10/12$.

The integrals $P_1(\tilde{t}_2)$ and $P_3(\tilde{t}_2)$ for
the argument of real interest,
$\tilde{t}_2 = (t^{a=1}_2) v_{2+} \in (\Lambda_S \otimes \R) \subset
(\widetilde{M} \otimes \R)$, is obtained by taking the limit
$u_2 \rightarrow t^{a=1}_2$, $\rho_2  \rightarrow t^{a=1}_2$ in their expressions in the chamber $\rho_2 \leq u_2 \leq 2\rho_2$.
Therefore,
\begin{align}
  P_1(t^{a=1}_2) & \; = \left[ \left(10-b_{\cal R} - \frac{7}{2}n_{1/2}
       + \frac{(248-18b_{\cal R}-29n_{1/2})}{6}\right) + \frac{4}{6}
        - \frac{10n_{1/2}}{6} \right](t^{a=1}_2), \nonumber \\
  & \; =  (52-4b_{\cal R} - 10n_{1/2}) (t^{a=1}_2), \\
  P_3(t^{a=1}_2) & \; = \left[ \left( \frac{20-4b_{\cal R}+2n_{1/2}}{4}
    + \frac{1-7n_{1/2}}{3}\right) + \frac{2(-8)}{12} - \frac{10}{12}n_{1/2}
    \right] \; (t^{a=1}_2)^3, \nonumber \\
   & \; = (4 -b_{\cal R} -n_{1/2}) (t^{a=1}_2)^3.
\end{align}
Those two results are used in section \ref{sssec:deg2-calc}.


%

\end{document}